\documentclass[final,5p,times,twocolumn,authoryear]{elsarticle}

\usepackage{amssymb}
\usepackage{amsmath}
\usepackage{CJKutf8}
\usepackage{lipsum}
\usepackage{subcaption}
\usepackage{placeins}
\usepackage{tabularx}
\usepackage{makecell}
\usepackage{xcolor}
\usepackage{hyperref}
\usepackage[T1]{fontenc}
\usepackage{newtxtext,newtxmath}

\usepackage[utf8]{inputenc} 

\usepackage{listings}
\usepackage{xcolor}

\colorlet{punct}{red!60!black}
\definecolor{background}{RGB}{245,245,245}
\definecolor{delim}{RGB}{20,105,176}
\colorlet{numb}{magenta!60!black}

\lstdefinelanguage{json}{
    basicstyle=\normalfont\ttfamily,
    numbers=left,
    numberstyle=\scriptsize,
    stepnumber=1,
    numbersep=8pt,
    showstringspaces=false,
    breaklines=true,
    frame=lines,
    backgroundcolor=\color{background},
    literate=
     *{0}{{{\color{numb}0}}}{1}
      {1}{{{\color{numb}1}}}{1}
      {2}{{{\color{numb}2}}}{1}
      {3}{{{\color{numb}3}}}{1}
      {4}{{{\color{numb}4}}}{1}
      {5}{{{\color{numb}5}}}{1}
      {6}{{{\color{numb}6}}}{1}
      {7}{{{\color{numb}7}}}{1}
      {8}{{{\color{numb}8}}}{1}
      {9}{{{\color{numb}9}}}{1}
      {:}{{{\color{punct}{:}}}}{1}
      {,}{{{\color{punct}{,}}}}{1}
      {\{}{{{\color{delim}{\{}}}}{1}
      {\}}{{{\color{delim}{\}}}}}{1}
      {[}{{{\color{delim}{[}}}}{1}
      {]}{{{\color{delim}{]}}}}{1},
}

\journal{arxiv}

\begin{document}


\begin{frontmatter}

	
	
	\title{KOKKAI DOC: An LLM-driven framework for scaling parliamentary representatives}
	
	
	\author[first]{Ken Kato}
	\affiliation[first]{organization={Divison of Engineering Science, University of Toronto}}
	\author[third]{Christopher Cochrane$^*$}
	\affiliation[third]{organization={Department of Political Science, University of Toronto Scarborough}}

	\begin{abstract}
	This paper introduces an LLM-driven framework designed to accurately scale the political issue stances of parliamentary representatives. By leveraging advanced natural language processing techniques and large language models, the proposed methodology refines and enhances previous approaches by addressing key challenges such as noisy speech data, manual bias in selecting political axes, and the lack of dynamic, diachronic analysis. The framework incorporates three major innovations: (1) de-noising parliamentary speeches via summarization to produce cleaner, more consistent opinion embeddings; (2) automatic extraction of axes of political controversy from legislators’ speech summaries; and (3) a diachronic analysis that tracks the evolution of party positions over time.
	
	We conduct quantitative and qualitative evaluations to verify our methodology. Quantitative evaluations demonstrate high correlation with expert predictions across various political topics, while qualitative analyses reveal meaningful associations between language patterns and political ideologies. This research aims to have an impact beyond the field of academia by making the results accessible by the public on teh web application: kokkaidoc.com. We are hoping that through our application, Japanese voters can gain a data-driven insight into the political landscape which aids them to make more nuanced voting decisions. 
	
	Overall, this work contributes to the growing body of research that applies LLMs in political science, offering a flexible and reliable framework for scaling political positions from parliamentary speeches. But also explores the practical applications of the research in the real world to have real world impact.

	\end{abstract}
	
	
	

\end{frontmatter}

\onecolumn
\thispagestyle{empty}

\clearpage

\tableofcontents
\newpage
\listoffigures
\newpage
\listoftables

\twocolumn
\pagenumbering{arabic}

\section{Introduction}
Quantifying and scaling the issue-specific stances of parliamentary representatives is essential for maintaining the transparency of democracy. It will give voters a more intuitive understanding of what the politicians are thinking without having to search through many lines of parliamentary speech records. Previously, political scientists have looked at data sources such as parliamentary voting data \citep{dw-nominate}, party manifestos \citep{CATALINAC_2018}, and more recently, parliamentary speeches \citep{Word-embeddings-for-analysis-of-ideological-placement}. Our method builds on these previous works and aims to provide a flexible and reliable way to scale the stances of parliamentary representatives using advanced NLP techniques and Large Language Models (LLMs).

The application of LLMs in political science has recently gained significant attention in the community (\cite{LLMs-and-political-science};\cite{kato2024lupinllmbasedpoliticalideology};\cite{llm-latent-position-of-politicians};\cite{li2024politicalllmlargelanguagemodels}). Due to the nature of the discipline relying predominantly on text data such as political speeches and legislative documents, it has been difficult for political scientists to gain insight from the vast amount of textual data available \citep{li2024politicalllmlargelanguagemodels}. However with the introduction of LLMs, political scientists have been empowered to process and analyze massive amounts of text data efficiently. In the context of scaling political ideologies, the works of  \citeauthor{kato2024lupinllmbasedpoliticalideology} and \citeauthor{llm-latent-position-of-politicians} apply LLMs to scale the political issue stance of parliamentary representatives but in different ways. 

The methods developed by \citeauthor{kato2024lupinllmbasedpoliticalideology} allow for a more flexible scaling of politicians on the axis of the researcher's choice and showed promising results in accurately estimating the issue-specific stance of representatives. However, it also fell short in the following aspects. 1) The speeches used to embed the issue positions of the representatives were still noisy and the parties still seemed too dispersed. 2) The manual selection of the axes of controversy by the researcher could introduce a layer of bias into the final estimation. 3) The analysis by \citeauthor{kato2024lupinllmbasedpoliticalideology} lacked any form of diachronic analysis which is essential to understand how a political system has evolved. Therefore, this research aims to extend the methods introduced by \citeauthor{kato2024lupinllmbasedpoliticalideology} in three major ways: 
\begin{enumerate}
    \item \textbf{De-noising through summarization}: In their work, \citeauthor{kato2024lupinllmbasedpoliticalideology} filters the raw political speeches collected from the Japanese Diet using a BERT-based classifier. In our method, we want to summarize the filtered opinion-based sentences using an LLM and then use the embedding of these summarizations. The idea is that this will de-noise the speeches further, allowing for a more accurate latent space representation of the legislator. 
    \item \textbf{Automatic extraction of the axis of controversy}: Previous methods by \citeauthor{kato2024lupinllmbasedpoliticalideology} and \citeauthor{llm-latent-position-of-politicians} required researchers to manually choose the axis of controversy they want to place the representatives on. In the work of \citeauthor{kato2024lupinllmbasedpoliticalideology} this was the acknowledgment of JSDF in the Japanese constitution, in the work of \citeauthor{llm-latent-position-of-politicians}, this was gun-control in the US. In this work, we would like to apply LLMs to parse through the speeches of representatives and extract possible axes of controversy from the corpora. 
    \item \textbf{Diachronic analysis of party positions on political axes}: While it is helpful to analyze the current ideological stance of representatives, examining how their position has evolved is another aspect political scientists have explored(\cite{Debating-Evil:-Using-Word-Embeddings-to-Analyze-Parliamentary-Debates}; \cite{Word-embeddings-for-analysis-of-ideological-placement}). In this research, we have explored how the ideological stance of parties have evolved over time on the extracted political topics. 
\end{enumerate}

Additionally, we validate our method by quantitatively comparing it with expert predictions of issue positions of parties and qualitatively evaluate the method by analyzing the common noun-phrases that are associated with different extracted clusters of politicians. We also conducted a diachronic analysis of the issue-specific positions of the parties to show how the stances of parties evolved over time and to demonstrate that our method is able to capture the evolution of the party positions on the political axes.
We believe that these extensions will allow for a more accurate and reliable scaling of political ideologies of parliamentary representatives with reduced human intervention and will provide a more comprehensive analysis of any political system where parliamentary speeches are available.

\section{Related Work}
In this section, we present a variety of related works ranging from earlier methods to quantitatively express political ideologies to modern methods using neural networks and LLMs. We also cover a range of works that analyze parliamentary speech data in general.

\subsection{Earlier methods: Roll-call analysis}

The Dynamically Weighted Nominal Three-Step Estimation (DW-NOMINATE) scores developed by \citeauthor{dw-nominate} remain the benchmark for comparing the spatial orientation of the legislator ideologies until today. It is an extension of the NOMINATE score \citep{spatial-model-for-legislative-roll-call} developed by the same authors a few years earlier, and they are based on the roll call votes of the representatives. Their algorithm aims to estimate the parameters of a utility function for each representative and expresses each representative as a point on the Euclidean space. Their work laid the foundation for the spatial expression of legislator ideologies and is a gold-standard reference for political scientists today.

Our method builds on this work by extending this sort of estimation beyond roll-call votes. Our method estimates the issue positions of parliamentary representatives based on their parliamentary speeches, which is more suitable for the context of Japanese politics where roll-call vote data is not as public as in the US. (Some votings happen based on stand up and sit down votes, which are not recorded in the same way as roll-call votes) Our method also allows political scientists to scale the representatives flexibly on the axes of their choice, which gives us a more granular, issue-specific analysis of the positions of the representatives. 


\subsection{Earlier methods: Word frequency based approach}
\subsubsection{Wordscores - \citealt{LAVER_BENOIT_GARRY_2003}}
One of the earlier methods utilizing text data to position political parties was developed by \citeauthor{LAVER_BENOIT_GARRY_2003}. Their method was one of the first to treat text data not as a discourse to be read, interpreted, and understood but as a collection of word data. Their method focused specifically on the frequency of the words in documents where the political positions were known (reference texts) and documents where political positions were unknown. (virgin texts) Their method used a 3-step process where: 1) They collected reference texts of which the political positions were known; 2) They generated word scores which represented the probability that we are reading a specific reference text given only that word; 3) Score the virgin texts using the word scores generated.

In their work, they show that the position estimations for British and Irish parties they obtain from the party manifestos are in agreement with previous position estimations and they export this methodology to the German political system as well. Furthermore, they apply the word scores they obtained from the reference party manifestos to legislative speeches to estimate the positions of individual Irish politicians as well as parties. 

\subsubsection{Wordfish - \citealt{wordfishing}}
Another early word-frequency-based approach is the wordfishing approach developed by \citeauthor{wordfishing} which is still widely used among political scientists to estimate policy positions from text data. It was created to overcome the limitations of the Wordscores approach such as 1) The need for reference texts that can be difficult to agree upon among political scientists; and 2) Its limited applicability to time-series estimation as the political lexicon is constantly changing. 
To overcome the aforementioned limitations of the Wordscore approach, the authors use a statistical model of word counts by assuming a Poisson distribution for word frequencies. This allows them to estimate the policy positions of text data without having to prepare anchor/reference texts. 

Their method has been used by various authors such as \citeauthor{sharks-minnows},\citeauthor{Curini_Hino_Osaka_2020} and \citeauthor{CATALINAC_2018} and has allowed political scientists to analyze policy positions from text data in various forms and languages.

However, recent works have pointed out the limitations of the word frequency-based approaches. For example, \citeauthor{Political-Text-Scaling-Meets-Computational-Semantics} point out that using frequency-based approaches for text analysis, semantically similar words can be treated as dissimilar words hence semantically similar sentences being treated as dissimilar sentences.

Our methodology overcomes this challenge by using LLMs to embed the parliamentary speech data. LLMs capture the semantic meaning of the text data and gives us a "richer" representation of the text data compared to the word frequency-based approaches.

\subsection{Embedding-based methods}
In the past, researchers have used embedding-based methods to analyze and spatially position legislative speeches. For example, \citeauthor{Debating-Evil:-Using-Word-Embeddings-to-Analyze-Parliamentary-Debates} used the Word2Vec algorithm developed by Google to analyze legislative speeches that deal with war criminals and Nazi collaborators in post-war Netherlands. They analyzed how the conversations evolved from focusing on the crimes these individuals had committed against the nation to the suffering of the victims during the period. They also analyzed how the conversations shifted from focusing on death penalties to focusing on imprisonment. In their work, \citeauthor{Debating-Evil:-Using-Word-Embeddings-to-Analyze-Parliamentary-Debates} come up with a workaround to use multiple word2vec models trained on different corpora to conduct a diachronic analysis between different periods. 

Another example of a text embedding-based approach to political positioning would the the work of \citeauthor{Political-Text-Scaling-Meets-Computational-Semantics}. In their work, they combined a symbolic text representation such as tf-idf or a semantic text representation such as word embeddings with an unsupervised, graph-based clustering algorithm such as PageRank or HFLP. Through this method, they challenge the traditional word frequency-based approaches discussed earlier and validate that their method can position political text better than the frequency-based methods in multiple languages.

\citeauthor{Embeddings-Based-Clustering-for-Target-Specific-Stances} use a different algorithm to embed political text. They use Google's Convolutional Neural Network (CNN) based multilingual text encoder to analyze the tweets of the Turkish public during the presidential and parliamentary election of 2018 regarding various topics. They undergo a multi-step process of manual labeling, label propagation, embedding, and clustering to reveal a significant degree of polarization of the public regarding topics such as Trump, Syria, and Erdogan. They also reveal the correlation across topics by looking at the mutual information metric between the clusters for different topics.

Instead of focusing on the embeddings of the text data such as legislative speech or party manifesto, \citeauthor{Word-embeddings-for-analysis-of-ideological-placement} focus on the embeddings of the indicator variables which represent the party affiliation of the text. They call this the party embeddings. In their work, they implement a shallow neural network consisting of one hidden layer that aims to predict the word at position \textit{t} given words appearing in a window $\Delta$ before and after the word. The model input consists of these surrounding words as well as the party indicator variable. After fitting the model to the parliamentary speech corpora, they got a model that is capable of embedding the party indicator variables in a meaningful manner. They then reduce the dimension of the embeddings using the PCA method and analyze what the dimensions semantically represent. They also verify that the political positioning they retrieved is aligned with the previous methods such as DW-NOMINATE and other expert surveys. Furthermore, they conduct an intriguing diachronic analysis of the US Senate as well as the British and Canadian parliaments and show how the ideological placement of the parties has evolved on different dimensions. 

While these embedding-based approaches proved to be powerful methods to capture a more nuanced representation of political entities, they have limitations. Most of the methods require a lot of manual labor in terms of data labeling, data cleaning, and model training. Our method aims to overcome these limitations by automating most of the steps and using pre-trained LLMs to embed the text data. Our method also allows for a more flexible scaling of the political entities on the axes of the researcher's choice.

\subsection{Use of LLMs}
At the time of writing this paper, we are only aware of a few works that explore the application of LLMs in political issue stance positioning. This section will introduce these works and their contributions. 
\subsubsection{Direct prompting methodologies}
\citeauthor{llm-latent-position-of-politicians} use a pair-wise comparison approach to position senators on ideological scales such as liberal-conservative, abortion, and gun control. In their method, they provide the GPT-3.5 model by OpenAI with two senators to compare and prompt it to give the senator who is more pro-choice, conservative, or supporting gun control. By repeating this match-up between senators, they can position the senators on a spectrum using the Bradley-Terry model. They show that while the \textbf{LaMP}(\textbf{La}nguage \textbf{M}odel \textbf{P}airwise comparison) scores obtained highly correlate with the DW-NOMINATE scores, the scores do not simply parrot the DW-NOMINATE scores. They argue that LaMP scores are more aligned with surveyed political scientists because LaMP scores do not solely rely on roll-call votes like DW-NOMINATE scores. instead, it takes into account how the politicians are depicted on mainstream media and other sources that were used to train the GPT-3.5 model. One major concern and limitation of this work is that this requires the LLM to "know" about the politicians that are being compared. As LLMs are not aware of the newest text content on the internet, there will be limitations on scaling politicians that have not served before or did not receive great media coverage.

In their work, \citeauthor{Scaling-Political-Texts-with-ChatGPT} use a prompt engineering approach to spatially position British parties and US senators on a left-right scale. They feed in British party manifestos or tweets from US senators into the GPT-4 model using the API service provided by OpenAI and task the LLM to score the given text from 0 to 100 on a left-right scale. They verify that the scores they obtained using this method correlate extensively with existing expert estimates as well as past research. The authors mention the benefit of their methodology over the work of \citeauthor{llm-latent-position-of-politicians} as their method does not require the LLM to "know" the political actors to estimate their political issue position as they are provided the tweets/manifestos of the political actors. They also attempt to make their method robust by averaging different iterations of prompting the LLMs to eliminate any variability due to the randomness in prompt responses.  

\subsubsection{Representation-based methodologies}
So far, the works introduced have mainly focused on one source of text data such as tweets, parliamentary speeches, or manifestos \citeauthor{Understanding-Politics-via-Contextualized}. extends this approach by suggesting a framework to combine political text from multiple sources. They suggest a unique neural architecture, "Compositional Reader", capable of processing multiple sources of text in one shot and successfully training it on two major tasks Authorship prediction and Referenced Entity Prediction. They show that this architecture can outperform the previous BERT-based architectures. They also show that the representations of politicians learned by the model are capable of capturing a greater depth of political context compared to previous methodologies as their representations do not simply capture the linguistic context but a broader political context. Although this method guarantees a more contextualized representation of political ideologies, one bottleneck with this method is that it requires a lot of investment in terms of computational resources and data collection, making the methodology less accessible to political science researchers who might not have the computational resource/know-how to collect the data from the wide variety of data sources.

The work by \citeauthor{kato2024lupinllmbasedpoliticalideology} can be seen as a hybrid approach to the above-outlined methods. It uses GPT4o-mini to generate the reference speeches that are used to create an axis to project the legislators onto but does not directly use GPT4o-mini to scale the representatives like \citeauthor{Scaling-Political-Texts-with-ChatGPT} and \citeauthor{llm-latent-position-of-politicians} did. This way, the LLM does not have to "know" the representatives, which was an issue with \citeauthor{llm-latent-position-of-politicians} and it does not ask for numerical outputs from the LLM which is known to be vulnerable towards numerical reasoning tasks. Their method "de-noises" parliamentary speeches through a fine-tuned BERT-based classifier and uses the embeddings of the filtered speeches to create a representation of the legislators. These representations are projected onto an axis composed of two opposing political views which are either generated political speeches or hand-picked political figures. Using the latent representations of the legislators can be seen as a similar approach to the work of \citeauthor{Debating-Evil:-Using-Word-Embeddings-to-Analyze-Parliamentary-Debates} or \citeauthor{Word-embeddings-for-analysis-of-ideological-placement} but this method does not require training of their model and hence can be seen as a more light-weight approach. They show that their method correctly places the different party members on the political spectrum although we observe some dispersion in the parties, indicating room for improvement in their methodology. This paper aims to build on this work by improving the accuracy of the ideological scaling as well as introducing a diachronic analysis of the party positions on the political axes.

\section{Parliamentary speech data}
\label{section:parliamentary speech data}

For this research, we extended the data collection pipeline developed by our previous work\citep{kato2024lupinllmbasedpoliticalideology} and collected speeches for representatives in the House of Representatives and the House of Councilors. These representatives were serving as of December 18th, 2024. Furthermore, to conduct a diachronic analysis of the issue positions of the parties, we have scraped representatives who served at any point after the year 2000.\footnote{Data source for historical House of Representatives \url{https://kokkai.sugawarataku.net/giin/rgiin.html}}\footnote{Data source for historical House of Councilors: \url{https://www.sangiin.go.jp/japanese/san60/giin/index.htm}} Similar to the previous work, we have created a list of query words we have used to query the Japanese parliamentary debate record API\citep{kokkai_diet_api}. See table \ref{tab:English-query} for the English translation of the query words used. These keywords were generated by GPT4o-mini and confirmed by us. An example prompt to generate the query words is available in appendix \ref{fig:parliamentary_api_example}. 

\begin{table}[htbp] 
\centering
\begin{tabularx}{.45\textwidth}{ c|X } 
\hline
Topic & Query Words \\  
\hline
Defence & Defense, Self-Defense Forces, Security, Arms, Military, Collective Self-Defense, Exclusive Defense, Defense Spending, Bases, National Defense \\ 
\hline
Nuclear Power &  Power plant, Nuclear power generation, Restart, Decommissioning, Nuclear fuel, Radioactivity, Radioactive waste, Energy policy, Base load power source, Fukushima Daiichi Nuclear Power Plant \\
\hline
Aging population & declining birthrate, fertility rate, childcare support, childcare leave, children on waiting lists, population decline, daycare centers, child allowance , family policy , childcare burden \\
\hline
Climate Change & Climate Change, Global Warming , Carbon Dioxide, Decarbonization, Carbon Neutral, Renewable Energy , Greenhouse Gas, Carbon Zero by 2050, Environmental Policy, Energy Transition\\
\hline
Economic policy & Economy, Business climate, GDP, Deflation, Inflation, Fiscal policy, Monetary policy,Employment , Income, Business investment, Consumption, Export/Import, Small business, Start-up, Minimum wage\\
\hline
\end{tabularx}
\caption{English translations of query words}\label{tab:English-query}
\end{table}

We have collected the speeches for the topics of defense, the aging population, nuclear power plants, climate change, and economic policies. See table \ref{tab:stats} for the data statistics. 

\begin{table}[htbp] 
\centering
\begin{tabularx}{.45\textwidth}{ X|X|X } 
\hline
House & Representatives / Party & Speeches / Speech segments\\  
\hline
Current Representatives & 465 members / 11 parties & 63633 speeches / 90575 extracted segments \\ 
\hline
Current Councilors & 240 members / 10 parties  & 30922 speeches / 44223 extracted segments \\
\hline
Total (including historical) & 1885 members & 207750 speeches / 304444 extracted segments\\
\hline
\end{tabularx}
\caption{Statistics on collected speech data}
\label{tab:stats}
\end{table}

Notice in table \ref{tab:stats} that we are distinguishing between \textbf{speeches} and \textbf{speech segments}. We define speech as an instance of a representative speaking in the parliament and speech segment to be period(\begin{CJK}{UTF8}{min}。\end{CJK}in Japanese) delimited sentence inside of a speech. We are distinguishing between them because using a fine-tuned BERT model based on cl-tohoku/bert-base-japanese-v3 \citep{Tohoku}, we are classifying each speech segment into opinion-based sentences, fact-based sentences, questions, descriptive sentences, and others and extracting only the opinion-based sentences. This model is made available on huggingface. \footnote{Huggingface model: \url{https://huggingface.co/kkatodus/jp-speech-classifier}} 

We used a hand-labeled dataset comprising 1,439 speech segments to fine-tune this classifier. \footnote{Code to create and fine-tune classifier available:\url{https://github.com/kkatodus/AI_projects/tree/main/jp_speech_categorization}}The number of extracted speech segments indicates how many were classified as opinion-based sentences. Also, note that we are not keeping instances of speeches that did not contain opinion-based sentences; therefore, we have more extracted segments than instances of speeches. You can see an example of an extracted opinion-based sentence in figure \ref{fig:extracted opinion}.
\begin{figure}

\begin{lstlisting}[language=json,firstnumber=1]
[
{
"speech_id": "121215254X0032...",
"house_name": "House of Councillors",
"meeting_name": "plenary session",
"date": "2023-10-25",
"speech_url": "https://kokkai.ndl.go.jp/txt/(...)",
"speaker_group": "LDP",
"extracted_opinions": [
	"The baton has been passed from the Abe and Kan administrations, which took a ..."
]
},
{
"speech_id": "121015254X00220...",
"house_name": "House of Councillors",
"meeting_name": "plenary session",
"date": "2022-10-06",
"speech_url": "https://kokkai.ndl.go.jp/txt/(...)",
"speaker_group": "LDP",
"extracted_opinions": [
	"Especially in Asia, Japan should take the lead as the standard-bearer of freedom and democracy, showing firmness in both security and economy."
]
},
]
\end{lstlisting}
\caption{Example of an extracted opinion-based speech segment(Translated to English)}
\label{fig:extracted opinion}
\end{figure}

\begin{figure}[h]
\centering
  \centering
  \includegraphics[width=1\linewidth]{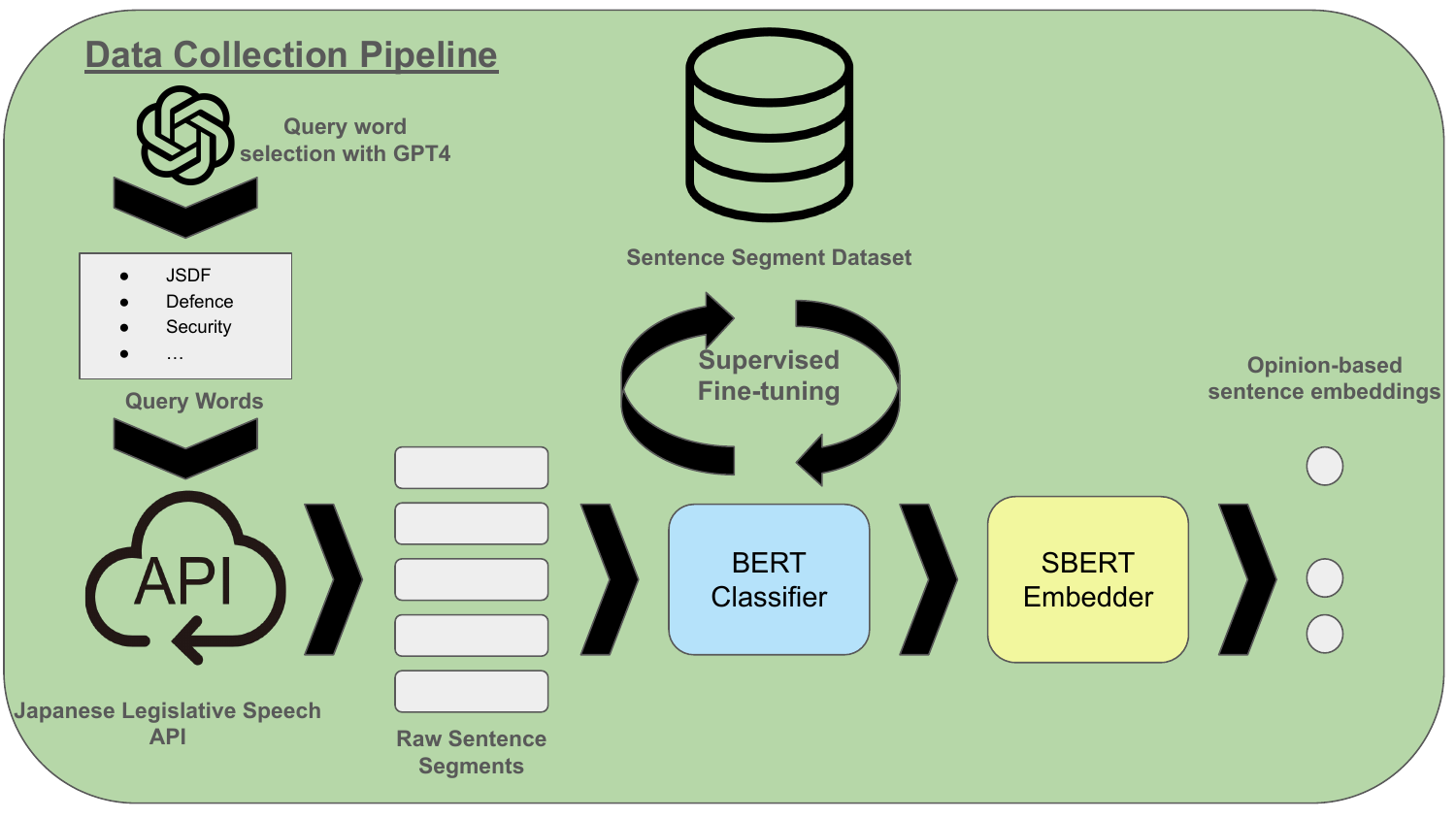}
  \caption{Graphical representation of the data collection and embedding pipeline}
  \label{fig:data-collection-phase}
\end{figure}

In this research, the extracted speech segments are summarized using GPT4o-mini and then embedded using the previous BERT model using Sentence BERT(SBERT) \citep{reimers2019sentencebert}. See the full data collection and embedding pipeline in figure \ref{fig:data-collection-phase}.

\section{Methodology}

Our method consists of 3 major aspects that build on the work of \citeauthor{kato2024lupinllmbasedpoliticalideology}. Those are:
\begin{enumerate}
    \item \textbf{De-noising through summarization}:  
    \item \textbf{Automatic extraction of the axis of controversy}
    \item \textbf{Diachronic analysis of party positions on political axes}
\end{enumerate}

This section will explain the previous method by \citeauthor{kato2024lupinllmbasedpoliticalideology} more in-depth and then move on to describe the motives behind the new methods as well as the technical details. 

\subsection{L(u)PIN method by \citeauthor{kato2024lupinllmbasedpoliticalideology}}
\label{section: lupin method}
The previous method comprises two key aspects:
\begin{enumerate}
    \item Filtration of parliamentary speech data
    \item Selection of reference points to form the axis of controversy and projection of representatives onto the axis
\end{enumerate}

\begin{figure}[h]
\centering
  \centering
  \includegraphics[width=1\linewidth]{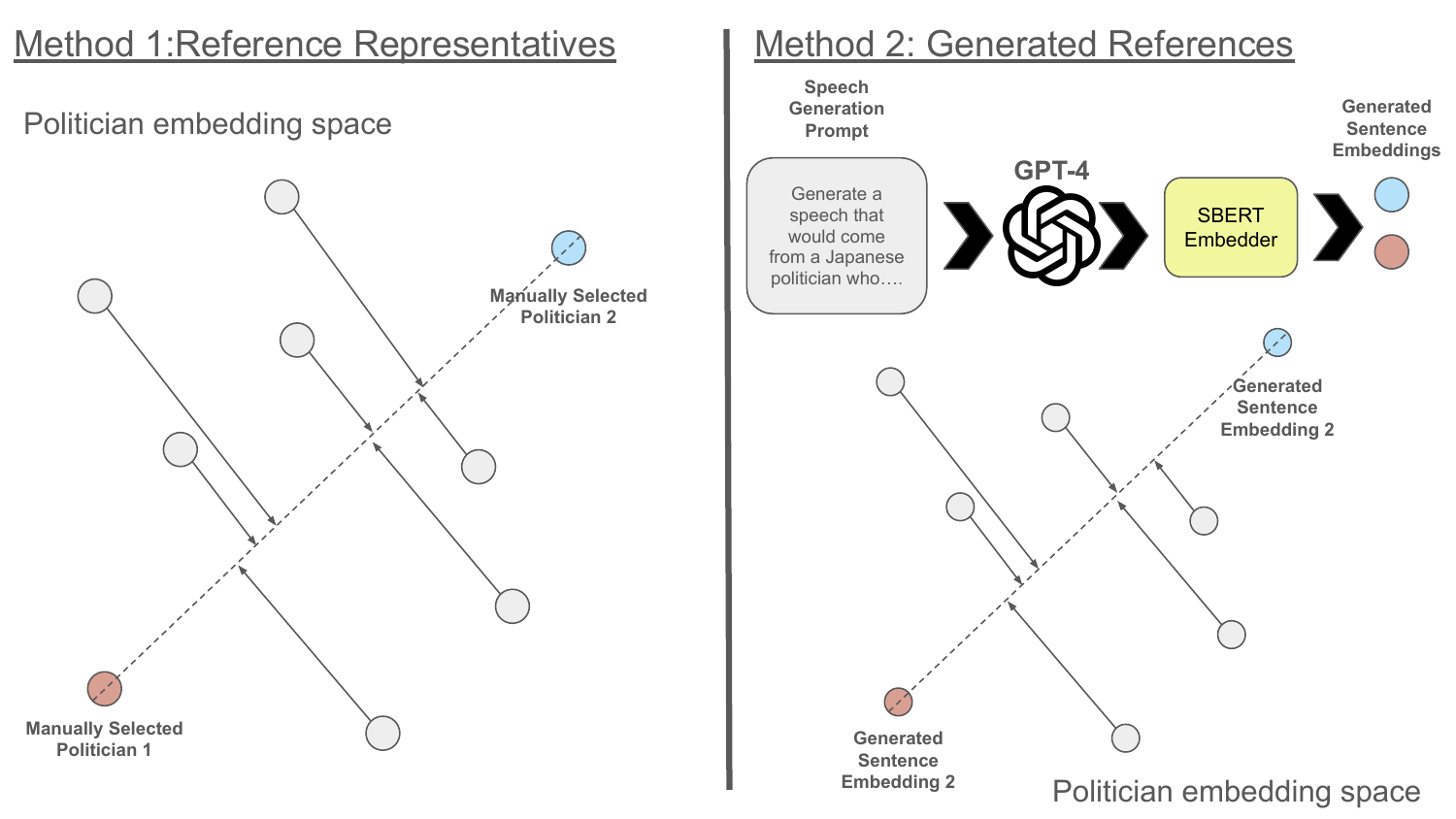}
  \caption{Graphical representation of the L(u)PIN method \citep{kato2024lupinllmbasedpoliticalideology}}
  \label{fig:lupin-projection}
\end{figure}

The first step is already mentioned in the previous section \ref{section:parliamentary speech data} so this section will focus more on the second step. The purpose of the second step is to express the ideological stance of a politician on a political issue using a scalar value. How can we achieve this reliably? We can either, 1) Select two politicians who have opposing views on the topic and measure everyone else relative to these two "anchors" or 2) Generate speeches that would come from politicians with opposing views on the political topic of interest and measure everyone else relative to these points. Figure \ref{fig:lupin-projection} compares and contrasts these two approaches. The main issue with the first approach is that it requires a lot of manual labor by the researcher and also a deep understanding of the political dynamics of the country to select the appropriate "anchor" representatives. This also becomes a huge risk in introducing bias from the researchers. Therefore we believe that generation of speeches and measuring the politicians relative to these points is a more reliable way to scale politicians and hence this research will extend the generative approach of this previous work.

\subsection{De-noising through summarization}
\label{section: denoising}
\begin{figure}[h]
\centering
  \centering
  \includegraphics[width=1\linewidth]{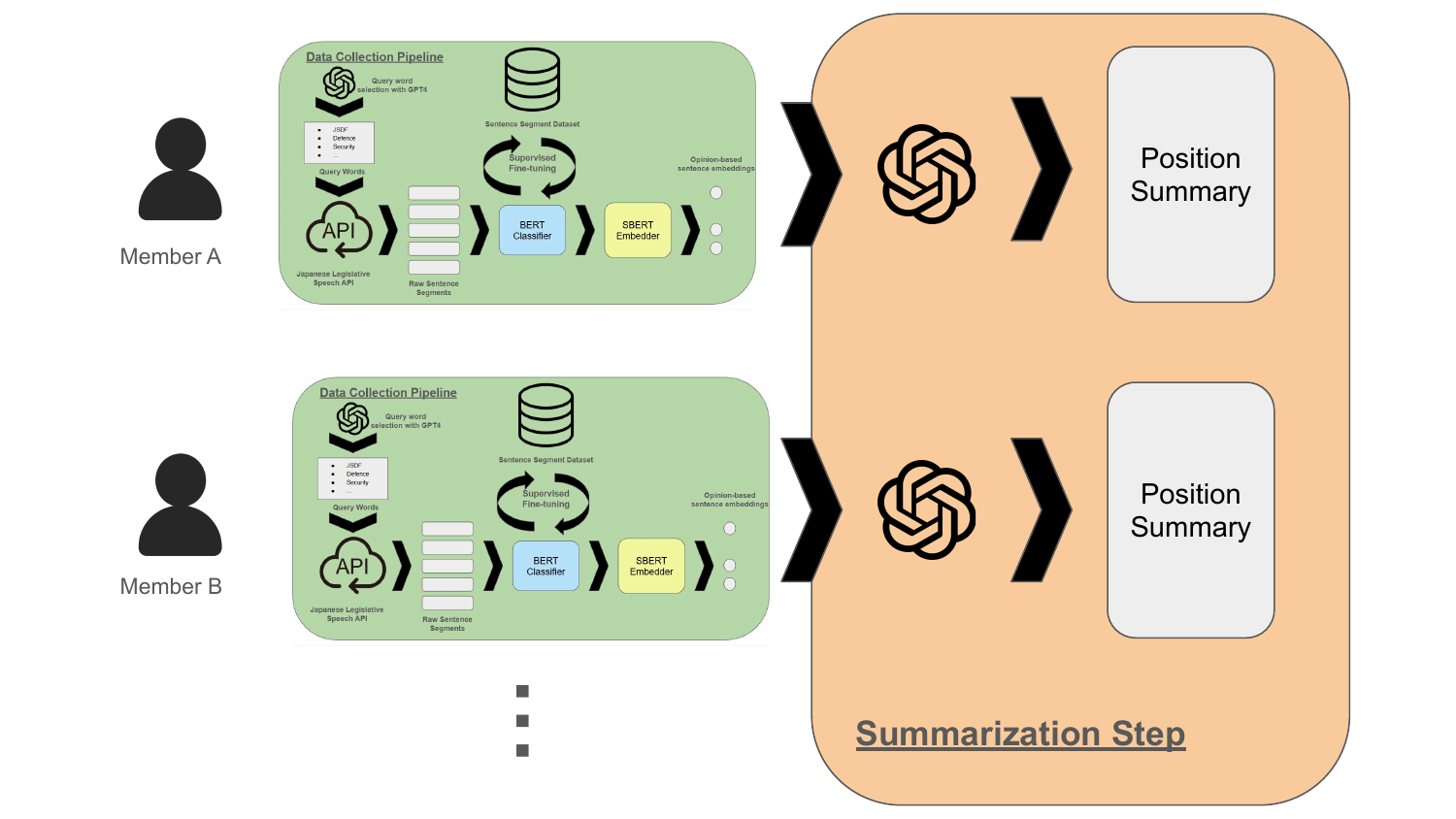}
  \caption{Graphical representation of summarization step}
  \label{fig:data-summarization}
\end{figure}

In the work by \citeauthor{kato2024lupinllmbasedpoliticalideology}, the results were still scattered despite filtration on the speech segments used to create the ideological embedding for the representatives. This indicated that noise was still introduced through tone, word choice, etc. See the results obtained through the previous method by \citeauthor{kato2024lupinllmbasedpoliticalideology} in appendix \ref{fig: Previous box plots}. To tackle this, we have introduced a summarization step by GPT4o-mini to make the sentences more consistent and comparable. See a visual representation of this step in figure \ref{fig:data-summarization}. 

To select the appropriate prompt to summarize the opinion-based speech segments, we empirically analyzed how different prompting styles affect the embeddings of the final summarization. See table \ref{tab:summary prompt styles} for the different prompting styles compared. We took a subset of our collected data and summarized it with the different prompting styles, embedded them using the SBERT model mentioned previously, and then used UMAP\citep{mcinnes2018umap-software} to reduce the dimensionality of the embeddings. 

\begin{table}[htbp] 
\centering

\begin{tabularx}{.45\textwidth}{ p{1.7cm}|X } 
\hline
Base prompt & Please infer and summarize this politician's stance on Japan's defense based on the politician's statements in the following Diet proceedings. 
\newline

$\langle$ Statements $\rangle$ \\\\  
\hline
With Context & Instructions:
\newline
Based on the politician's statements in the following Diet proceedings, infer and summarize this politician's stance on Japan's defense. 
\newline

Overview:
\newline
This summary is intended for those who have little knowledge of Japanese politics. Please summarize in a way that is easy to understand even for those who are not interested in politics.
'''''''''''''''''''''''''''''''''''''''''''''''
\newline
$\langle$ Statements $\rangle$ \\ \\ 
\hline
With Context and Few shot  & Instructions:
\newline
Based on the politician's statements in the following Diet proceedings, infer and summarize this politician's stance on Japan's defense. 
\newline
Overview:
\newline
This summary is intended for those who have little knowledge of Japanese politics. Please summarize in a way that is easy to understand even for those who are not interested in politics.
'''''''''''''''''''''''''''''''''''''''''''''''
\newline
Example 1:
\newline
$\langle$ Example Speech $\rangle$
\newline
Summary Example:
\newline
$\langle$ Summary Example $\rangle$
\newline
Example 2:
\newline
$\langle$ Example Speech $\rangle$
\newline
Sample summary:
\newline
$\langle$ Summary Example $\rangle$
\newline
'''''''''''''''''''''''''''''''''''''''''''''''
\newline
$\langle$ Statements $\rangle$ \\
\hline
\end{tabularx}
\caption{Comparison of different prompting styles for summarizing opinion-based speech segments(Translated)}
\label{tab:summary prompt styles}
\end{table}

\begin{figure}[h]
\centering
\begin{subfigure}{0.22\textwidth}
  \centering
  \includegraphics[width=1\linewidth]{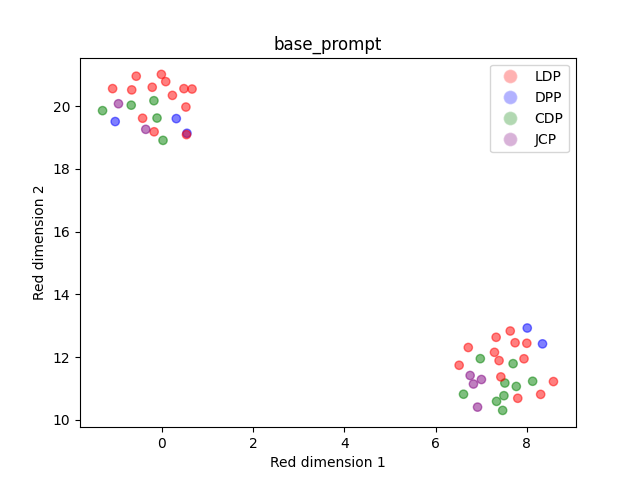}
  \caption{Base prompt: UMAP plot of summaries}
\end{subfigure}
\begin{subfigure}{0.22\textwidth}
  \centering
  \includegraphics[width=1\linewidth]{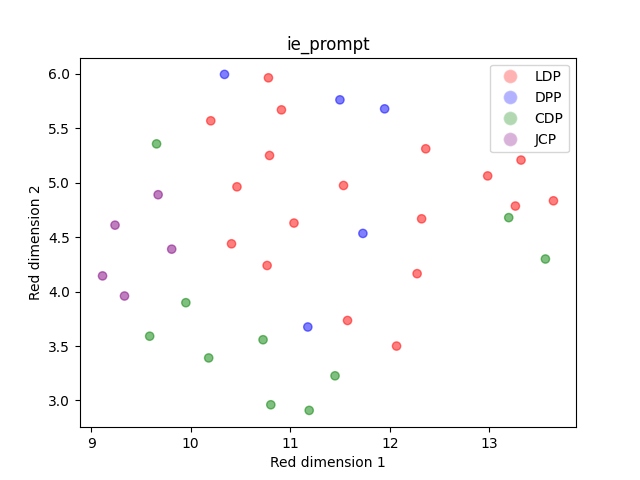}
  \caption{With Context: UMAP plot of summaries}
\end{subfigure}
\begin{subfigure}{0.22\textwidth}
  \centering
  \includegraphics[width=1\linewidth]{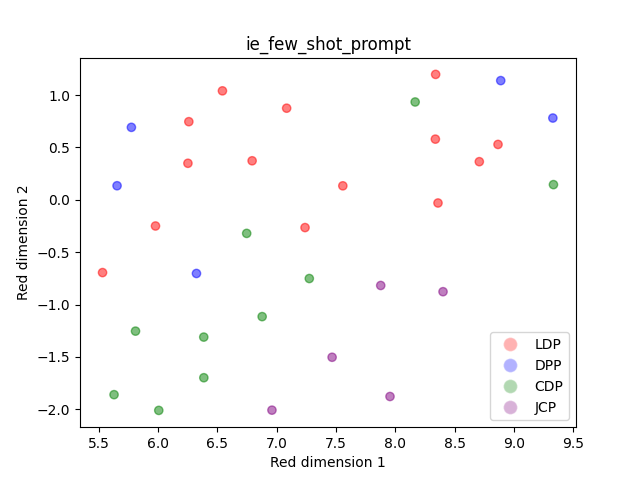}
  \caption{With Context and Few shot: UMAP plot of summaries}
\end{subfigure}
\caption{UMAP plot of summaries generated by different prompting styles. The colors represent the party affiliation of the politicians. The axes are the reduced dimensions of the SBERT embeddings}
\label{fig:UMAP prompt comparison}
\end{figure}

As you can see in figure \ref{fig:UMAP prompt comparison}, with the base prompt, we have introduced a noise in some form as we now have two clusters that seemingly are unrelated to party affiliation. After some investigation, we noticed that these two clusters were formed because of the inconsistent formatting of the reply by GPT4o-mini. One cluster represented a more formatted/bulleted reply while the other cluster represented a more essay-style reply. See in appendix \ref{fig:formatted reply GPT4o-mini summary} and \ref{fig:essay reply GPT4o-mini summary} for a comparison in reply formats. Therefore it was important for us to create a prompt that would provide a consistent reply format to make sure we do not add additional noise to the embeddings rather than de-noising the speeches. We found that our context + few shot prompt provided the most consistent reply format and hence we decided to use this prompt for the summarization step.

\subsection{Automatic extraction of the axis of controversy}
\label{section: extraction of axis of controversy}

\begin{figure}[h]
\centering
  \centering
  \includegraphics[width=1\linewidth]{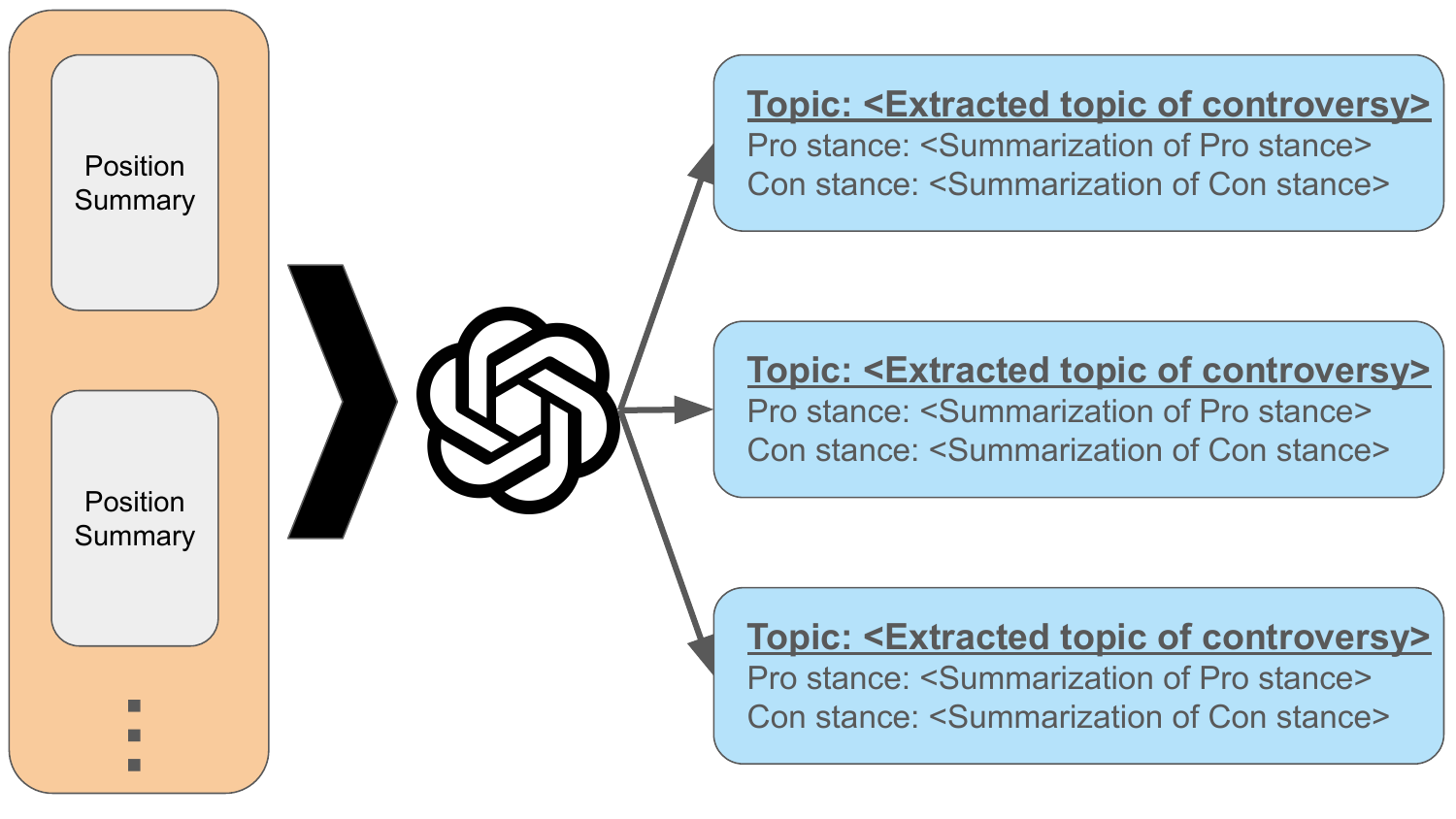}
  \caption{Graphical representation of axis extraction using GPT4o-mini}
  \label{fig:axis-extraction}
\end{figure}

One key bottleneck to scaling the method of \citeauthor{kato2024lupinllmbasedpoliticalideology} is the manual selection of the axis of political controversy. In their paper, they described a process of axis creation where the researchers selected two politicians that are known to be opposite ends of a political debate (e.g. they selected Tomomi Inada and Akira Kasai to evaluate the other politicians on the pro/con axis for the acknowledgment of JSDF in the Japanese constitution.) or generated speeches that would come from politicians on both ends of the spectrum using GPT4o-mini. Although effective, this method requires extensive human intervention, making it resource-intensive and susceptible to biases. Furthermore, we will miss many interesting debates or controversies which the researchers are unaware of. We therefore use GPT4o-mini to extract the "axis" of political debate from the stance summaries we generated in the previous step. See figure \ref{fig:axis-extraction} for a visual representation of this step. The summarized stances of politicians are fed into GPT4o-mini with the prompt shown in figure \ref{fig:axes detection} to generate axes within the broad topic of interest. 

Using this method, we were able to extract axes of political controversy for each of the topics of defense, aging population, climate change, and economic policy. In figure \ref{fig: raw axes extraction} you can see the raw reply from GPT4o-mini to extract the axes of controversy and see figure \ref{fig: extracted axes defense},\ref{fig: extracted axes economy},\ref{fig: extracted axes nuclear power}, \ref{fig: extracted axes aging population},  \ref{fig: extracted axes climate change} for the extracted axes from each topic processed to JSON format. Note that while we did scale the politicians on all of the extracted axes, some of the extracted axes were not suitable for scaling as they were very vague or did not directly relate to the topic of interest. Therefore in section \ref{section:results}, we will only show the results where we decided that the extracted axes were appropriate.

\begin{figure}[htbp]
    \centering
    \begin{quote}
The following is a summary text extracted from the proceedings of the Diet, identifying the issues and describing the opinions in favor of and against each.

\textbf{Issue}: The Self-Defense Forces should be clearly stated in the Constitution.

\textbf{For}: The existence of the Self-Defense Forces is already accepted by the public, and its inclusion in the Constitution will strengthen international credibility and legal stability.

\textbf{Oppose}: The SDF do not need to be explicitly stated in the Constitution, and the existing legal system is sufficient.
Specifying the SDF in the Constitution may create legal challenges and undermine the spirit of the exclusive defense of Japan.

\textbf{Issue}: Increasing the defense budget

\textbf{For}: Japan needs to strengthen its defense capabilities in order to respond to changes in the security environment surrounding Japan. Specifically, it is necessary to introduce advanced technologies such as missile defense and cyber security.

\textbf{Oppose}: Increased defense spending could put pressure on social security and education budgets. In addition, it is difficult to gain public understanding to allocate reconstruction funds and tax hikes to defense spending.

\textbf{Issue}: Exercise of the right of collective self-defense

\textbf{For}: In order to respond to changes in the international security environment, it is necessary to modernize defense policy by clearly stipulating the exercise of the right of collective self-defense in the Constitution.

\textbf{Oppose}: The exercise of the right of collective self-defense deviates from the principle of exclusive defense, and constitutional amendments and changes in interpretation should be pursued with caution. It could increase the risk of war.

\textbf{Issue}: Strengthening food security

\textbf{Yes}: Increasing food self-sufficiency and strengthening the agricultural production base is important for national security. It is necessary to increase domestic production and stabilize people's lives.

\textbf{Oppose}: There is a fear that agricultural support will be put on the back burner due to the expansion of the defense budget. In addition, improving self-sufficiency requires not only state initiative, but also private-sector strength.
       
    \end{quote}
    \caption{Raw reply from GPT4o-mini for axes extraction prompt for the topic of defence(Translated)}
    \label{fig: raw axes extraction}
\end{figure}

These extracted axes are used to generate reference summaries from a politician with an ideological stance of interest. Referring to the work of \citeauthor{kato2024lupinllmbasedpoliticalideology}, they show that generating sentences that would come from a politician with a specific ideological stance and using the embedding of such speeches is an effective method of scaling politicians on an axis of our choice. In figure \ref{fig: Previous umap}, we show a figure from the paper where they show that selecting representatives that are for/against the acknowledgment of the JSDF in the Japanese constitution represented in the latent space is comparable to generating speeches that would come from such representatives using LLM. This is shown qualitatively by comparing the lines drawn between the reference politicians and generated reference speeches and confirming that they are parallel in the latent space.

\subsection{Diachronic analysis of party positions on political axes}
\label{section: diachronic analysis}
\begin{figure}[h]
\centering
  \centering
  \includegraphics[width=1\linewidth]{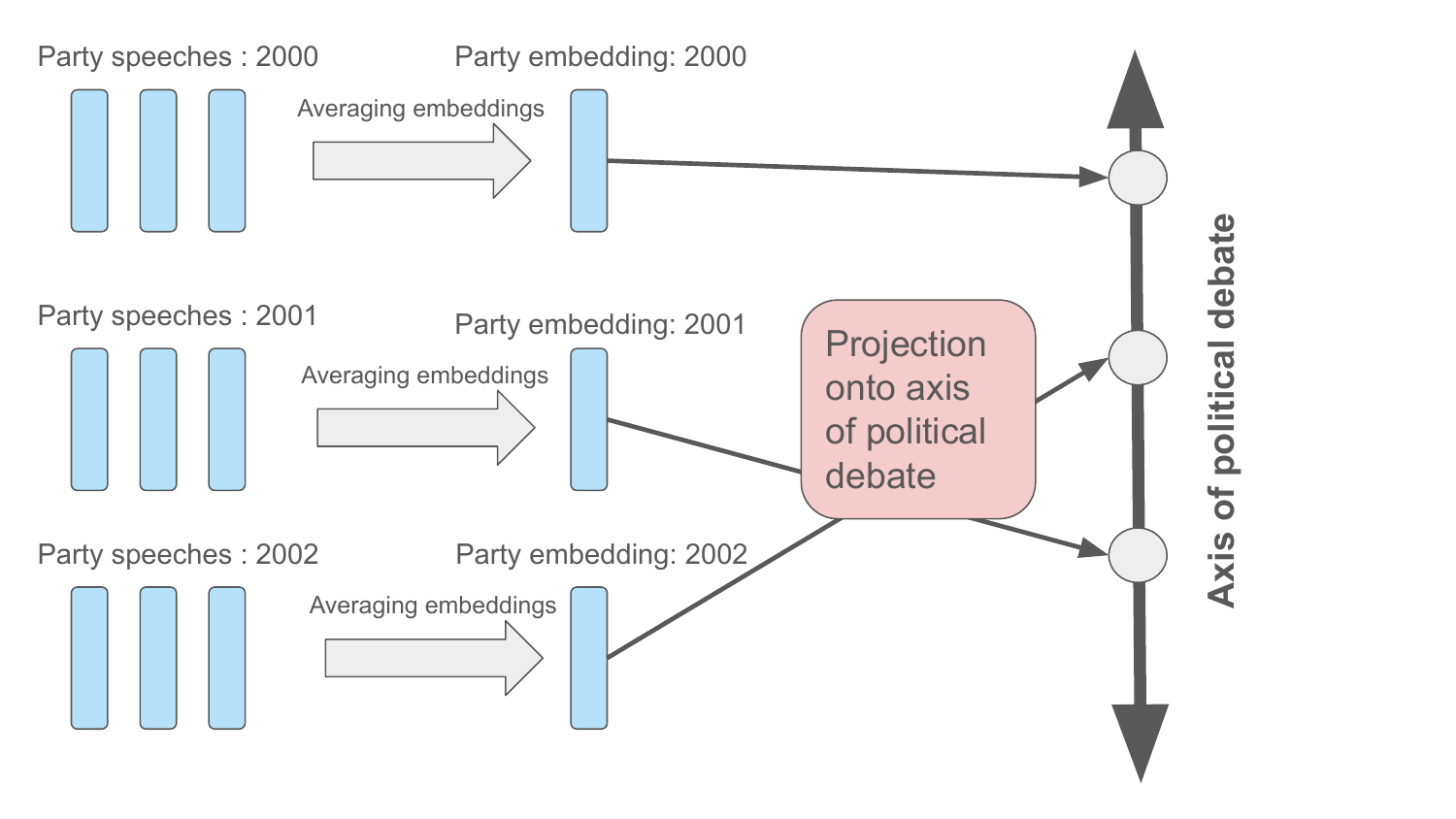}
  \caption{Graphical representation of diachronic analysis of party positions}
  \label{fig:diachronic analysis}
\end{figure}

\begin{figure}[h]
	\centering
	  \centering
	  \includegraphics[width=1\linewidth]{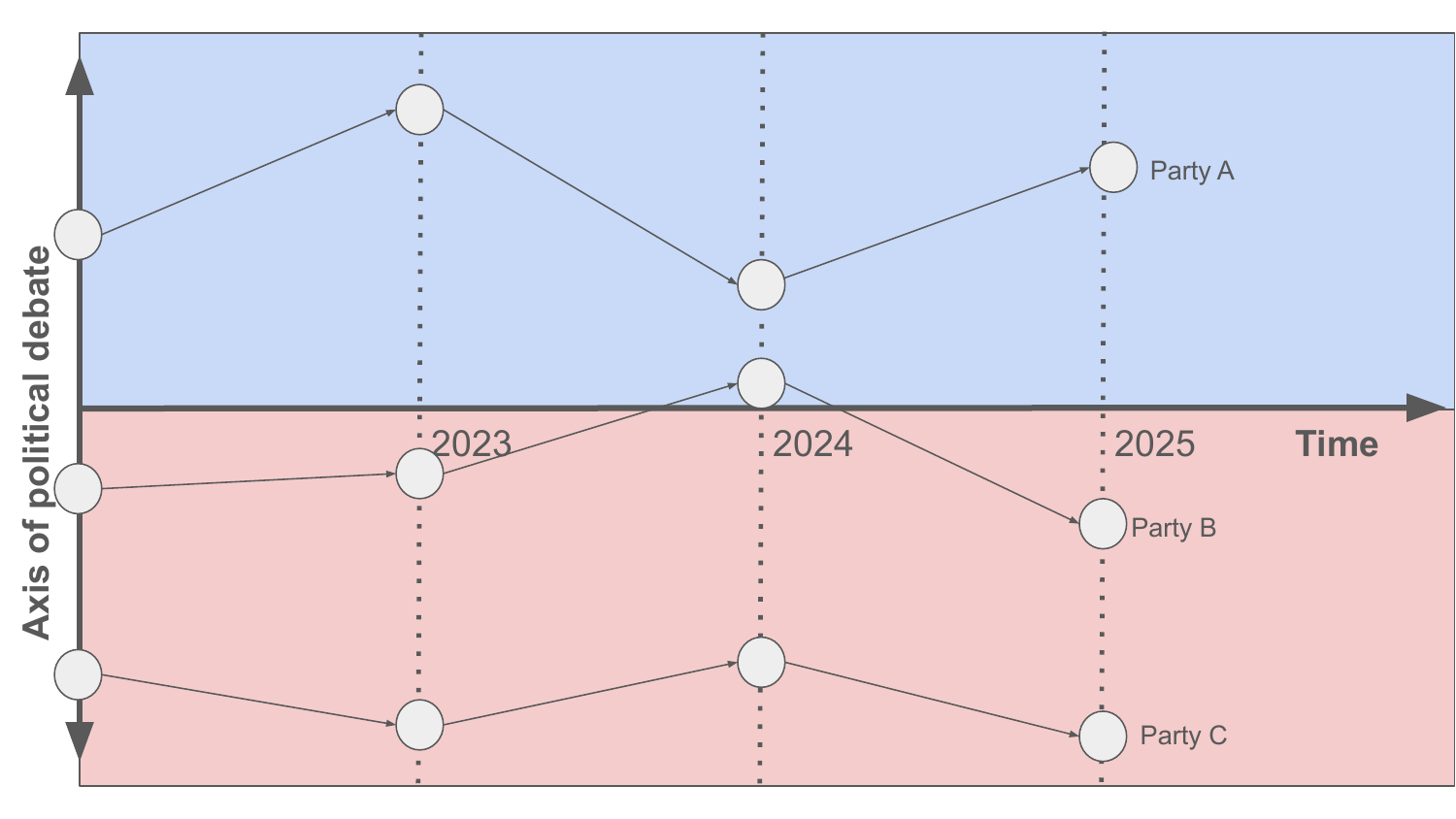}
	  \caption{Graphical representation of the output of diachronic analysis}
	  \label{fig:diachronic}
\end{figure}

The final aspect of our methodology is to conduct a diachronic analysis of party positions on political axes. This is inspired by the work of \citeauthor{Word-embeddings-for-analysis-of-ideological-placement} where they show how the ideological placement of the parties has evolved over time in the US Senate as well as the British and Canadian parliaments. We will use the scaling methodology of \citeauthor{kato2024lupinllmbasedpoliticalideology} to scale the party positions for the years between 2000-2024 by creating a latent space representation of the parties by taking the average of all the embeddings of all the opinion-based sentences of the party members for different years. We will then project these latent representations onto the axes of controversy extracted in the previous step. A visual representation of this analysis is shown in figure \ref{fig:diachronic analysis} and \ref{fig:diachronic}. Note that we have conducted this analysis on parties that were continuously active between 2000 and 2024. These are the Liberal Democratic Party(LDP), Japanese Communist Party(JCP) and Komeito. 

\subsection{Computation and costs}
We spent 103.39 USD on OpenAI API usage and the entire research was conducted on a single aorus 15p laptop with an intel i7-11800H CPU and an NVIDIA GeForce RTX 3070 Laptop GPU. 

\FloatBarrier

\section{Results}
\label{section:results}
In this section, we will discuss the results obtained through implementing the aforementioned methodology of scaling and analyzing the political stance of parliamentary representatives. This section comprises two parts where in the first part we will show the results of scaling the current serving representatives on the axes of controversy extracted in section \ref{section: extraction of axis of controversy} and in the second part we will show the results of the diachronic analysis of party positions on the political axes in the period of 2000-2024.

\subsection{Scaling the representatives}
\subsubsection{Topic: Defence}
\begin{figure}[h]
\centering
    \begin{subfigure}{0.22\textwidth}
      \centering
      \includegraphics[width=1\linewidth]{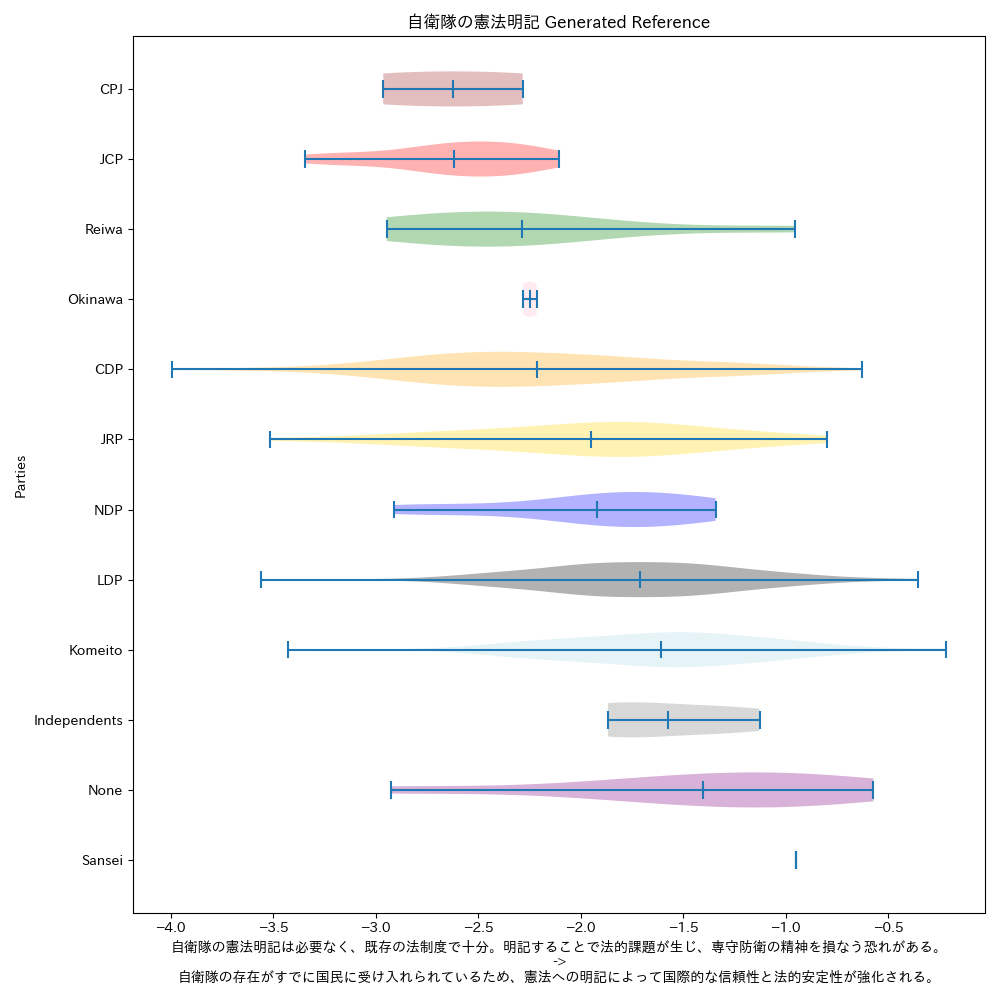}
      \caption{Violin plot of scaled representatives}
    \end{subfigure}
    \begin{subfigure}{0.22\textwidth}
      \centering
      \includegraphics[width=1\linewidth]{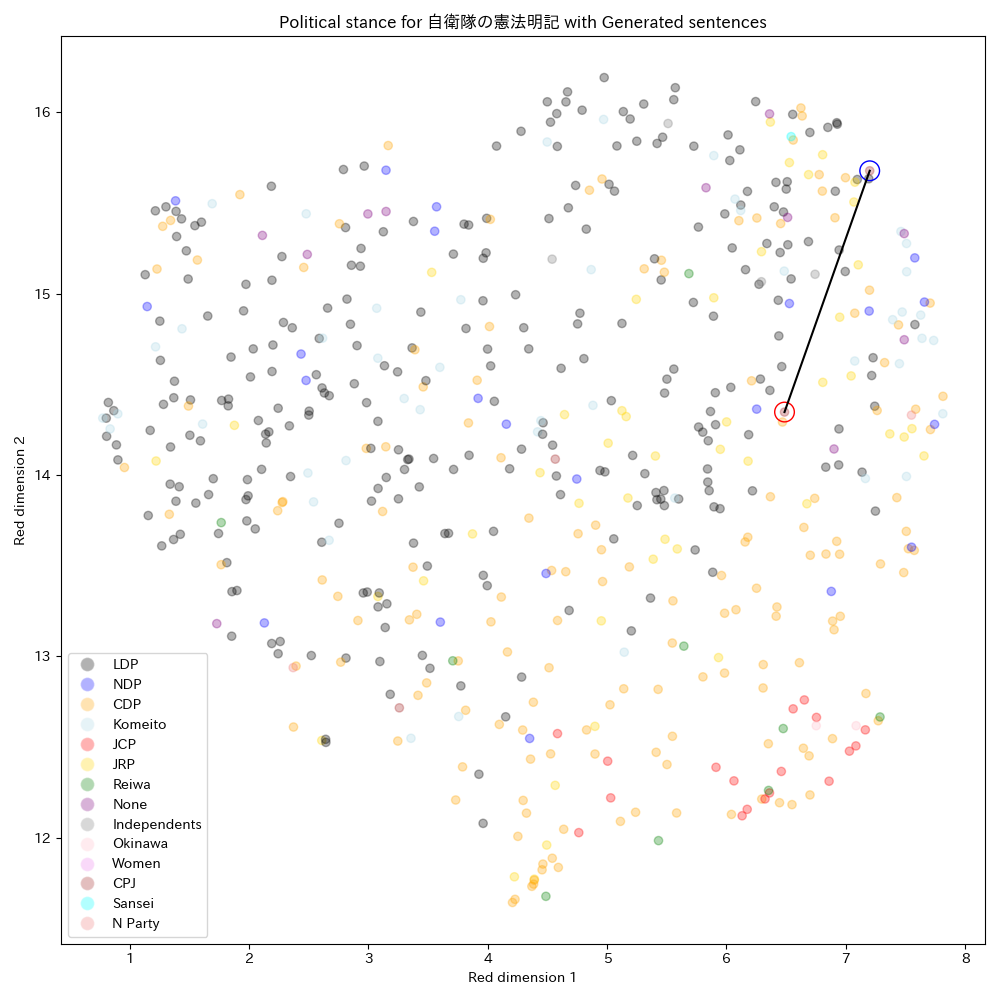}
      \caption{UMAP visualization of representatives and axes of reference}
    \end{subfigure}
\caption{Figures relating to the topic of whether the JSDF should be explicitly stated in the constitution.
Left figure: Violin plot of scaled representatives
Right figure: UMAP visualization of representatives and axes of reference. Representatives who have a pro stance located on the upper side of the UMAP visualization and representatives who have an anti stance located on the lower side.}
\label{fig: results-defence-constitution}
\end{figure}

\begin{figure}[h]
\centering
    \begin{subfigure}{0.22\textwidth}
      \centering
      \includegraphics[width=1\linewidth]{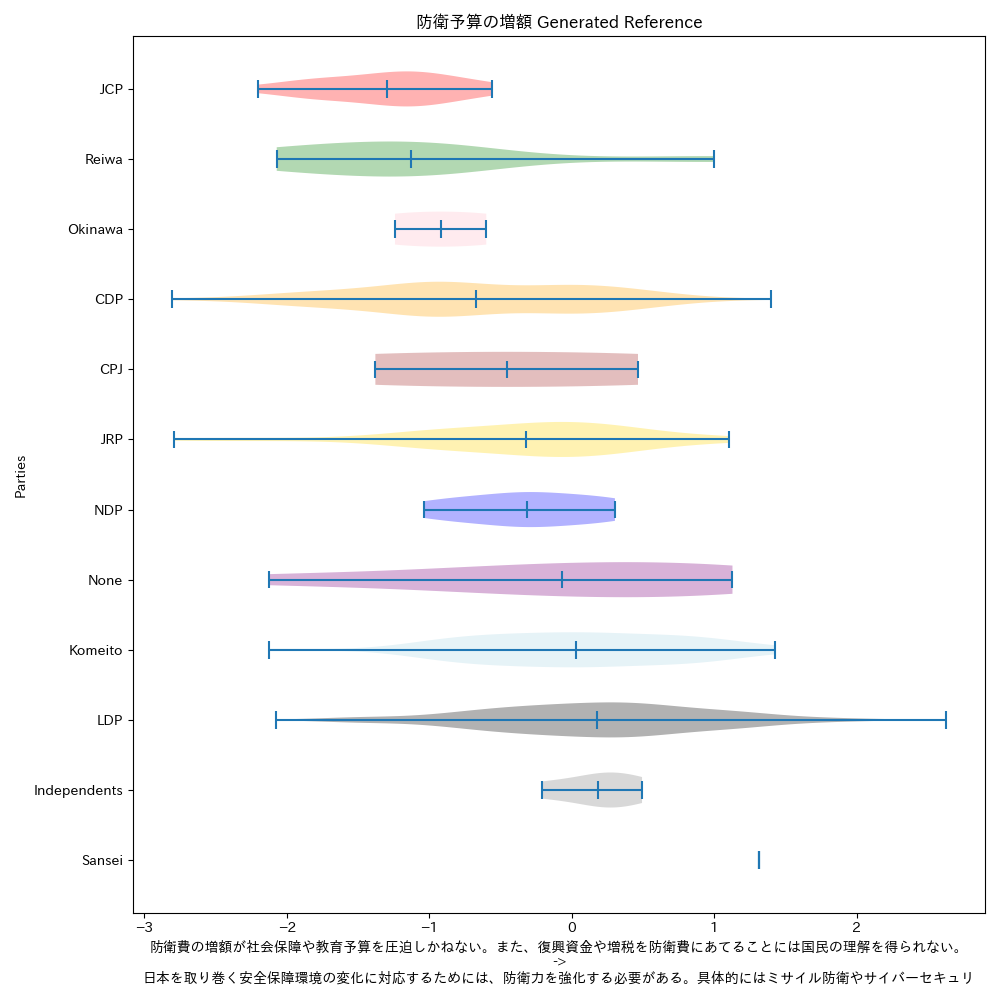}
      \caption{Violin plot of scaled representatives}
    \end{subfigure}
    \begin{subfigure}{0.22\textwidth}
      \centering
      \includegraphics[width=1\linewidth]{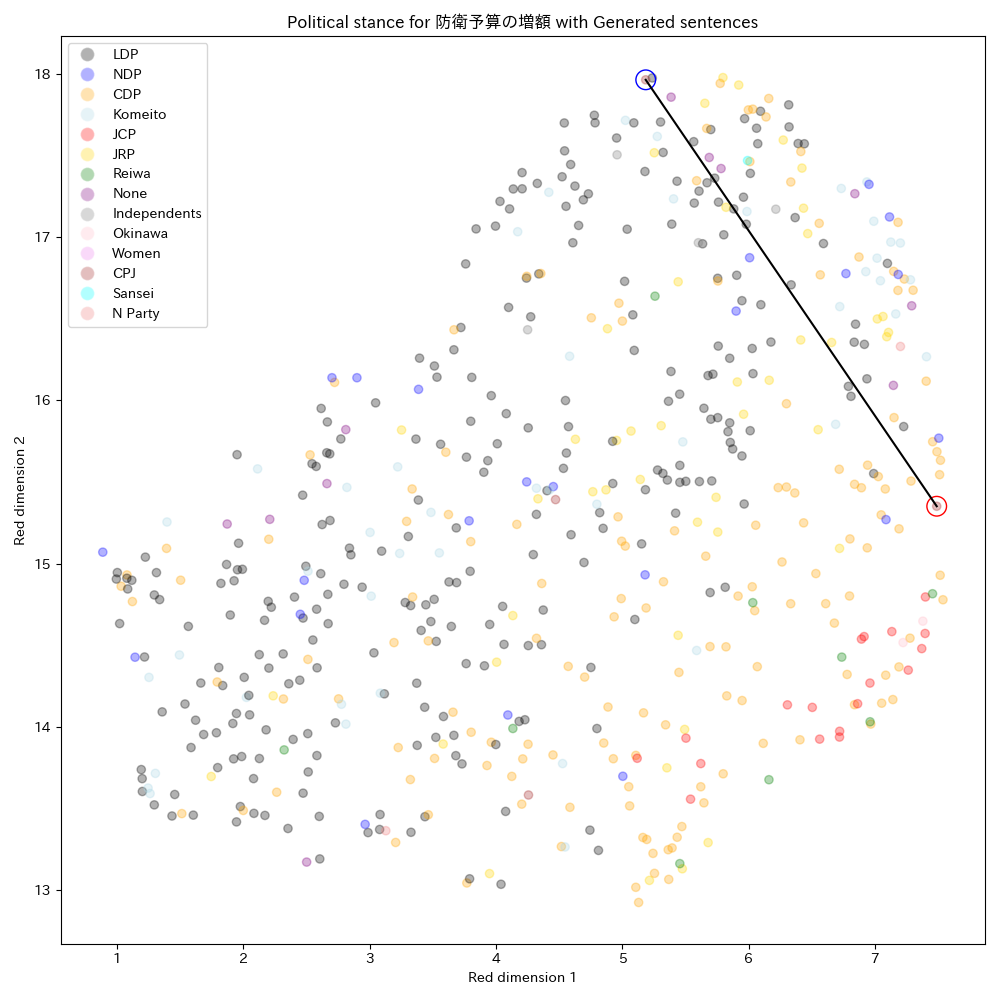}
      \caption{UMAP visualization of representatives and axes of reference}
    \end{subfigure}
\caption{Figures relating to the topic of whether Japan should increase its defense budget.
Left figure: Violin plot of scaled representatives
Right figure: UMAP visualization of representatives and axes of reference. Representatives who have a pro stance located on the upper left side of the UMAP visualization and representatives who have an anti stance located on the lower right side.}
\label{fig: results-defence-budget}
\end{figure}

\begin{figure}[h]
\centering
    \begin{subfigure}{0.22\textwidth}
      \centering
      \includegraphics[width=1\linewidth]{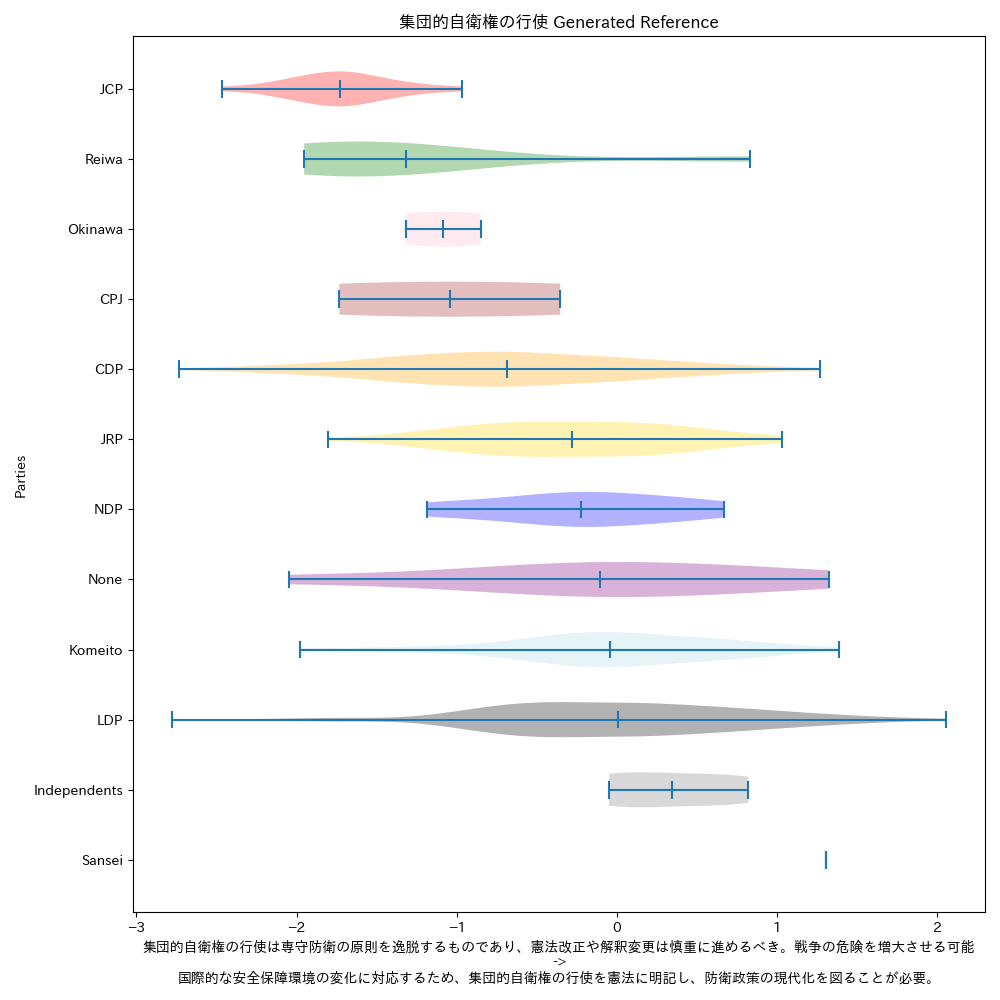}
      \caption{Violin plot of scaled representatives}
    \end{subfigure}
    \begin{subfigure}{0.22\textwidth}
      \centering
      \includegraphics[width=1\linewidth]{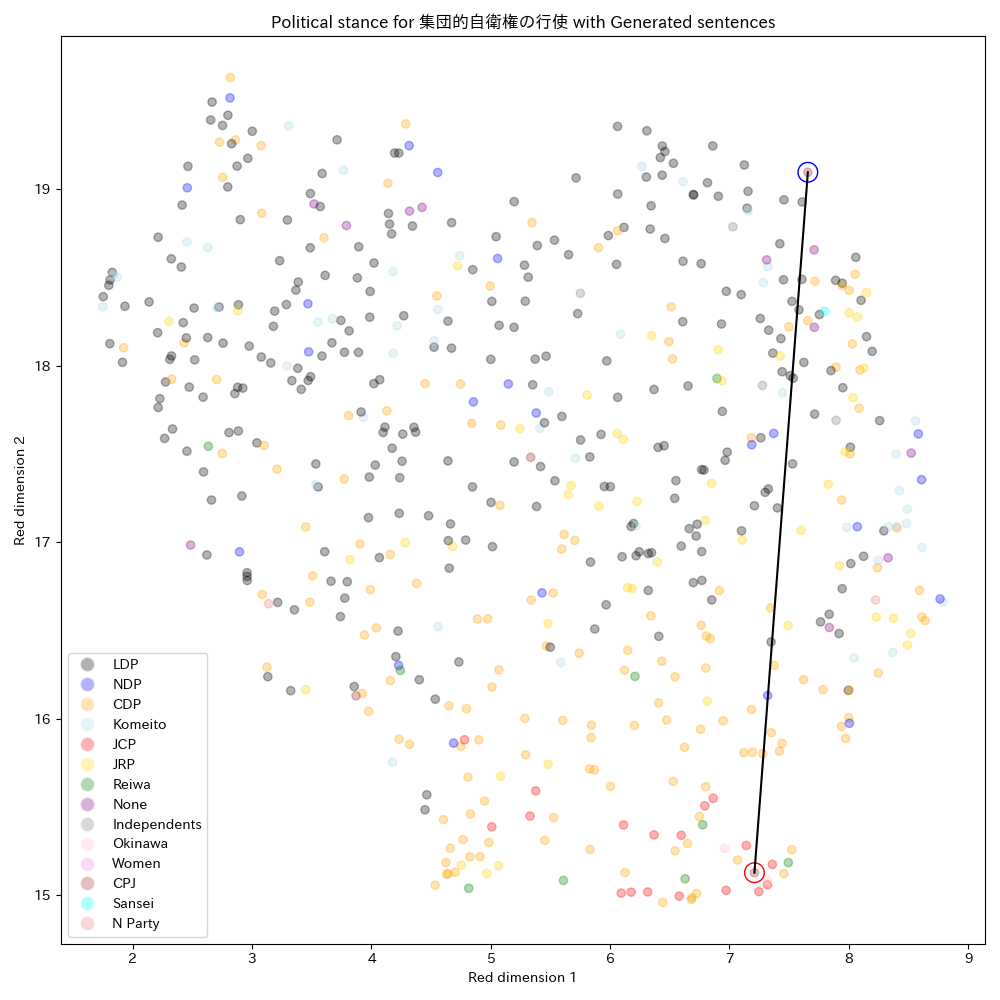}
      \caption{UMAP visualization of representatives and axes of reference}
    \end{subfigure}
\caption{Figures relating to the topic of whether the JSDF should be allowed to exercise collective self-defense.
Left figure: Violin plot of scaled representatives
Right figure: UMAP visualization of representatives and axes of reference. Representatives who have a pro stance located on the upper side of the UMAP visualization and representatives who have an anti stance located on the lower side.}
\label{fig: results-defence-collective}
\end{figure}

We have extracted the main points of controversy for this topic and scaled politicians on these axes. (see figure \ref{fig: extracted axes defense} for full list of topics) As mentioned previously, we have not scaled the politicians on all of the issues, but rather on issues which we deemed suitable to scale the representatives on. See figures \ref{fig: results-defence-constitution}, \ref{fig: results-defence-budget}, \ref{fig: results-defence-collective} for the results. The left figures show the violin plots of the scaled representatives and the right figures show the UMAP visualization of the representatives and the axes of reference. The axis we extracted and scaled the politicians on is shown on the right figures as a black line. 

For the topic of defence, we see a similar pattern across the axes of projection. The parties who are more aligned with right ideologies in Japanese politics(pro acknowledgement of JSDF in constitution, pro increase in defense budget, pro collective self-defense) are positioned on the right side of the axes of projection. The parties who are more aligned with left ideologies in Japanese politics(anti acknowledgement of JSDF in constitution, anti increase in defense budget, anti collective self-defense) are positioned on the left side of the axes of projection. This distinction is visible in the UMAP projections as well where right-leaning parties (LDP, Komeito, NDP, etc.) are clustered closely and the left-leaning parties (CDP, JCP, etc.) are also clustered closer to eachother. This is consistent with the political dynamics in Japan where there is a clear polarization between the right and left ideologies. Take a look at figure \ref{fig: results-defence-constitution} for instance, it is remarkable that Komeito and LDP who have been in the same coalition government for a long time are positioned closely on the pro side of the axis of projection. On the other hand, JCP and CDP who are considered to be more cautious towards the acknowledgement of the JSDF are positioned closely on the anti side of the axis of projection.


\FloatBarrier
\subsubsection{Topic: Nuclear Power}

\begin{figure}[h]
\centering
    \begin{subfigure}{0.22\textwidth}
      \centering
      \includegraphics[width=1\linewidth]{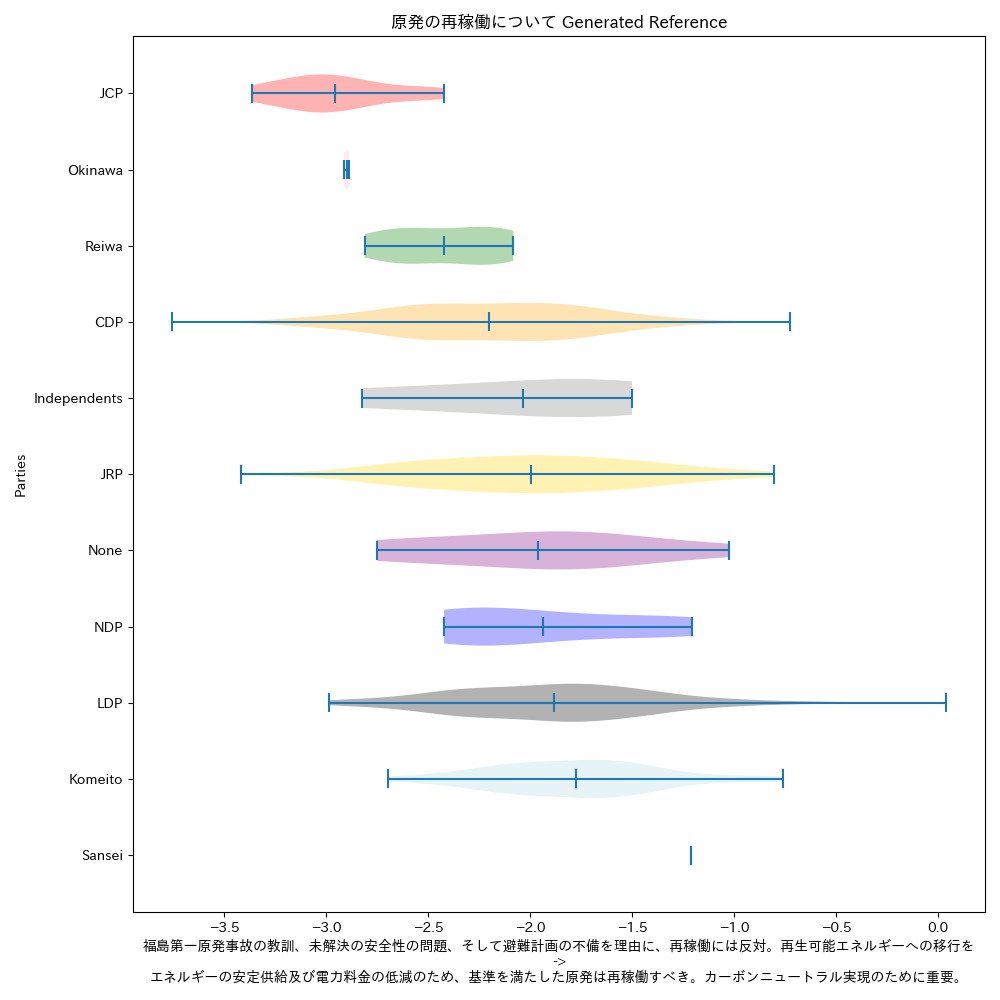}
      \caption{Violin plot of scaled representatives}
      \label{fig:sub1}
    \end{subfigure}
    \begin{subfigure}{0.22\textwidth}
      \centering
      \includegraphics[width=1\linewidth]{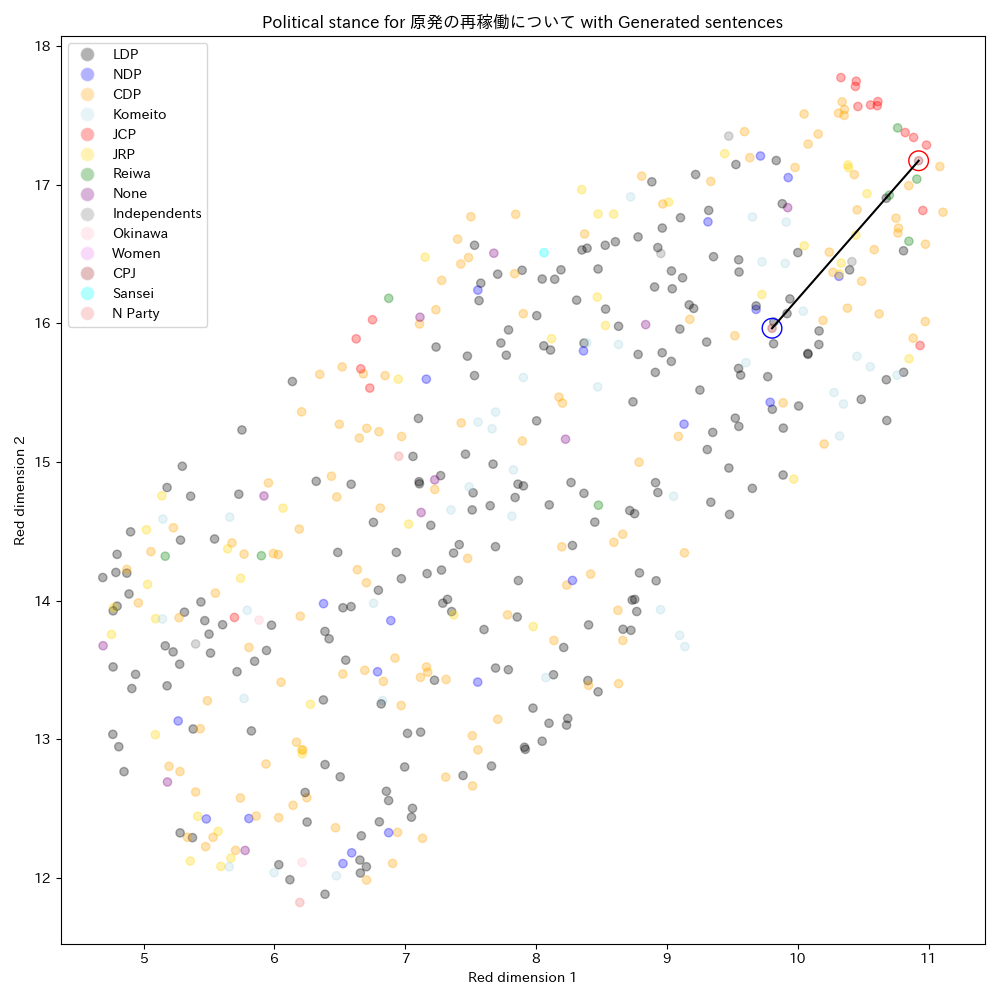}
      \caption{UMAP visualization of representatives and axes of reference}
    \end{subfigure}
\caption{Figures relating to the topic of whether Japan should restart nuclear power plants. Left figure: Violin plot of scaled representatives Right figure: UMAP visualization of representatives and axes of reference. Representatives who have a pro stance located on the lower left side of the UMAP visualization and representatives who have an anti stance located on the upper right side.}
\label{fig: restarting-nuclear}
\end{figure}

The topic of nuclear power in Japan is a highly controversial one. The Fukushima Daiichi nuclear disaster in 2011 has left a lasting impact on the public perception of nuclear power in Japan. The results show a clear distinction between the pro-nuclear and anti-nuclear stances. The parties that are more aligned with the pro-nuclear stance are positioned on the right side of the axes of projection and the parties that are more aligned with the anti-nuclear stance are positioned on the left side of the axes of projection. Figure \ref{fig: restarting-nuclear} for instance, shows a clear distinction and distance between the LDP and JCP who are polar opposites on the topic of restarting nuclear power plants. It is also remarkable that the CDP, who are considered to be also cautious towards restarting nuclear power plants, but is not as extreme as the JCP, is positioned between the JCP and LDP.

\begin{figure}[h]
\centering
    \begin{subfigure}{0.22\textwidth}
      \centering
      \includegraphics[width=1\linewidth]{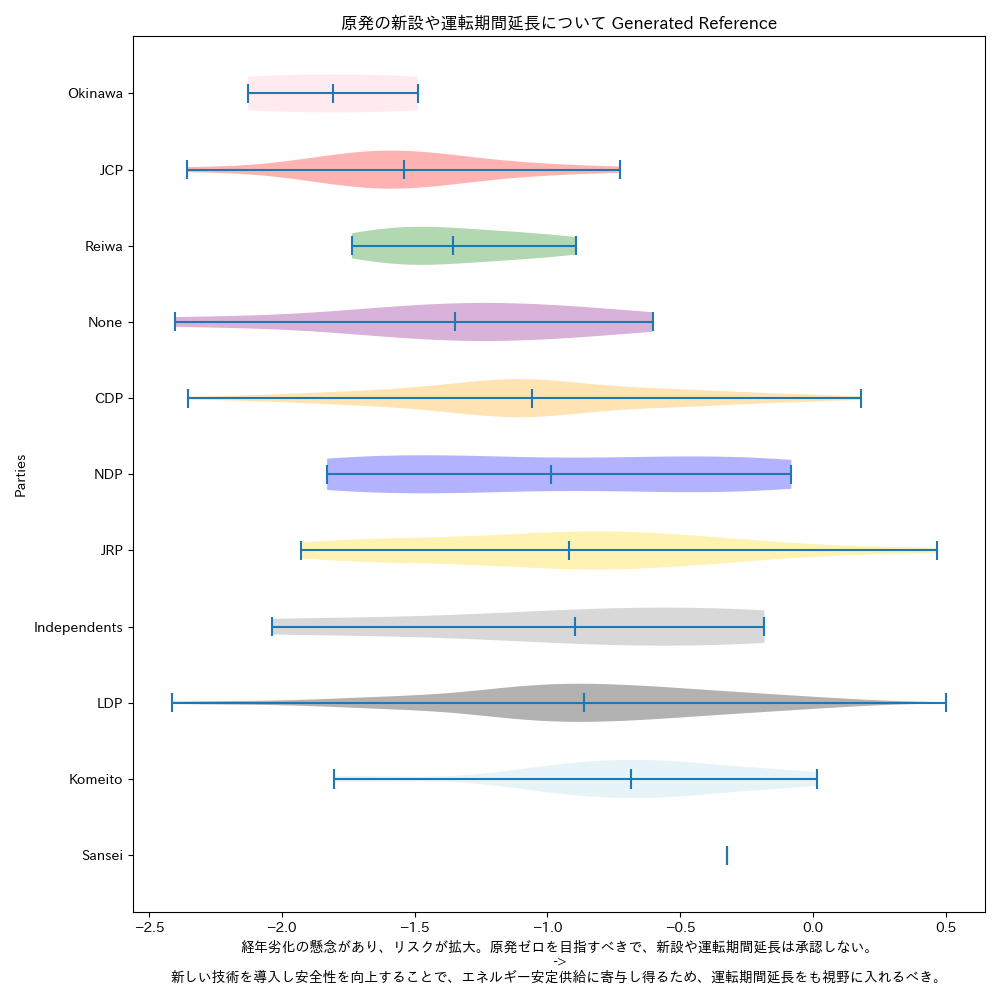}
      \caption{Violin plot of scaled representatives}
    \end{subfigure}
    \begin{subfigure}{0.22\textwidth}
      \centering
      \includegraphics[width=1\linewidth]{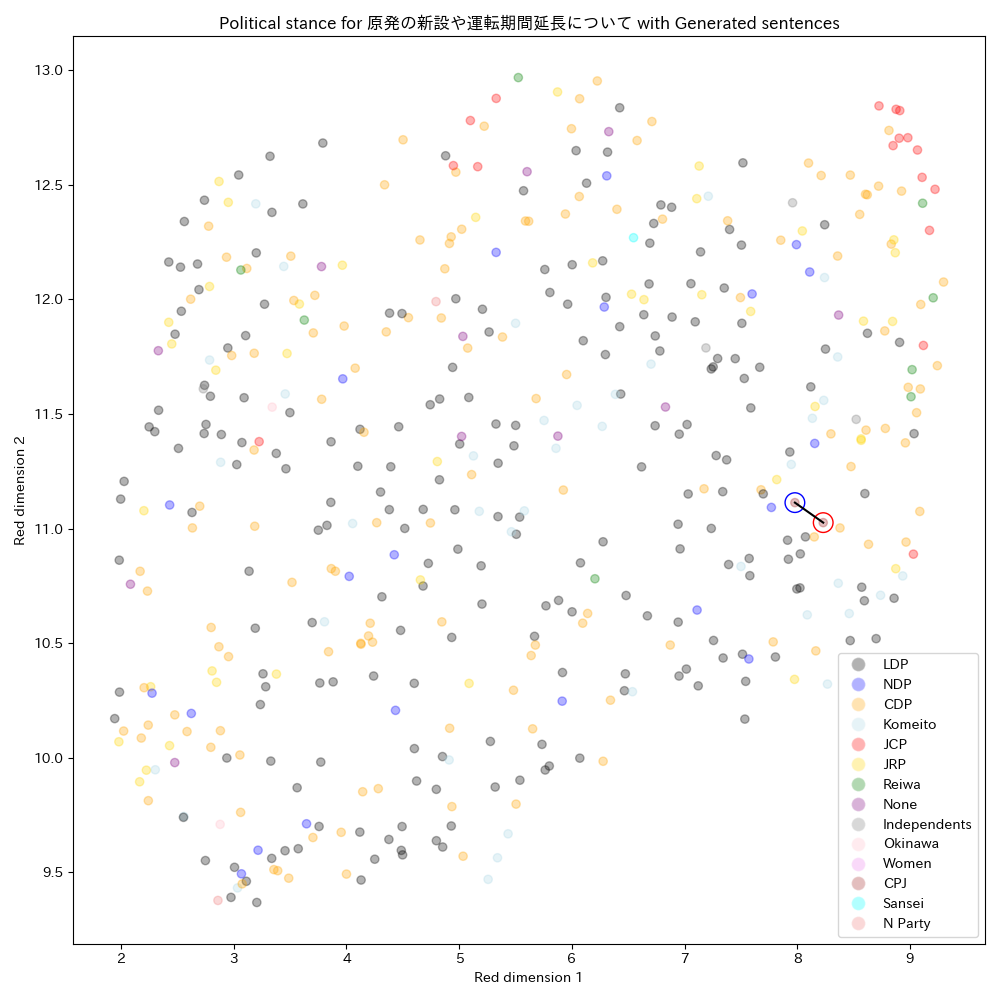}
      \caption{UMAP visualization of representatives and axes of reference}
    \end{subfigure}
\caption{Figtures relating to the topic of whether Japan should build new nuclear power plants and extend the lifetime of existing plants. Left figure: Violin plot of scaled representatives Right figure: UMAP visualization of representatives and axes of reference. Representatives who have a pro stance located on the left side of the UMAP visualization and representatives who have an anti stance located on the right}
\label{fig: results-nuclear-new}
\end{figure}

\begin{figure}[h]
\centering
    \begin{subfigure}{0.22\textwidth}
      \centering
      \includegraphics[width=1\linewidth]{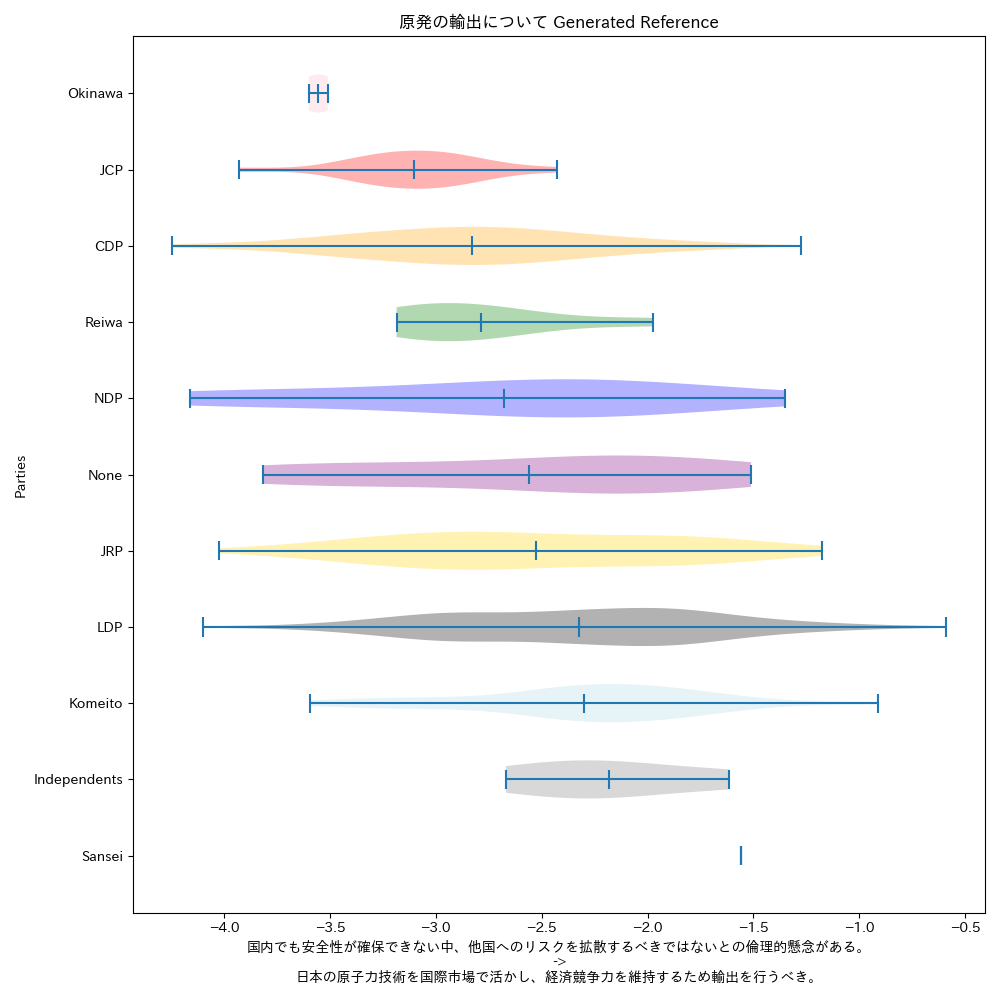}
      \caption{Violin plot of scaled representatives}
    \end{subfigure}
    \begin{subfigure}{0.22\textwidth}
      \centering
      \includegraphics[width=1\linewidth]{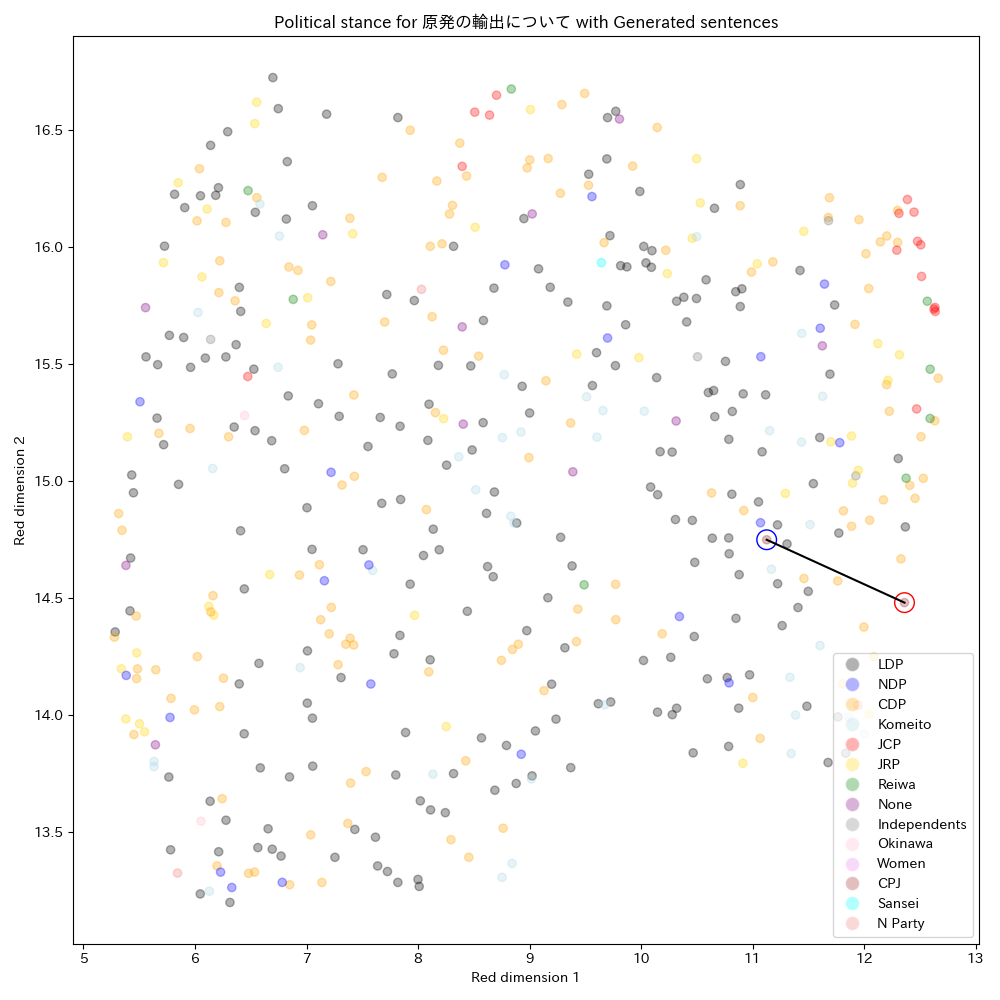}
      \caption{UMAP visualization of representatives and axes of reference}
    \end{subfigure}
\caption{Figures relating to the topic of whether Japan should export nuclear power plants. Left figure: Violin plot of scaled representatives Right figure: UMAP visualization of representatives and axes of reference. Representatives who have a pro stance located on the left side of the UMAP visualization and representatives who have an anti stance located on the right side.}
\label{fig: results-nuclear-export}
\end{figure}

\FloatBarrier

\subsubsection{Topic: Economy}

\begin{figure}[h]
\centering
    \begin{subfigure}{0.22\textwidth}
      \centering
      \includegraphics[width=1\linewidth]{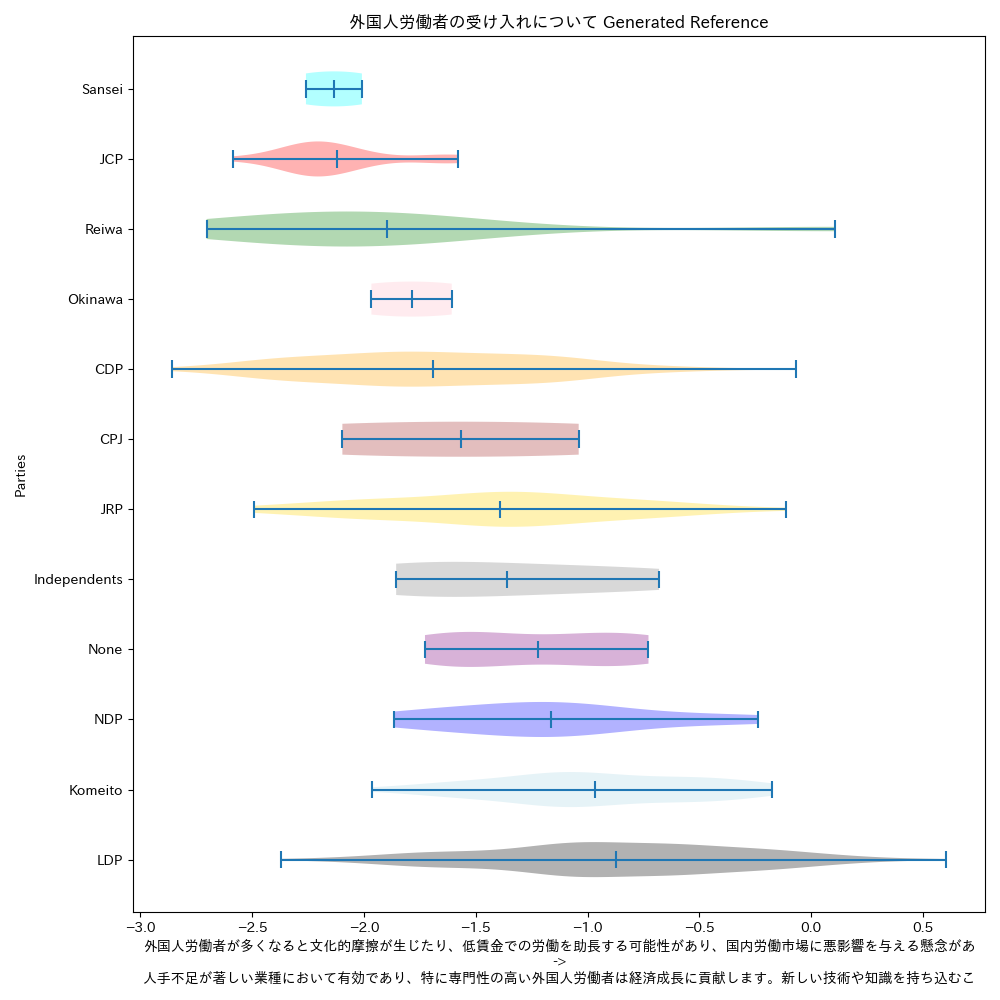}
      \caption{Violin plot of scaled representatives}
    \end{subfigure}
    \begin{subfigure}{0.22\textwidth}
      \centering
      \includegraphics[width=1\linewidth]{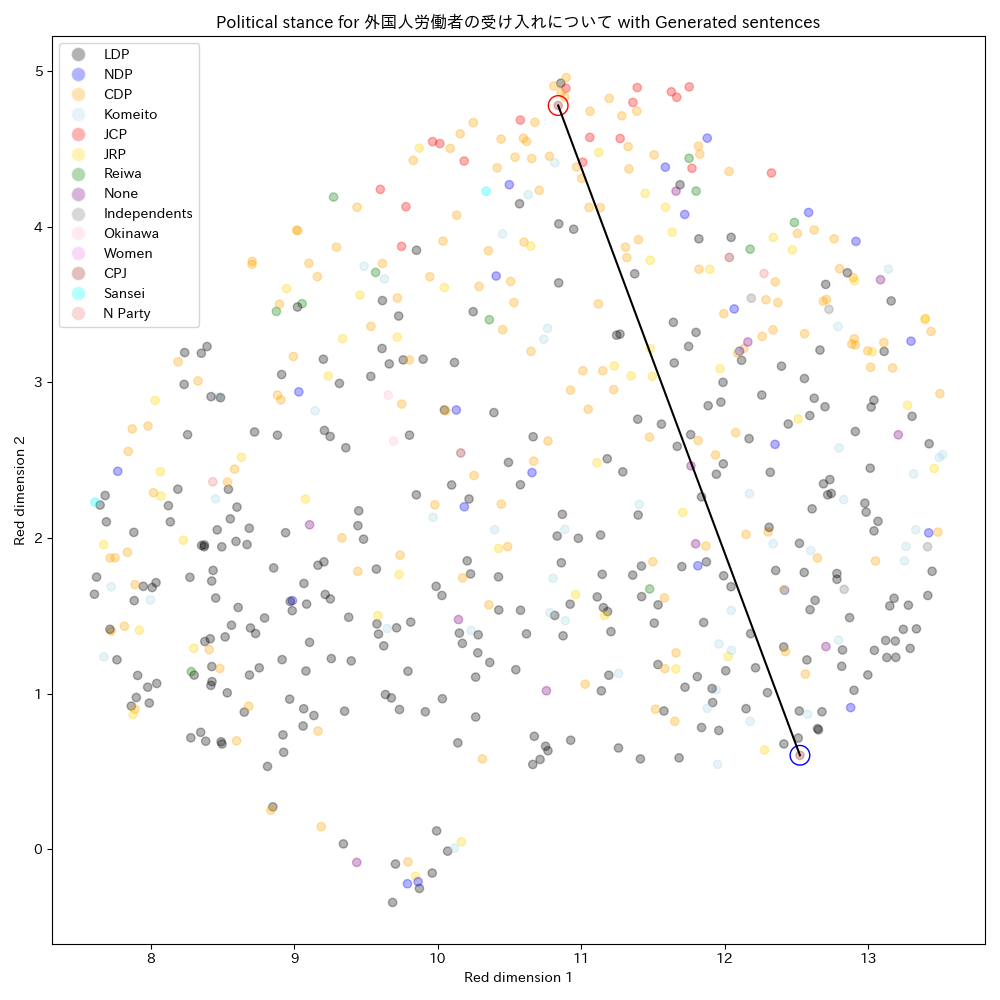}
      \caption{UMAP visualization of representatives and axes of reference}
    \end{subfigure}
\caption{Figures relating to the topic of accepting foreign workers in Japan. Left figure: Violin plot of scaled representatives Right figure: UMAP visualization of representatives and axes of reference. Representatives who have a pro stance located on the upper side of the UMAP visualization and representatives who have an anti stance located on the lower side.}
\label{fig: results-economy-foreign}
\end{figure}

\begin{figure}[h]
\centering
    \begin{subfigure}{0.22\textwidth}
      \centering
      \includegraphics[width=1\linewidth]{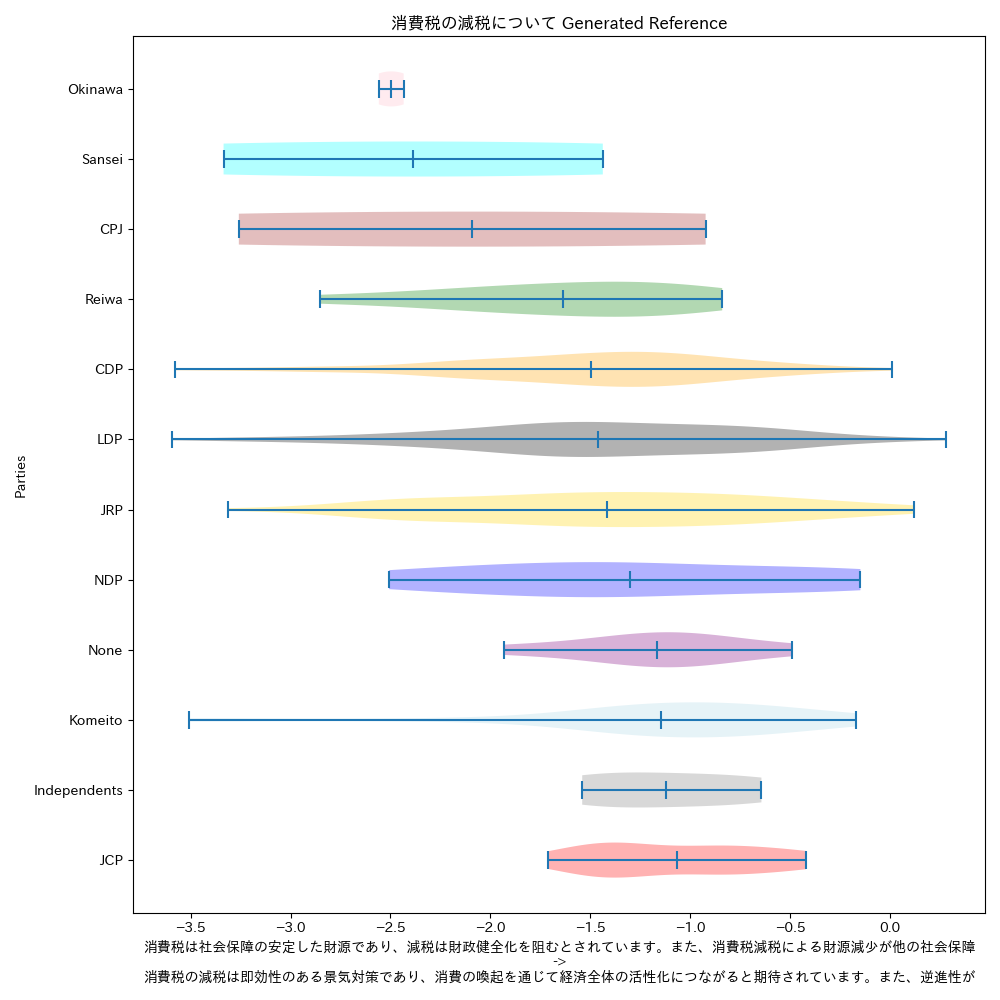}
      \caption{Violin plot of scaled representatives}
    \end{subfigure}
    \begin{subfigure}{0.22\textwidth}
      \centering
      \includegraphics[width=1\linewidth]{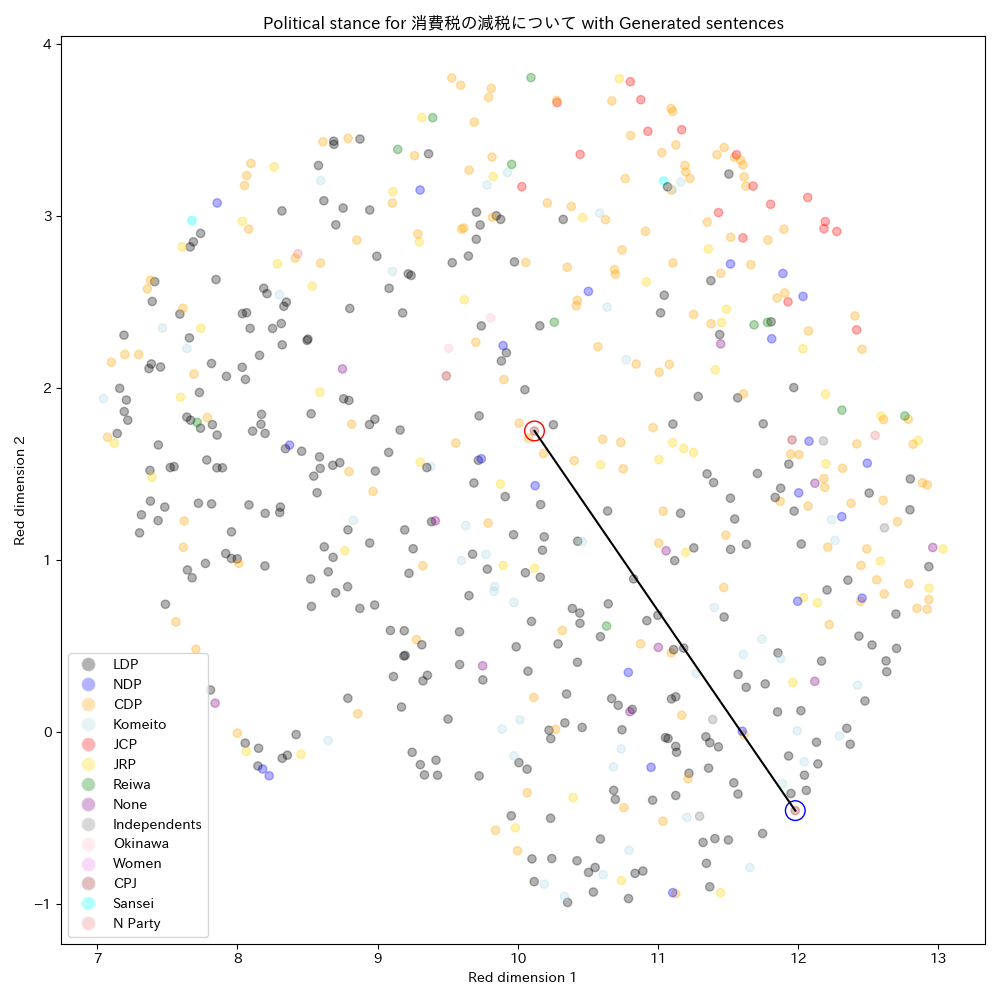}
      \caption{UMAP visualization of representatives and axes of reference}
    \end{subfigure}
\caption{Figures relating to the topic of reducing the consumption tax rate. Left figure: Violin plot of scaled representatives Right figure: UMAP visualization of representatives and axes of reference. Representatives who have a pro stance lower right on the upper side of the UMAP visualization and representatives who have an anti stance located on the upper left.}
\label{fig: results-economy-consumption-tax-rate}
\end{figure}

\begin{figure}[h]
\centering
    \begin{subfigure}{0.22\textwidth}
      \centering
      \includegraphics[width=1\linewidth]{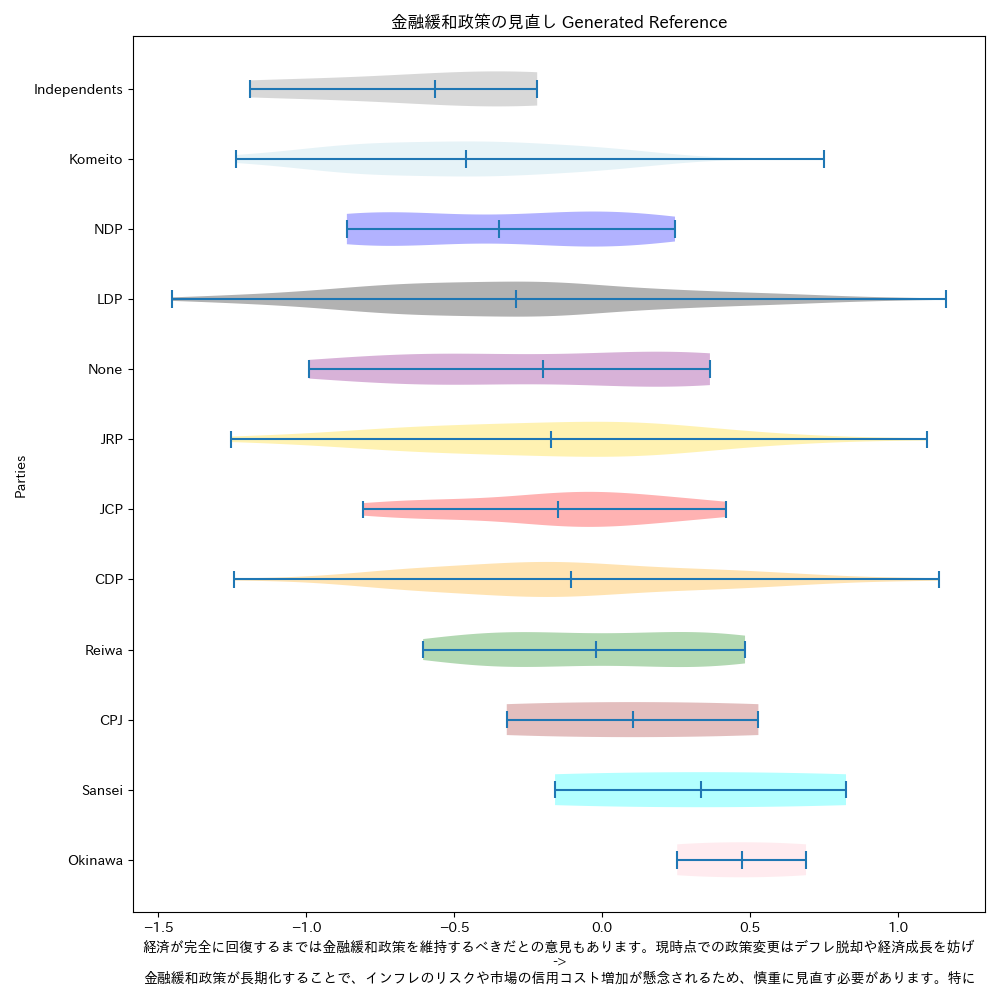}
      \caption{Violin plot of scaled representatives}
    \end{subfigure}
    \begin{subfigure}{0.22\textwidth}
      \centering
      \includegraphics[width=1\linewidth]{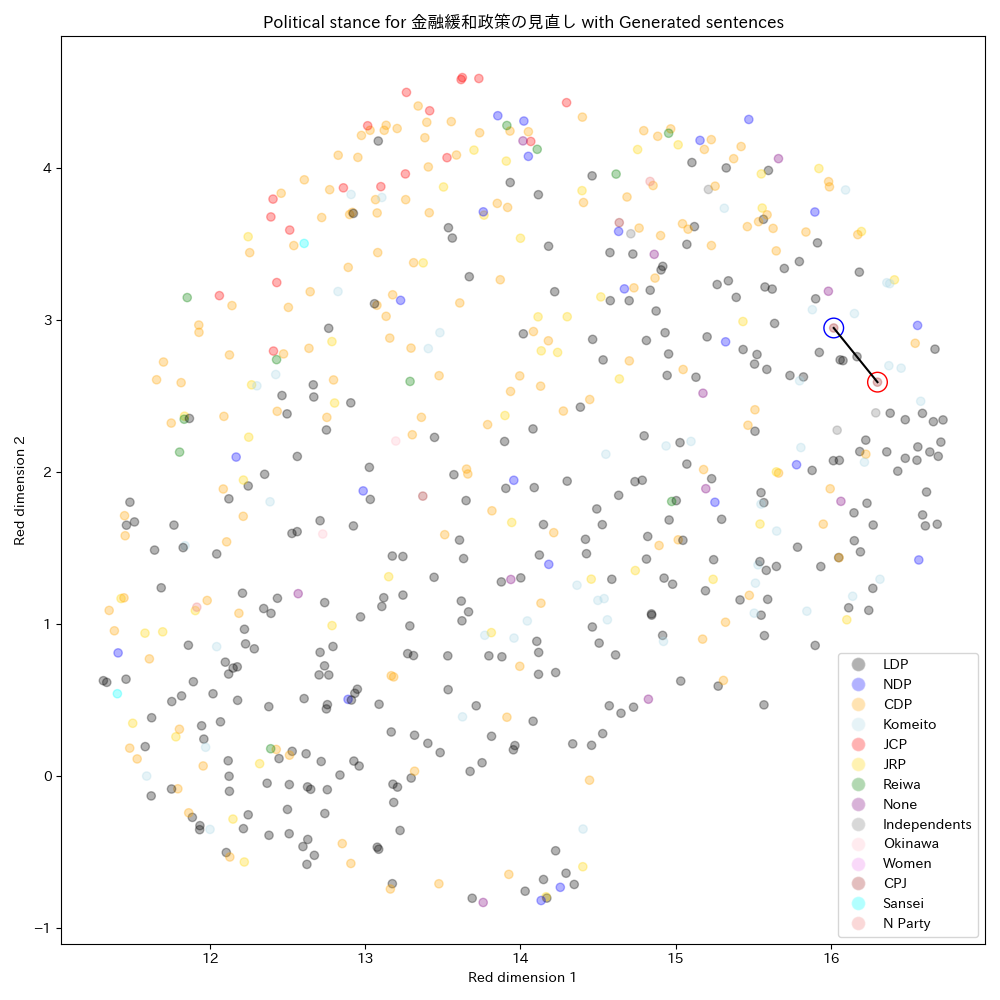}
      \caption{UMAP visualization of representatives and axes of reference}
    \end{subfigure}
\caption{Figures relating to the topic of reconsidering the financial easing policies. Left figure: Violin plot of scaled representatives Right figure: UMAP visualization of representatives and axes of reference. Representatives who have a pro stance located on the upper side of the UMAP visualization and representatives who have an anti stance located on the lower side.}
\label{fig: results-economy-financial-easing}
\end{figure}

There are several axes of controversy in the topic of Japanese economy. With its aging population, Japan is lacking the necessary work force to keep its economy running and hence the extent of accepting foreign workers is a important topic of interest. The consumption tax rate is another topic that is highly debated in Japan. Japan's economy has been stagnant for the past 30 years with the average income of the Japanese people unchanged. Hence a hike in consumption tax rate will have a significant impact on the people. The Bank of Japan has been trying to stimulate the economy through a negative interest policy but this has not been very effective so far. Hence the reconsideration of the financial easing policies is another topic of interest and controversy. Parties such as LDP and Komeito are more cautious about lowering the consumption tax rate (figure \ref{fig: results-economy-consumption-tax-rate}) and changing the financial easing policies (figure \ref{fig: results-economy-financial-easing}) while parties such as CDP and JCP are more in favor of these changes. The topic of foreign workers is more nuanced with parties such as LDP and Komeito are more in favor of accepting foreign workers with the current immigration system while parties such as CDP and JCP point out the exploitation of foreign workers in Japan and call for a reformation in the Japanese immigration system. (figure \ref{fig: results-economy-foreign})

\FloatBarrier

\subsubsection{Topic: Aging Population}

\begin{figure}[h]
\centering
    \begin{subfigure}{0.22\textwidth}
      \centering
      \includegraphics[width=1\linewidth]{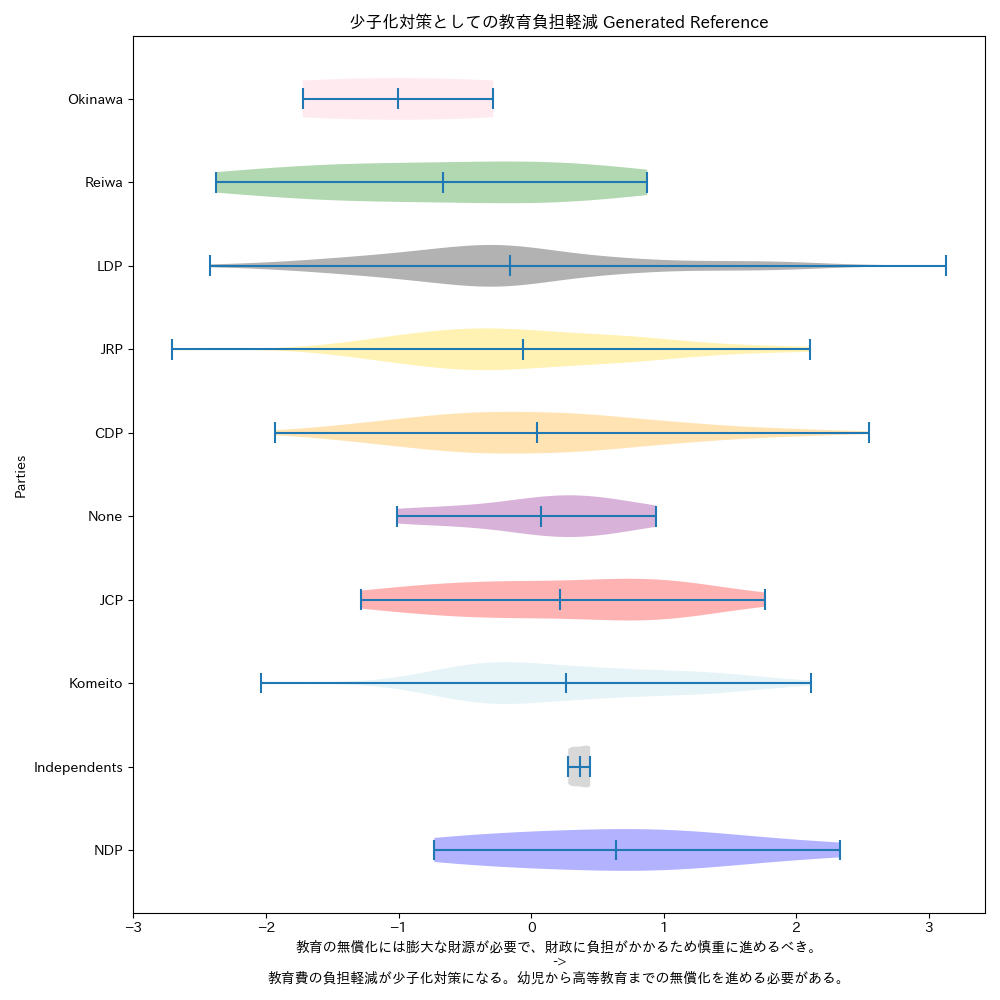}
      \caption{Violin plot of scaled representatives}
    \end{subfigure}
    \begin{subfigure}{0.22\textwidth}
      \centering
      \includegraphics[width=\linewidth]{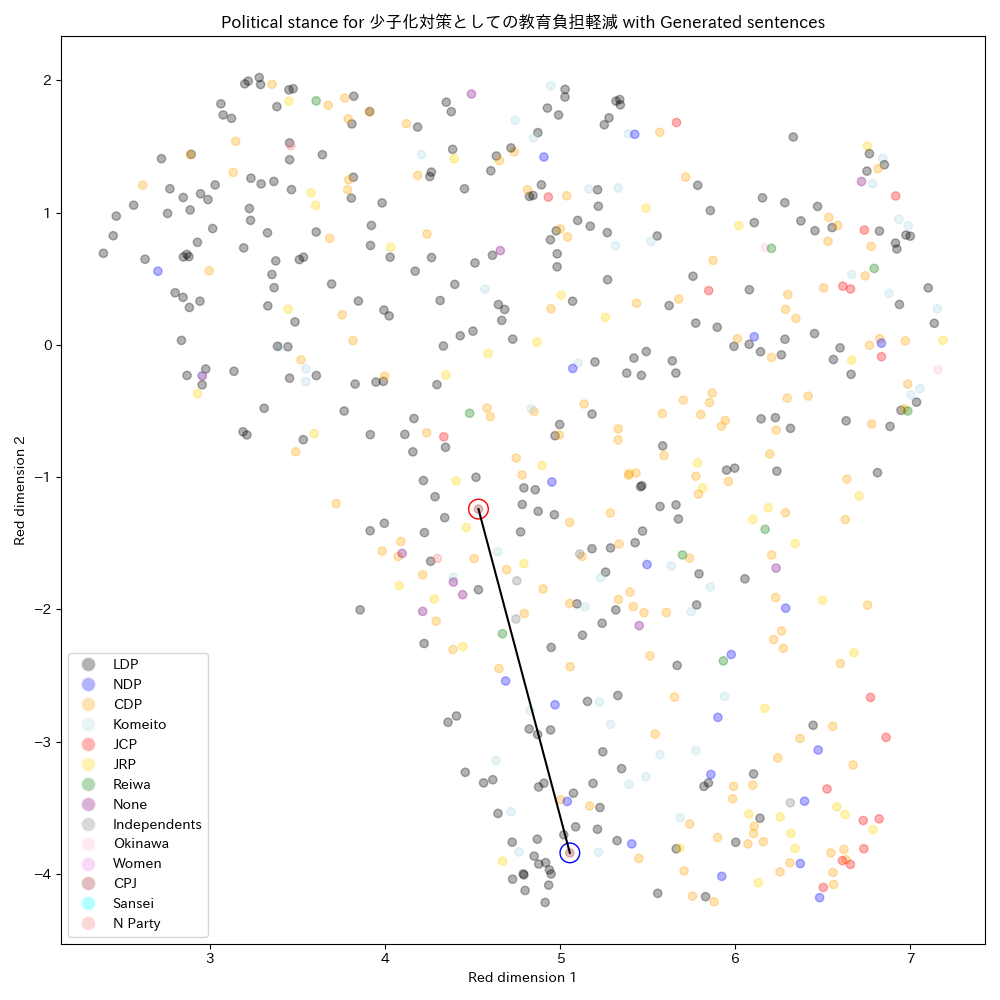}
      \caption{UMAP visualization of representatives and axes of reference}
    \end{subfigure}
\caption{Figures relating to the topic of decreasing the education cost for parents. Left figure: Violin plot of scaled representatives Right figure: UMAP visualization of representatives and axes of reference. Representatives who have a pro stance located on the lower side of the UMAP visualization and representatives who have an anti stance located on the upper side.}
\label{fig: results-aging-education-cost}
\end{figure}

In Japan, the aging population is a major concern. The birth rate is declining and the population is aging and shrinking rapidly. This is causing a shrinking Japanese economy and productivity. Hence, policies addressing this issue are highly debated. Through our pipeline, we have identified a major axes of controversy that relates to decreasing the education cost for parents.(figure \ref{fig: results-aging-education-cost}) While LDP is more cautious about increased government spending on education, parties such as NDP, Komeito and JCP call for increased government spending on education. This is especially interesting because through our analysis, we have found that Komeito usually aligns with their coalition partner LDP on most issues but for the issue of decreasing the education cost for parents, Komeito is more aligned with the left-leaning parties. This is consistent with the party's background of being a religious party which values traditional family values and hence they are supportive of policies that would help families.

\FloatBarrier

\subsection{Diachronic analysis of party positions on political axes}
\label{section: diachronic}
\begin{figure}[h]
	\centering
		\begin{subfigure}{0.48\textwidth}
		  \centering
		  \includegraphics[width=\textwidth]{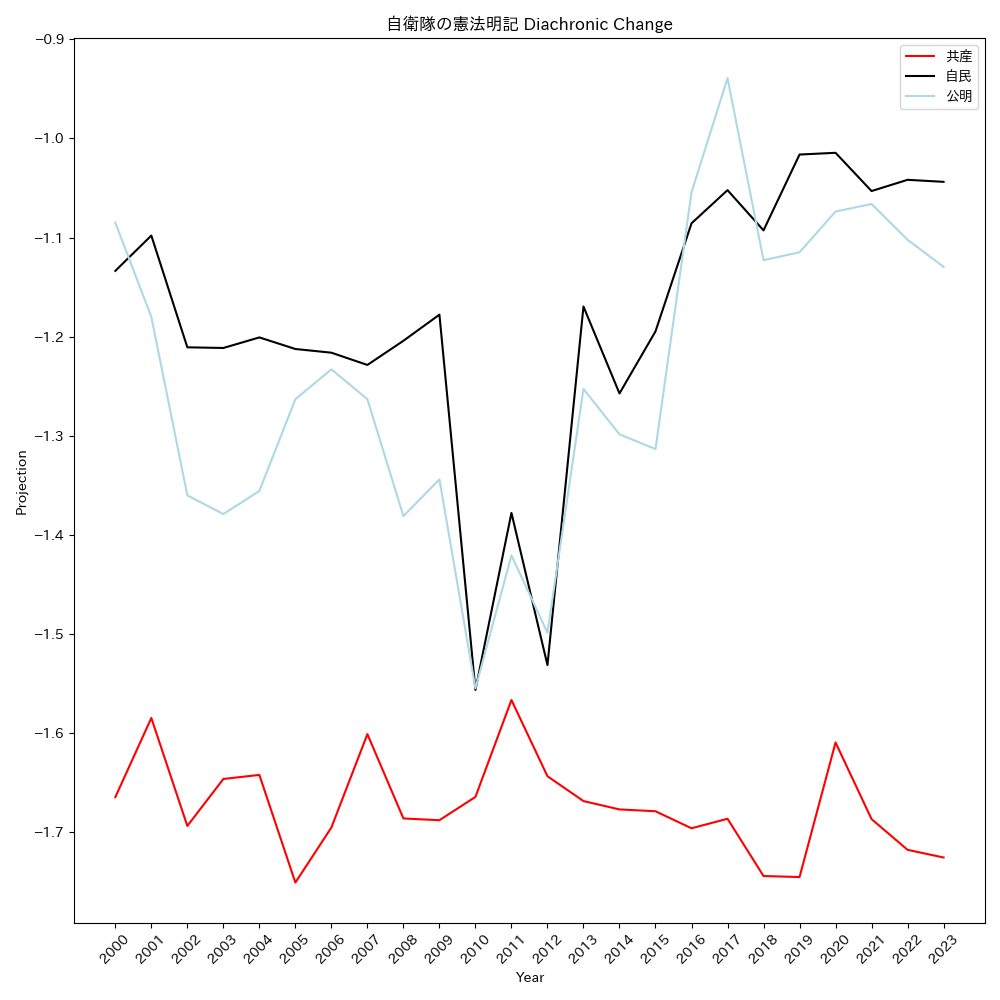}
		  \caption{Visualization of mean party issue position for years between 2000-2024 \\\hspace{\textwidth}Black: LDP, Red: JCP, Lightblue: Komeito}
		  \label{fig:sub1}
		\end{subfigure}
		\hfill
		\begin{subfigure}{0.48\textwidth}
		  \centering
		  \includegraphics[width=\textwidth]{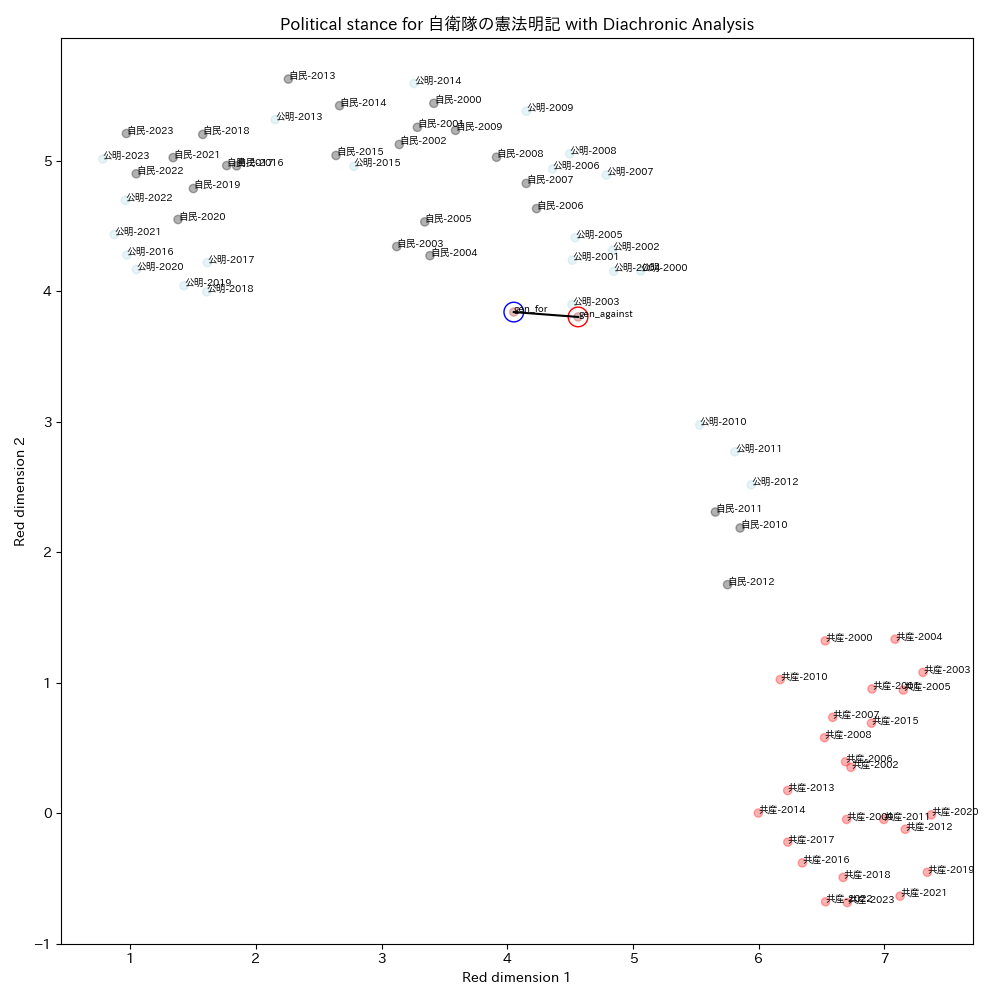}
		  \caption{UMAP visualization of issue position of parties for years between 2000-2024 \\\hspace{\textwidth} 
		  Black: LDP, Red: JCP, Lightblue: Komeito\\\hspace{\textwidth}
		  Blue circle:Pro Red circle: Con}
		  \label{fig:sub2}
		\end{subfigure}
	\caption{Acknowledgement of JSDF in the Japanese constitution}
	\label{fig: results-diachronic-defence-constitution}
\end{figure}

\begin{figure}[h]
	\centering
		\begin{subfigure}{0.48\textwidth}
		  \centering
		  \includegraphics[width=\textwidth]{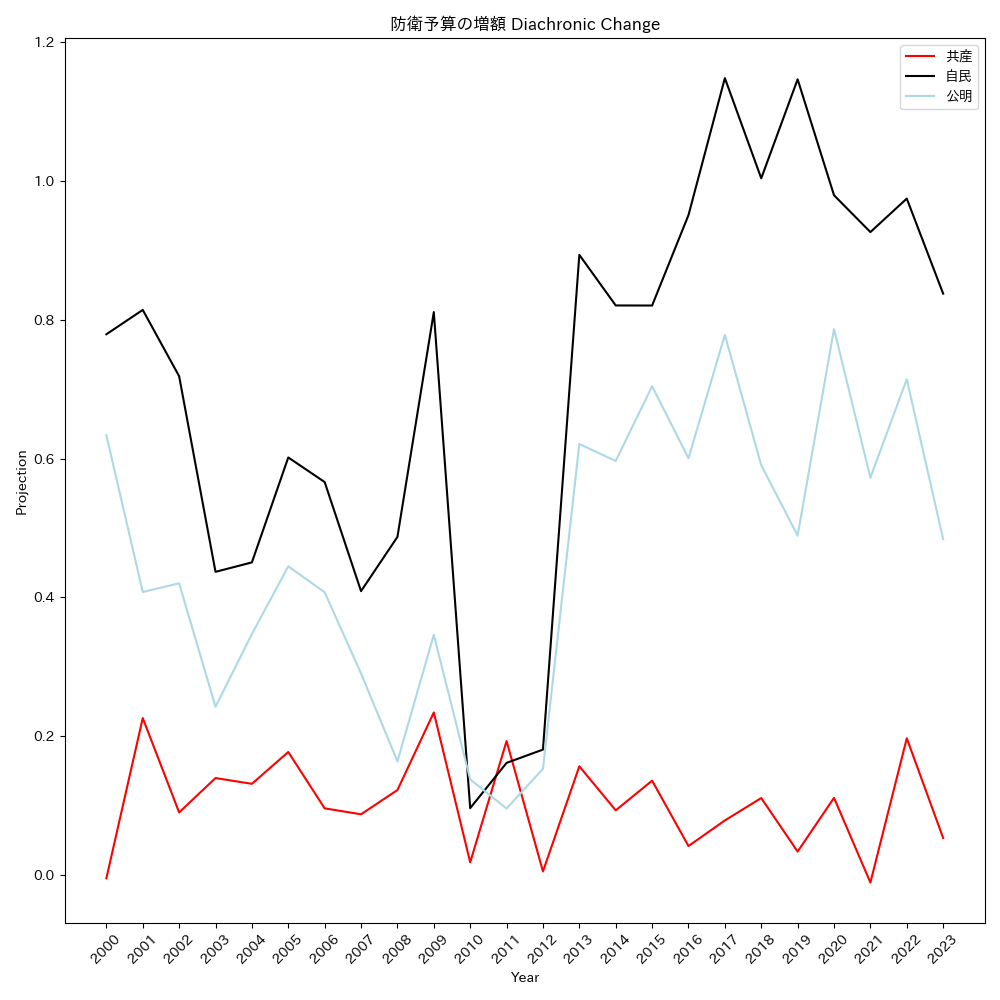}
		  \caption{Visualization of mean party issue position for years between 2000-2024 \\\hspace{\textwidth} Black: LDP, Red: JCP, Lightblue: Komeito}
		  \label{fig:sub1}
		\end{subfigure}
		\hfill
		\begin{subfigure}{0.48\textwidth}
		  \centering
		  \includegraphics[width=\textwidth]{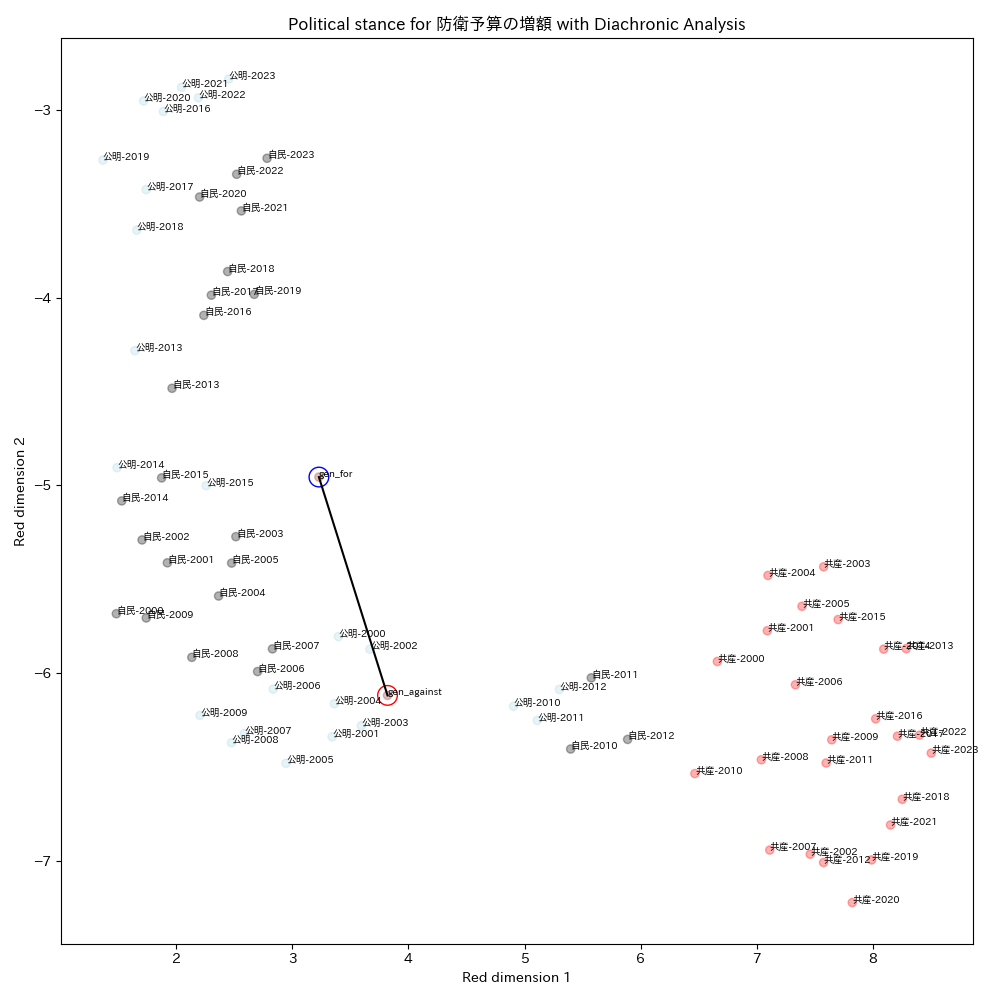}
		  \caption{UMAP visualization of issue position of parties for years between 2000-2024 \\\hspace{\textwidth} 
		  Black: LDP, Red: JCP, Lightblue: Komeito\\\hspace{\textwidth}
		  Blue circle:Pro Red circle: Con }
		  \label{fig:sub2}
		\end{subfigure}
	\caption{Increase in defense budget}
	\label{fig: results-diachronic-defense-budget}
\end{figure}

\begin{figure}[h]
	\centering
		\begin{subfigure}{0.48\textwidth}
		  \centering
		  \includegraphics[width=\textwidth]{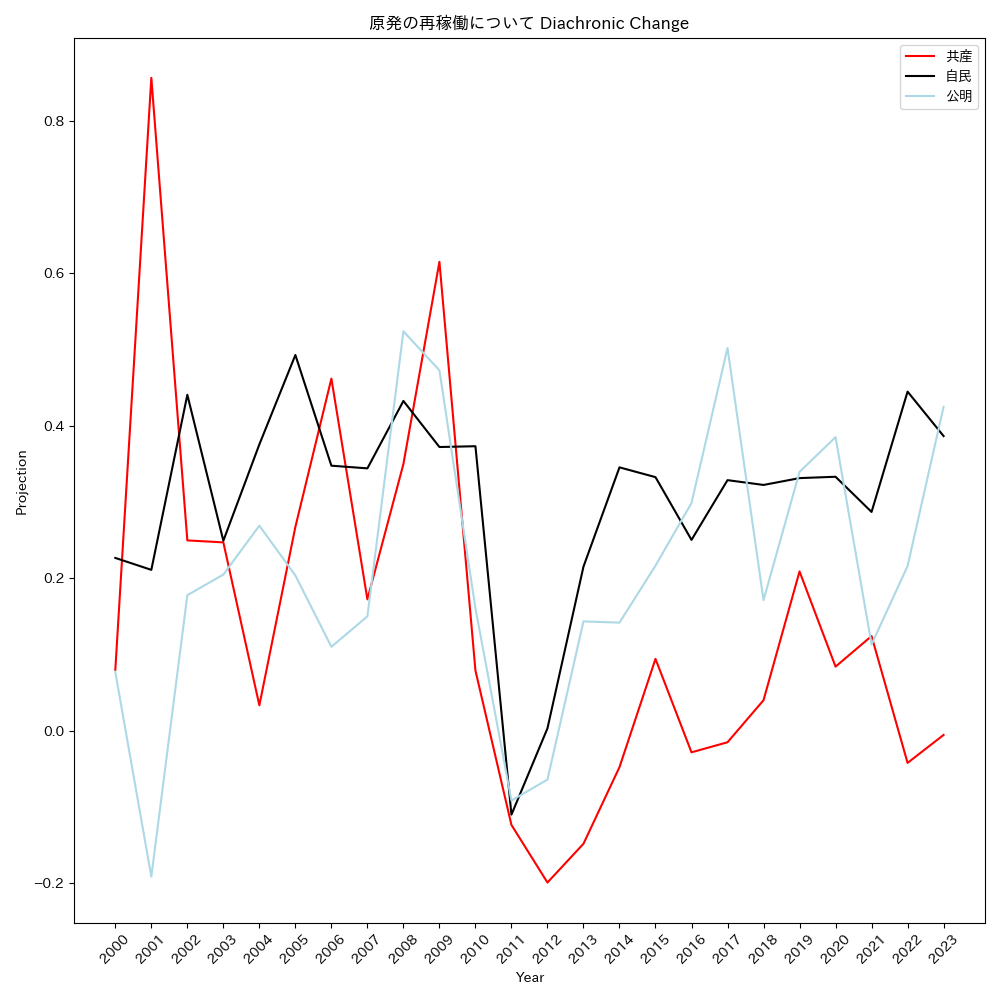}
		  \caption{Visualization of mean party issue position for years between 2000-2024 \\\hspace{\textwidth}  Black: LDP, Red: JCP, Lightblue: Komeito}
		  \label{fig:sub1}
		\end{subfigure}
		\hfill
		\begin{subfigure}{0.48\textwidth}
		  \centering
		  \includegraphics[width=\textwidth]{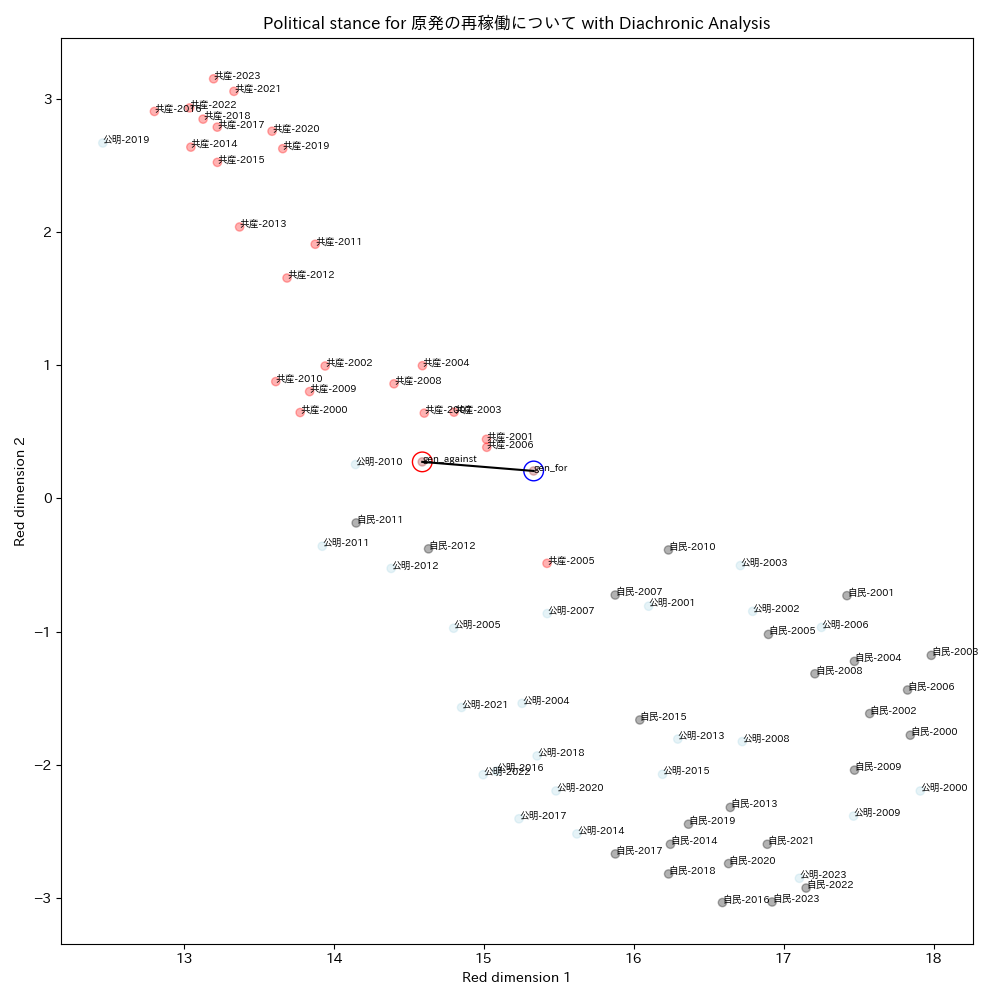}
		  \caption{UMAP visualization of issue position of parties for years between 2000-2024 \\\hspace{\textwidth} 
		  Black: LDP, Red: JCP, Lightblue: Komeito\\\hspace{\textwidth}
		  Blue circle:Pro Red circle: Con}
		\end{subfigure}
	\caption{Restarting nuclear powerplants}
	\label{fig: results-diachronic-nuclear-restart}
\end{figure}

\begin{figure}[h]
	\centering
		\begin{subfigure}{0.48\textwidth}
		  \centering
		  \includegraphics[width=\textwidth]{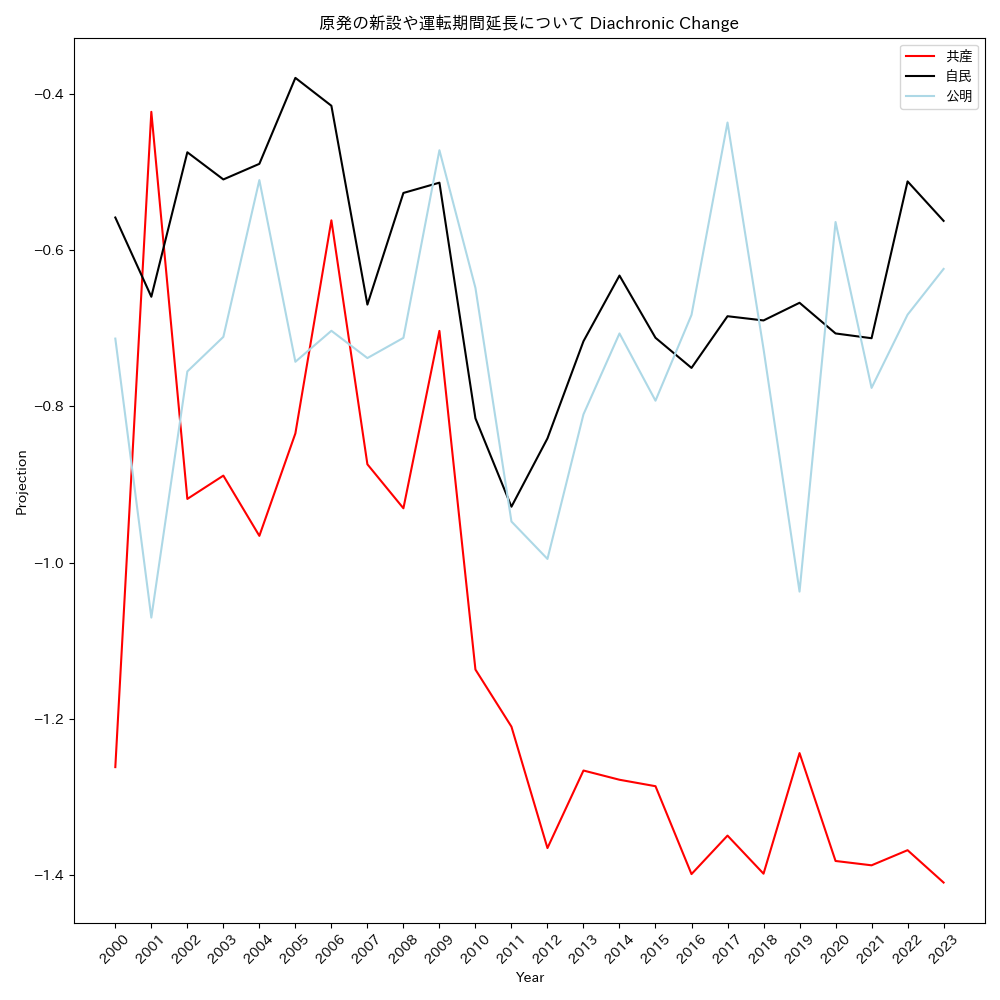}
		  \caption{Visualization of mean party issue position for years between 2000-2024 \\\hspace{\textwidth} Black: LDP, Red: JCP, Lightblue: Komeito}
		  \label{fig:sub1}
		\end{subfigure}
		\hfill
		\begin{subfigure}{0.48\textwidth}
		  \centering
		  \includegraphics[width=\textwidth]{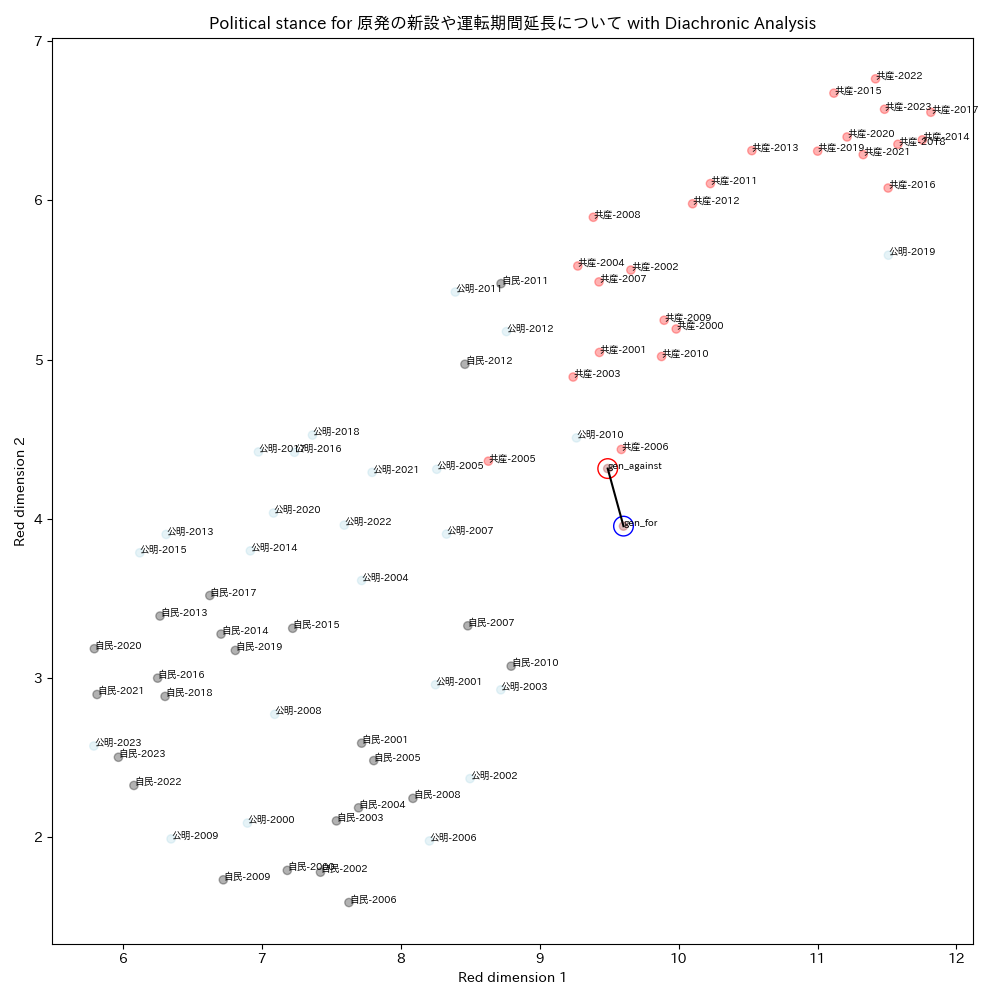}
		  \caption{UMAP visualization of issue position of parties for years between 2000-2024\\\hspace{\textwidth} 
		  Black: LDP, Red: JCP, Lightblue: Komeito\\\hspace{\textwidth}
		  Blue circle:Pro Red circle: Con}
		  \label{fig:sub2}
		\end{subfigure}
	\caption{Building new nuclear powerplants or extending the lifetime of existing ones}
	\label{fig: results-diachronic-nuclear-new}
\end{figure}

We have conducted a diachronic analysis of mean party issue positions of LDP, JCP and Komeito. The statistics on the number of speech segments as well as the number of representatives used for the analysis is summarized in table \ref{tab:nuclear_speech} and \ref{tab:defense_speech}. See figures \ref{fig: results-diachronic-defence-constitution}, \ref{fig: results-diachronic-defense-budget}, \ref{fig: results-diachronic-nuclear-restart}, \ref{fig: results-diachronic-nuclear-new} for the results. The left figures show the mean party issue position for the years between 2000-2024 and the right figures show the UMAP visualization of the issue position of parties for years between 2000-2024. We will point out some interesting observations and trends in section \ref{section: event-based}, where we will link the observation with events that has occured at different time periods. 

\onecolumn

\clearpage
\twocolumn

\section{Discussion}
In this section, we will be discussing the 3 ways we have used to evaluate our methodology. 
\begin{itemize}
	\item \textbf{Quantitative evaluation:} We have compared the ordering of the parties we have obtained from our methodology to the ordering obtained by experts in regards to different topics. 
	\item \textbf{Qualitative evaluation:} We have analyzed the common noun phrases in the different clusters of representatives to understand and interpret the stance of politicians in each cluster. 
	\item \textbf{Event-based evaluation:} We have analyzed our results for the diachronic analysis of party positions on topics of nuclear power and defence and identified historical events and occurrences that have influenced the party positions.
\end{itemize}

\subsection{Quantitative evaluation: Comparison against expert estimations}
\begin{figure}[ht]
    \centering
    \begin{subfigure}[t]{0.22\textwidth}
        \centering
        \includegraphics[width=\linewidth]{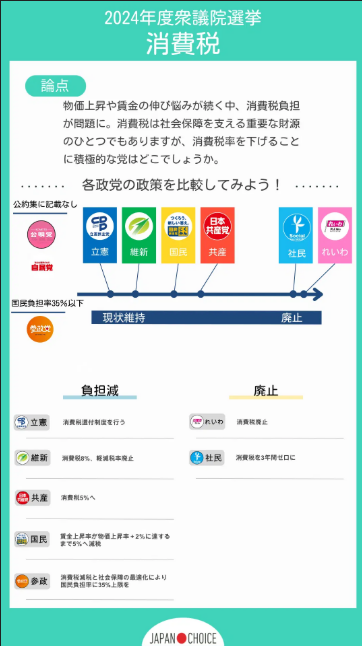}
        \caption{Consumption tax rate}
        \label{fig:consumptiontax}
    \end{subfigure}\hfill
    \begin{subfigure}[t]{0.22\textwidth}
        \centering
        \includegraphics[width=\linewidth]{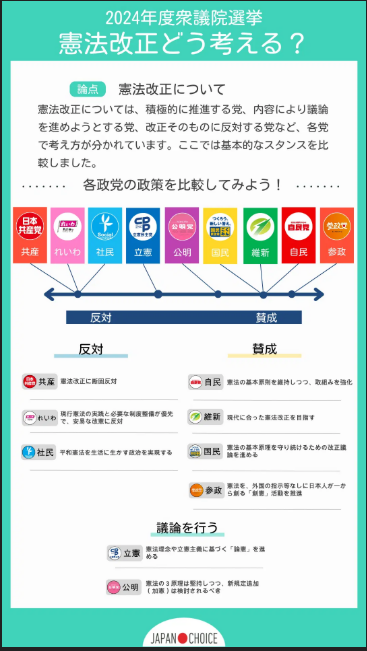}
        \caption{Changing the Japanese constitution to acknowledge the JSDF}
        \label{fig:jsdfconsitution}
    \end{subfigure}\hfill
    \begin{subfigure}[t]{0.22\textwidth}
        \centering
        \includegraphics[width=\linewidth]{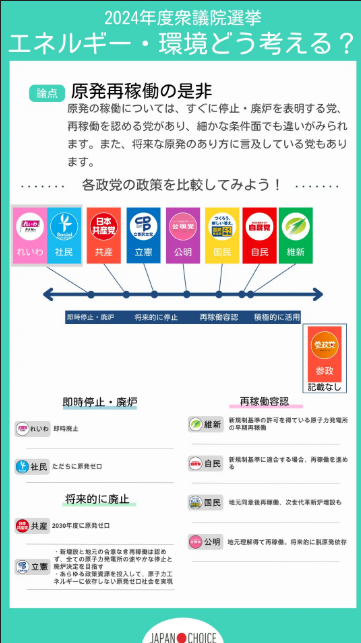}
        \caption{Using nuclear power in Japan}
        \label{fig:nuclear}
    \end{subfigure}\hfill
    \begin{subfigure}[t]{0.22\textwidth}
        \centering
        \includegraphics[width=\linewidth]{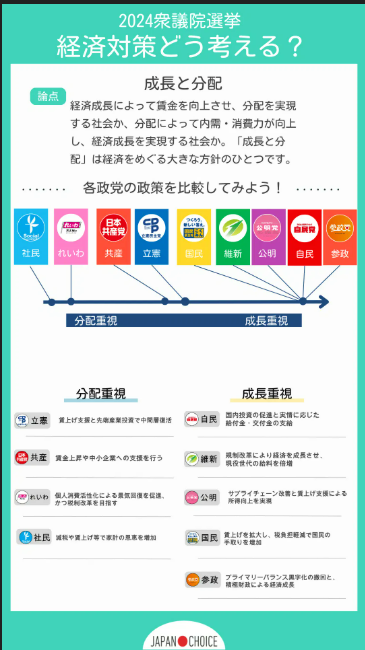}
        \caption{Growth or distribution of wealth}
        \label{fig:growthordistrib}
    \end{subfigure}
    \caption{Party estimations by Mielka \citep{Mielka}}
    \label{fig:mielka-figs}
\end{figure}

To our knowledge, there is no existing dataset/research which quantifies the political stance of Japanese parliamentary representatives on various topics. Therefore, we have compared the ordering of the parties we have obtained from our methodology to the ordering estimated by experts in regards to different topics. For the expert estimations, we used the results of \citeauthor{Mielka} who are a NPO in Japan that focuses on political transparency and dissimation of political information. In a reply to our inquiry, they shared that in order to place the parties on a spectrum of political disagreement, they have at least 3 experts discussing the placement of the parties and decide on the final placement of the parties. Figure \ref{fig:mielka-figs} shows the estimations of the party positions on different topics by Mielka.

\subsubsection{Metrics of comparison}

In order to compare the ordering of parties, we have used 3 methods of comparison to conduct a rounded evaluation of our results. Those are:
\begin{itemize}
	\item Spearman's rank correlation coefficient\citep{spearman1904}
	\item Kendall's tau rank correlation coefficient\citep{kendall1938}
	\item Longest Common Subsequence(LCS) ratio
\end{itemize}

The equation for Spearman's rank correlation coefficient is given by:

\begin{center}
	\[	\rho = 1 - \frac{6\sum d_i^2}{n(n^2-1)} \]
	
\end{center}

where $d_i$ is the difference between the ranks of the parties in the two orderings and $n$ is the number of parties. The value of $\rho$ ranges from -1 to 1. A value of 1 indicates that the two orderings are identical, 0 indicates no correlation, and -1 indicates that the two orderings are in reverse order. 

The equation for Kendall's tau rank correlation coefficient is given by:

\begin{center}
	\[	\tau = \frac{P-Q}{\frac{1}{2}n(n-1)} \]
\end{center}

where $P$ is the number of concordant pairs, $Q$ is the number of discordant pairs, and $n$ is the number of parties. The value of $\tau$ ranges from -1 to 1. A value of 1 indicates that the two orderings are identical, 0 indicates no correlation, and -1 indicates that the two orderings are in reverse order.

The Longest Common Subsequence(LCS) ratio measures the ratio between the length of the sequence and the longest common subsequence of the two orderings. The value of the LCS ratio ranges from 0 to 1. A value of 1 indicates that the two orderings are identical, 0 indicates no common subsequence. The formula is shown below:
\begin{center}
  \[	\mathrm{LCS\ ratio} = \frac{2 \times \mathrm{LCS}}{n_1 + n_2} \]
\end{center}

where $n_1$ and $n_2$ are the number of parties in the two orderings and $\mathrm{LCS}$ is the length of the longest common subsequence of the two orderings.

An example case of the comparison is the topic of using nuclear power in Japan. The order obtained by Mielka experts is(most against to most for usage of nuclear power):
\begin{center}
\textbf{SDP}, \textbf{Reiwa}, \textbf{JCP}, \textbf{CDP}, \textbf{Komeito}, \textbf{NDP}, \textbf{LDP}, \textbf{JRP}
\end{center}

The order obtained by our methodology is (we consider the ordering of the average positions of the parties):

\begin{center}
	\textbf{JCP}, \textbf{Okinawa}, \textbf{Reiwa}, \textbf{CDP}, \textbf{JRP}, \textbf{NDP}, \textbf{LDP}, \textbf{Komeito}
\end{center}

Now using the 3 metrics mentioned, we find that the measurements are:
\begin{itemize}
	\item Spearman's $\rho$ = 0.9642
	\item Kendall's $\tau$ = 0.9047
	\item LCS ratio = 0.7142
\end{itemize} 
This indicates that the orderings are highly correlated and that our obtained results are aligned with expert predictions. 
Note that we have ignored parties that are not contained in both of the orderings which could be due to 
\begin{itemize}
	\item The party not having seats in the parliament but Mielka still including that party in their estimations
	\item The party having seats in the parliament but Mielka not including those in their estimation
\end{itemize}

See table \ref{table:spearman} for the correlation values for different topics. (Note: we were not able to validate all our scaled results as Mielka does not scale the parties for all of our topics.)

\begin{table}[h]
\centering
\begin{tabularx}{\columnwidth}{|X|X|X|X|}
\hline
\textbf{Topic} & \textbf{Spearman's $\rho$}& \textbf{Kendall's $\tau$} &\textbf{LCS ratio}\\ \hline
(Ours) Acknowledge JSDF in constitution & 0.9286 & 0.8571& 0.625\\ \hline
\citep{kato2024lupinllmbasedpoliticalideology} Acknowledge JSDF in constitution & 0.8857 & 0.7333& 0.5\\ \hline
(Ours) Restarting nuclear power plants & 0.9642 & 0.9047& 0.7142\\ \hline
\citep{kato2024lupinllmbasedpoliticalideology} Restarting nuclear power plants & 0.4857 & 0.3333& 0.8333\\ \hline

Reducing the consumption tax rate & -0.7 & -0.6 & 0.8\\ \hline
\end{tabularx}
\caption{Correlation metrics for different topics}
\label{table:spearman}
\end{table}

For the topic of reducing the consumption tax rate, we see that the Spearman's $\rho$ and Kendall's $\tau$ are negative. This is because the Reiwa party is placed on the opposite side of the ordering. However, all of the other parties are placed in the same order as the expert predictions and hence the LCS ratio score is high. For the topic of acknowledgement of JSDF in constitution and restarting nuclear power plants, we have compared our correlation values to the correlation values obtained by the ordering of \citeauthor{kato2024lupinllmbasedpoliticalideology}. We see that our methodology obtains a higher correlation value when compared to the expert predictions. Overall, the above results indicate that our methodology is able to robustly scale the political stance of parliamentary representatives on different topics.

\subsection{Qualitative evaluation: Common noun phrases in the different groups of scaled representatives}
\label{section: qual-eval}
\begin{figure}
\centering
  \centering
  \includegraphics[width=0.8\linewidth]{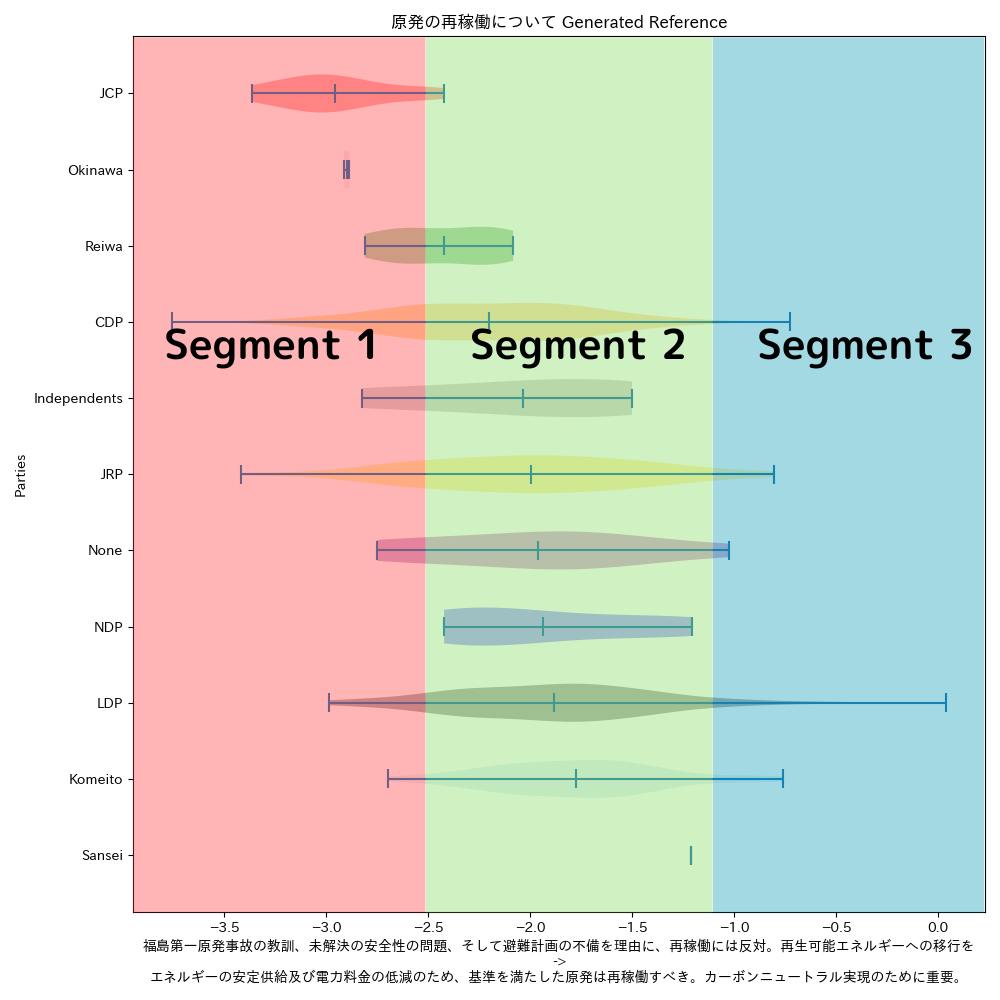}
  \caption{Visual representation of splitting the representatives into 3 clusters}
  \label{fig:3clusters}
\end{figure}

In order to evaluate the scaled results qualitatively, we have split the representatives into 3 clusters based on their scaled positions. We simply split the whole range of the scale into 3 segments and allocated the representatives that fall into the segment to the respective segments. A visual representation is shown in figure \ref{fig:3clusters}.

We then calculated the Pointwise Mutual Information(PMI)\citep{church-hanks-1990-word} of the noun phrases in the speeches of the representatives and the clusters of the representatives. The PMI is calculated as follows:
\begin{center}
	\[	\mathrm{PMI}(x,y) = \log_2 \left( \frac{P(x,y)}{P(x)P(y)} \right) \]
\end{center}

where $P(x,y)$ is the probability of the noun phrase $x$ and the cluster $y$ co-occurring, $P(x)$ is the probability of the noun phrase $x$ occurring and $P(y)$ is the probability of the cluster $y$ occurring. The PMI score indicates how much more likely the noun phrase $x$ and the cluster $y$ co-occur than if they were independent. A high PMI score indicates that the noun phrase is highly associated with the cluster. See below tables \ref{table:PMI-nuclear}, \ref{table:PMI-constitution} and \ref{table:PMI-taxrate} for examples of the noun phrases we have identified to be highly associated with the different clusters of representatives for the topics: Restarting Nuclear Power Plants in Japan, Changing the Japanese constitution to acknowledge the JSDF and Reducing the consumption tax rate respectively.

\subsubsection{Key Observations}
Below we outline some key observations from the PMI analysis:
\begin{itemize}
	\item \textbf{Restarting Nuclear Power Plants in Japan:}
		\begin{itemize}
			\item The representatives in the first segment(against) are associated with phrases like "Renewable energy emphasis", "High cost", "Anti-nuclear", "Fukushima nuclear plant", "Earthquake safety measures" etc. This indicates that the representatives who belong to this segment often mention the Fukushima nuclear disaster and are generally against the usage of nuclear power. 
			\item The representatives in the middle segment mention a mix of terms such as "Lack of transparency", "Trial", "Temporary housing", "Contaminated water treatment" and "Energy self-sufficiency rate". This might indicate that while the representatives in this segment are aware of the risks associated with nuclear power and do mention the aftermath of the Fukushima disaster, they are also aware of the imporant role nuclear power could play in Japan's energy self-sufficiency.
			\item The representatives in the last segment(for) mention terms such as "Industrial Policy", "Power sector", "Innovative nuclear technology", "Decarbonization", "Safety first" and "Need to reduce dependency". This indicates that the representatives in this segment are aware of the importance of safety in nuclear power plants but also emphasize the important role nuclear power plays in decarbonization and reducing dependency on other countries for energy.
		\end{itemize}
	\item \textbf{Changing the Japanese constitution to acknowledge the JSDF:}
		\begin{itemize}
			\item The representatives in the first segment(against) are associated with phrases like "Hazardous waste", "Land use regulation bill", "Opposition", "Article 9 of the Constitution", "Constitutional reinterpretation" and "Possession of counterattack capability", etc. These words convey that this group if hesitant to changing the constitutional interpretation of the JSDF and are concerned about the militarization of Japan.
			\item The representatives in the middle segment mention a mix of terms such as "Diplomatic dialogue", "Defense capability enhancement", "Land, sea and air", "Military integration","Cyber attack", etc. These words indicacte that while this segment still has concerns about militarization and prioritize diplomatic dialogues, they also acknowledge the importance of defence capability enhancement and strengthening Japan against cyberattacks from other countries. 
			\item The representatives in the last segment(for) seem to be mentioning more technical terms. They are associated with words such as "Unmanned asset defence capability", "Analog communication", Unmanned equipment", "Competitiveness enhancement". This indicates that this segment is more focused on the technology necessary to strengthen Japan's defence capabilities and are in favor of changing the constitution to acknowledge the JSDF which would solidify the status of the JSDF in Japan.
		\end{itemize}
	\item \textbf{Reducing the consumption tax rate:}
	\begin{itemize}
		\item The representatives who are against the reduction of consumption tax rate are associated with terms such as "Public broadcasting", "Public school", "Employment system reform", "Ministry of defence", etc. This indicates that this segment is more concerned about the deterioration of public services due to reduced tax revenue. This segment seems to be emphasizing employment system reforms more to boost the economy rather than reducing the consumption tax rate.
		\item The representatives in the middle segment are assocaited with a mix of terms such as "Grant-type scholarship", "Regional cities", "Economic security", "Small and micro enterprises", etc. These terms indicate that this group broadly mentions the macroeconomic implications of reducing the consumption tax rate.
		\item The representatives who are for reducing the consumption tax rate are associated with terms sucha as "Young and child-rearing generation", "Common people", "Domestic consumption", "Demand expansion", "Rising energy costs", etc. These words indicate that this segment is more focused on reducing the financial pressure on the younger generation and common consumers to boost the demand for goods in the economy. 
	\end{itemize}
\end{itemize}

Overall the PMI analysis has provided us with a deeper understanding of the stance of the representatives in the different clusters. The words that are associated with the different clusters were aligned with the issue positioning of the clusters which indicates that our scaling methodology is able to accurately capture the stance of the representatives on different topics.

\begin{table*}[htbp]
\centering
\renewcommand{\arraystretch}{1.5}%
\begin{tabularx}{\textwidth}{|>{\centering\arraybackslash}X|>{\centering\arraybackslash}X|>{\centering\arraybackslash}X|}
\hline
\textbf{Segment 1(against)} & \textbf{Segment 2(middle)} & \textbf{Segment 3(for)} \\ \hline
\begin{tabular}[c]{@{}l@{}}
Renewable energy emphasis \\ High cost \\ Nuclear-powered ships \\ Forestry revitalization \\ Its preparation \\ Textbook \\ Appeal \\ Nuclear-related companies \\ Secrecy \\ Industrial accident \\ Three times \\ Compensation scheme \\ Failure to function properly \\ Anti-nuclear \\ Japan-U.S. alliance \\ Corporate culture \\ Earthquake safety measures \\ Support \\ Recently \\ Nuclear issue \\ Prime Minister’s Office \\ Shika Nuclear Power Plant \\ Doubts \\ Fukushima nuclear plant \\ Pointing out ignorance of lessons learned \\ Information concealment 
\end{tabular} 
& 
\begin{tabular}[c]{@{}l@{}}
	Interests \\ Lack of transparency \\ Necessary personnel \\ Increase in imports \\ Energy self-sufficiency rate \\ Trial \\ Case law \\ Temporary housing \\ Roadmap \\ Nuclear facilities \\ Specific initiatives \\ Small modular reactor \\ Others \\ Industrial promotion \\ Carbon neutrality \\ Contaminated water treatment \\ Evacuation shelter \\ Fire brigade \\ Regional nuclear disaster prevention-\\ council \\ Hamaoka Nuclear Power Plant \\ Police \\ A municipality \\ Base \\ New industry \\ Regional development measures
\end{tabular} 
& 
\begin{tabular}[c]{@{}l@{}}
	Industrial policy \\ Sense of fulfillment \\ Power sector \\ Uniqueness \\ Local companies \\ Innovative nuclear technology \\ Nuclear damage compensation system \\ CO2 emission reduction \\ Decommissioning business \\ Safety first \\ Highest priority on safety \\ Decarbonization \\ Hokkaido \\ Nuclear power industry \\ Need to reduce dependency \\ Worldwide \\ Nuclear power plant operation period \\ Completion of decommissioning \\ Nuclear abolition \\ Small modular reactor \\ Carbon dioxide emissions \\ Nuclear engineers \\ Expanding options \\ Regional disaster prevention plan \\ Resident support
\end{tabular} \\ \hline
\end{tabularx}
\caption{Example of noun phrases that have high PMI with different clusters of representatives for the topic: Restarting Nuclear Power Plants in Japan}
\label{table:PMI-nuclear}
\end{table*}

\begin{table*}[htbp]
\centering
\renewcommand{\arraystretch}{1.5}%
\begin{tabularx}{\textwidth}{|>{\centering\arraybackslash}X|>{\centering\arraybackslash}X|>{\centering\arraybackslash}X|}
\hline
\textbf{Segment 1(against)} & \textbf{Segment 2(middle)} & \textbf{Segment 3(for)} \\ \hline
\begin{tabular}[c]{@{}l@{}}
	Hazardous waste \\ Land use regulation bill \\ This bill \\ Opposition \\ Government \\ Fiscal reconstruction \\ Iraq \\ Cautious \\ Government \\ International peace cooperation activities \\ Inter-ministerial \\ Childcare \\ Additional financial resources \\ Special committee \\ No upper limit \\ Full-spec \\ Germany \\ Article 9 of the Constitution \\ Alternative \\ Method \\ Official documents \\ Current administration \\ Constitutional reinterpretation \\ Possession of counterattack capability \\ Asian countries \\ Constitutional interpretation change
\end{tabular} 
& 
\begin{tabular}[c]{@{}l@{}}
	Diplomatic dialogue \\ Cautious \\ Defense capability enhancement \\ Land, sea, and air \\ Strategic document \\ Military integration \\ Government-sponsored bid rigging \\ News \\ Capitalist \\ Rebellion \\ Foreign capital regulation \\ Cyber attack \\ Uniformed personnel \\ Election \\ Backup base \\ Constitutional amendment debate \\ Production base \\ Island regions \\ Emergency response \\ Local government \\ Significant budget increase \\ State of emergency \\ Debt \\ Consent \\ Final \\ Disaster relief \\ Water source area \\ Directly under the Prime Minister
\end{tabular} 
& 
\begin{tabular}[c]{@{}l@{}}
	United Kingdom \\ Unmanned asset defense capability \\ Analog communication \\ Public transportation \\ Aircraft manufacturing industry \\ Civil sector \\ Unmanned equipment \\ Marine survey \\ Civilian technology \\ Code of ethics \\ Methane hydrate \\ Technology security \\ Broad sense \\ Competitiveness enhancement \\ Like-minded countries \\ Human \\ Multilateral cooperation \\ Stability assurance \\ Critical materials \\ World without \\ AI \\ Nuclear deterrence \\ Vulnerability \\ Three principles on arms exports \\ Economic revitalization \\ Marine resources \\ Those unfamiliar \\ Production capacity
\end{tabular} \\ \hline
\end{tabularx}
\caption{Example of noun phrases that have high PMI with different clusters of representatives for the topic: Changing the Japanese constitution to acknowledge the JSDF}
\label{table:PMI-constitution}
\end{table*}

\begin{table*}[htbp]
\centering
\renewcommand{\arraystretch}{1.5}%
\begin{tabularx}{\textwidth}{|>{\centering\arraybackslash}X|>{\centering\arraybackslash}X|>{\centering\arraybackslash}X|}
\hline
\textbf{Segment 1(against)} & \textbf{Segment 2(middle)} & \textbf{Segment 3(for)} \\ \hline
\begin{tabular}[c]{@{}l@{}}
	Public broadcasting \\ Public school \\ Care work \\ Living expenses \\ Employment system reform \\ Young seafarers \\ Stationed military forces \\ Procurement system \\ Employment security \\ Military priority \\ Equal or greater \\ Shipbuilding industry \\ Land acquisition \\ Defense industry \\ Care \\ Ministry of Defense \\ Professional occupation \\ Africa \\ Healthcare workers \\ Fiscal stability \\ Careful explanation \\ Ministry of Land, Infrastructure,-\\ Transport and Tourism \\ Credit guarantee system \\ Investment agreement \\ Medical institutions \\ Casino legalization \\ Inclusive society \\ Male \\ Disaster recovery
\end{tabular} 
& 
\begin{tabular}[c]{@{}l@{}}
	Grant-type scholarship \\ Support \\ International situation \\ Labor contract law \\ Technology transfer \\ Regional cities \\ Economic security \\ Problem solving \\ Support \\ Regional economy \\ Small and micro enterprises \\ Poverty issue \\ Hometown tax system \\ Universal service \\ Passive smoking \\ Fixed-sum benefit \\ Domestic vaccine \\ Private sector-led \\ At that time \\ Low-income support \\ Private high school \\ Driver \\ International cooperation \\ Infectious disease law \\ Ruling party \\ Environmental regeneration \\ Regional characteristics \\ Supplementary budget proposal \\ Ukraine situation
\end{tabular} 
& 
\begin{tabular}[c]{@{}l@{}}
	Animal welfare \\ There \\ Additional child allowance \\ Young and child-rearing generation \\ Common people \\ Domestic consumption \\ Game meat \\ Demand expansion \\ Import price \\ Food labeling \\ Asset deflation \\ Carbon pricing \\ Rising energy costs \\ Based on remarks \\ New business model \\ Fuel price \\ Both sides \\ Consumer behavior \\ Consumer price index \\ Post-COVID \\ Business succession tax system \\ Both parties \\ Public-private partnership \\ Between companies \\ Based on remarks \\ Loan system \\ Merit \\ Energy cost \\ Mid-sized enterprises
\end{tabular} \\ \hline
\end{tabularx}
\caption{Example of noun phrases that have high PMI with different clusters of representatives for the topic: Reducing the consumption tax rate}
\label{table:PMI-taxrate}
\end{table*}
\FloatBarrier

\subsection{Event-based evaluation: Diachronic analysis of party positions on political axes}
\label{section: event-based}
As shown previously, we have conducted a diachronic analysis on mean issue positions of parties on topics of nuclear power and defence. This section will introduce some key events which we believe explains the trends we have seen. 

\subsubsection{Defence}
\begin{figure}
	\centering
	\includegraphics[width=\linewidth]{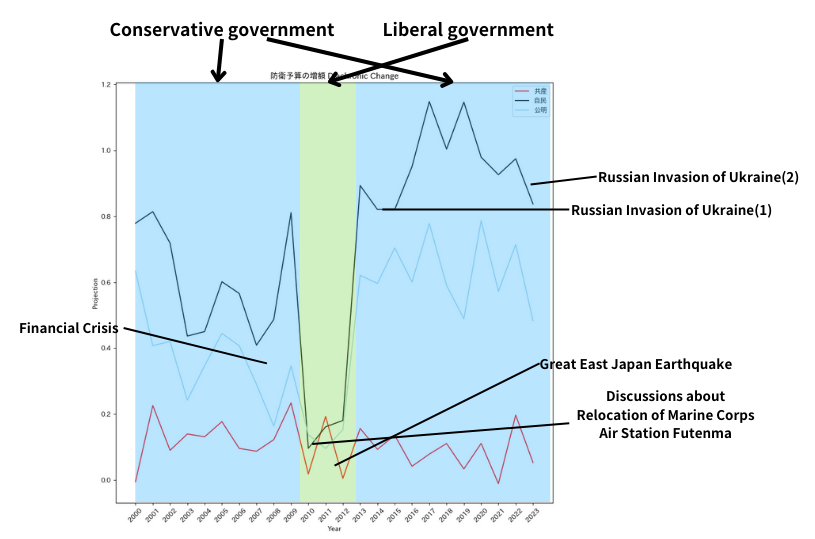}
	\caption{Labelled diachronic analysis of party positions on the topic of increasing defence budget}
	\label{fig:diachronic-defence}
\end{figure}

See figure \ref{fig:diachronic-defence} for the diachronic analysis of party positions on topic of increasing defence budget with the key events. We see that the LDP and Komeito have been in favor of increasing defence budget until 2009 with Komeito being more cautious towards it. However in the period 2009-2012, we see that both these parties have shifted their stance to be more against increasing the defence budget. This aligns with the period where the Democrats were in power indicating that the LDP and Komeito adopted a stance that was more against the budgeting decisions of the Democrats. During this time, there also was discussion on the Relocation of the US Marine Corps Air Station Futenma and the Democrats came under great scrutiny for their handling of the issue. This also could have influenced the stance of the LDP and Komeito to be more against the increased defence budget. Another factor that might have contributed to the shift is the Great East Japan Earthquake in 2011 which might have shifted the focus of the parties to more pressing issues. Lastly, after 2014 we see that LDP has shifted its stance to be more in favor of increasing the defence budget. This aligns with the Russian invasion of Crimea in 2014 which triggered great concern in Japan. 

\subsubsection{Nuclear Power}
\begin{figure}
	\centering
	  \centering
	  \includegraphics[width=\linewidth]{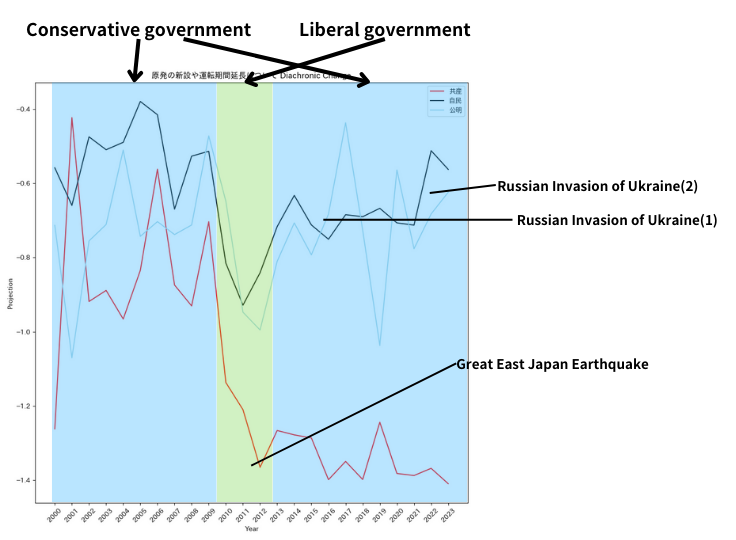}
	  \caption{Labelled diachronic analysis of party positions on the topic of building new nuclear powerplants or extending the lifetime of existing ones}
	  \label{fig:diachronic-nuclear}
\end{figure}

See figure \ref{fig:diachronic-nuclear} for the diachronic analysis of party positions on topic of building new nuclear powerplants or extending the lifetime of existing ones with the key events. We see that the LDP and Komeito have been in favor of building new nuclear powerplants or extending the lifetime of existing ones until 2011. However, after the Fukushima nuclear disaster in 2011, we see that the LDP and Komeito have shifted their stance to be more cautious towards nuclear power. This cautious attitude lasted until around 2013 where we see a rebound in pro-nuclear stance by the LDP and Komeito, possibly fuelled by the energy crisis that Japan faced due to the shutdown of nuclear powerplants as well as the Russian invasion of Crimea in 2014 and invasion of mainland Ukraine in 2022. The JCP has been against building new nuclear powerplants or extending the lifetime of existing ones from around 2009 onwards. 

Overall, our diachronic analysis of issue positioning of Japanese parties were aligned with our understanding of the events that have influenced Japanese politics during the timeperiod. This indicates that our methodology can also be used to conduct a dynamic analysis of shifts in party positions on different political axes and is not limited to a static analysis concerned only with the current positions of the parties.

\FloatBarrier

\section{Presenting the results to the public}
\label{section: kokkaidoc}
The authors have created a web application to make this research public and to empower the voters to make more nuanced voting decisions. As mentioned previously, quantifying and scaling the political issue stance of parliamentary representatives is essential for the transparency and productivity of democracy and it was important for us to create a platform where the research can be leveraged practically by the public. The website, reachable at \url{kokkaidoc.com} already makes the findings of \citeauthor{kato2024lupinllmbasedpoliticalideology} public and publishes parliamentary voting results as well as the current distribution of electoral districts. We are currently working on updating the website to include the findings from this research.

\begin{figure}[h]
\centering
    \begin{subfigure}{0.22\textwidth}
      \centering
      \includegraphics[width=1\linewidth]{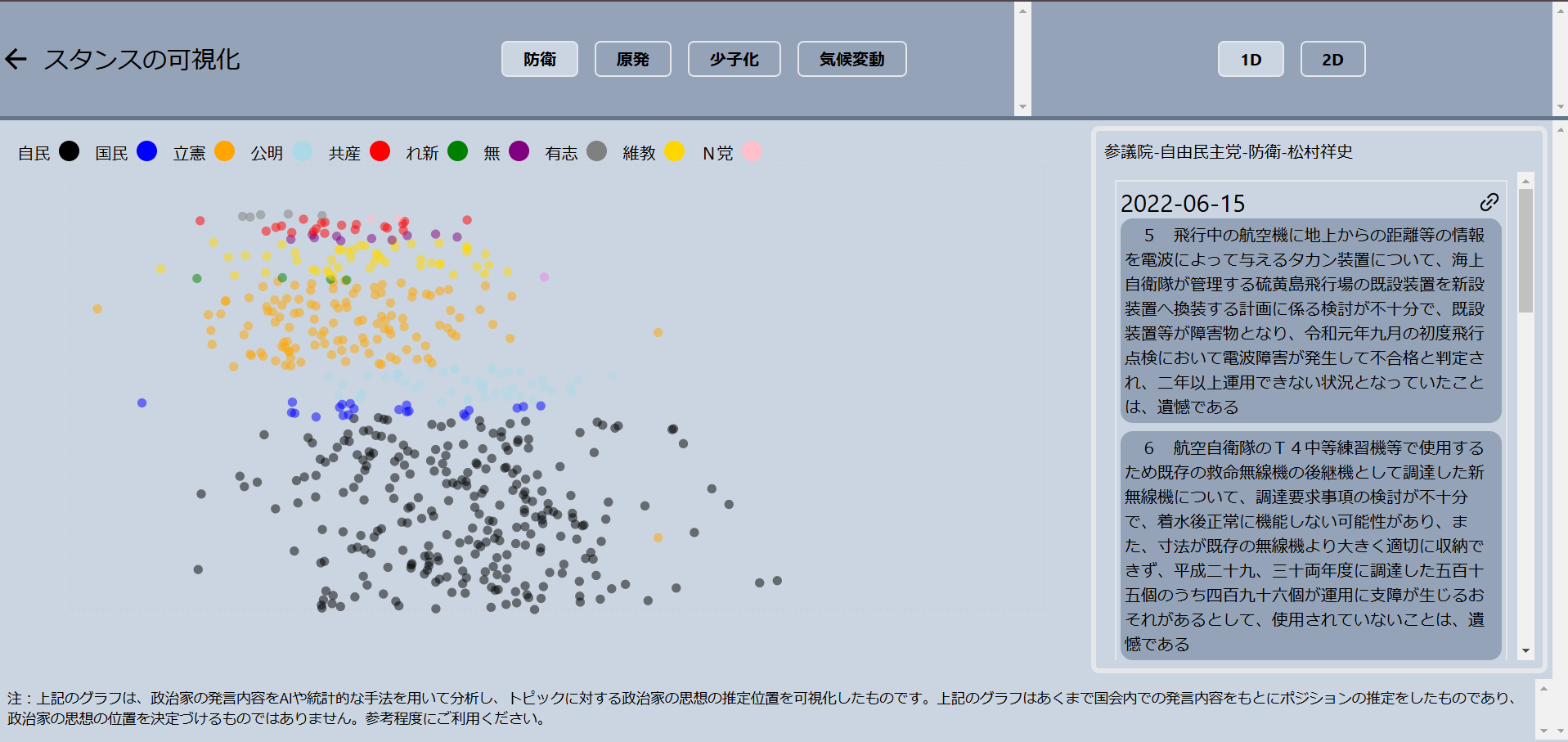}
      \caption{Scaling results presented on \url{kokkaidoc.com}}
    \end{subfigure}
    \begin{subfigure}{0.22\textwidth}
      \centering
      \includegraphics[width=1\linewidth]{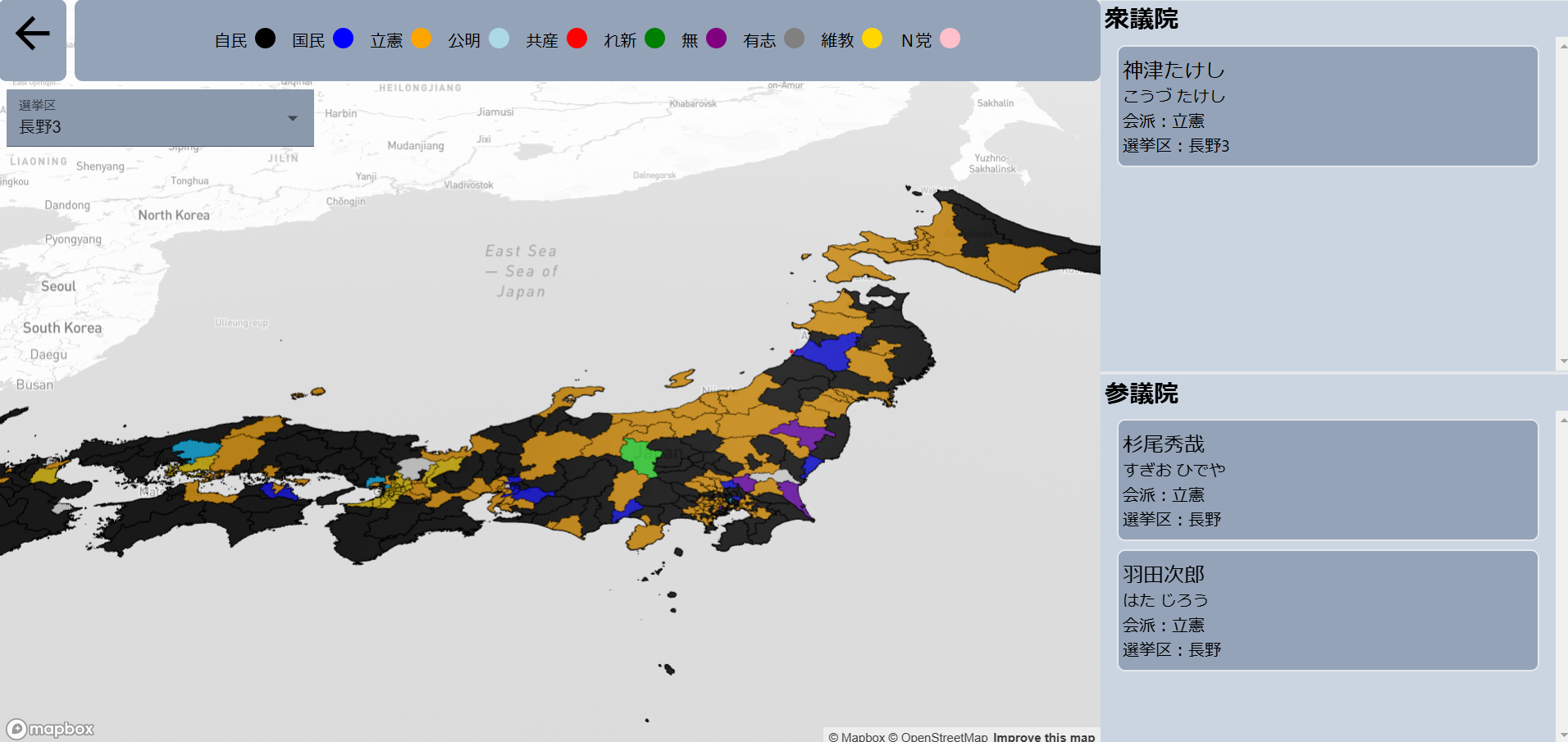}
      \caption{Electoral district map presented on \url{kokkaidoc.com}}
    \end{subfigure}
    \begin{subfigure}{0.22\textwidth}
      \centering
      \includegraphics[width=1\linewidth]{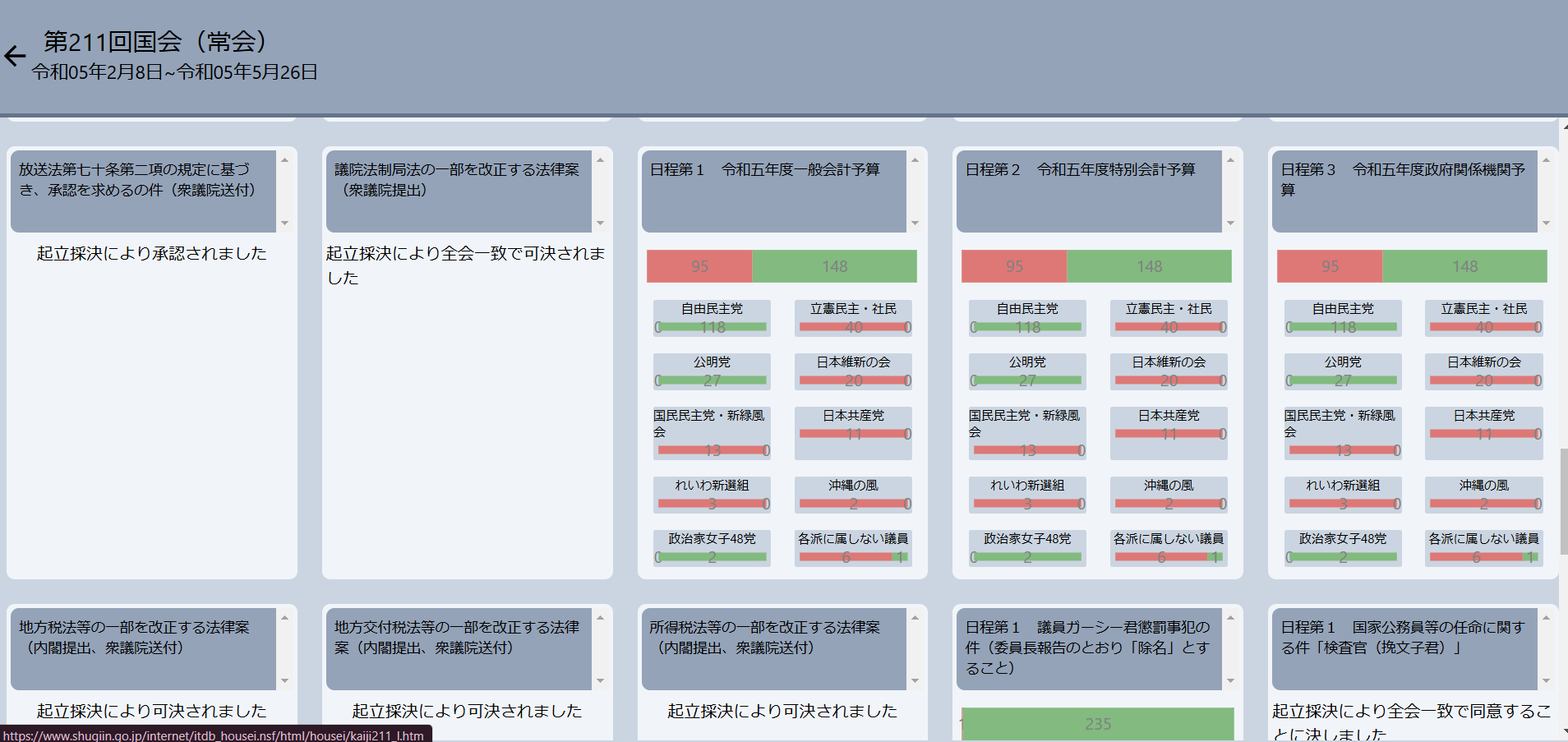}
      \caption{Voting results presented on \url{kokkaidoc.com}}
    \end{subfigure}
\caption{\url{kokkaidoc.com}}
\label{fig: kokkaidoc-figs}
\end{figure}

\section{Conclusion}
This research presents an LLM-driven framework for scaling the political issue stance of parliamentary representatives, addressing key limitations in previous methodologies. By implementing speech summarization for noise reduction, automatic extraction of axes of controversy, and diachronic analysis of ideological shifts, our approach provides a more refined, rounded and scalable method for political stance estimation. 

During our validation process, we showed quantitatively that our estimations align with expert prediction while qualitatively, we identified the semantics associated with the different clusters of representatives. These two evaluations indicated that the proposed methodology is robust and accurate in scaling the political issue stance of parliamentary representatives. Furthermore, our diachronic analysis of party positions aligned closely with political events that have influenced Japanese politics, demonstrating the effectiveness of our method in capturing dynamic shifts in party positions.

Our research also went beyond the academic realm by creating a web application to make this research public to be practically leveraged by the public. This application will empower Japanese voters in the future to make a more informed and nuanced voting decision in future elections.

Future work could extend this approach to broader political systems and explore ways to extract more meaningful axes of controversy. By continuously improving the framework, we aim to contribute to a more transparent and data-driven understanding of parliamentary ideologies.

\FloatBarrier
\clearpage
\appendix
\section{Previous work}

\FloatBarrier
\begin{figure}
\centering
    \begin{subfigure}{0.3\textwidth}
      \centering
      \includegraphics[width=1\linewidth]{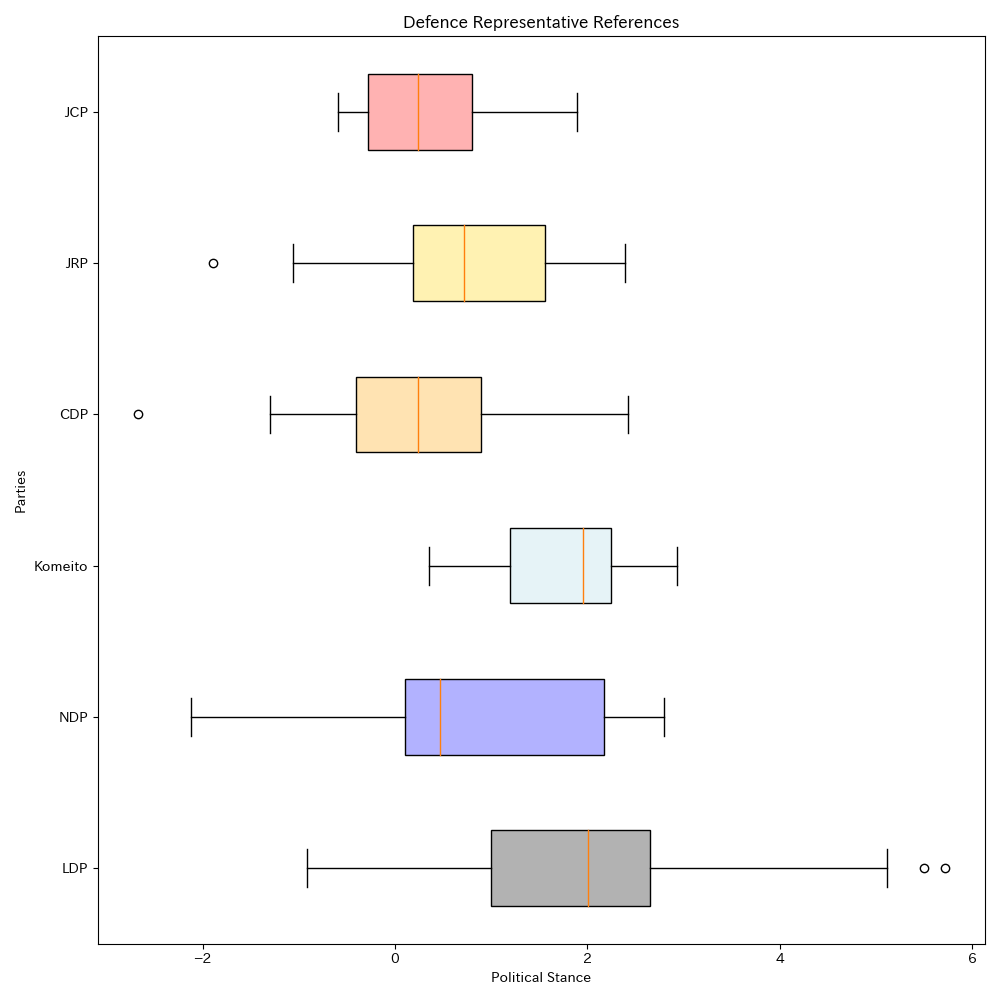}
      \caption{Box plot of result obtained by \citeauthor{kato2024lupinllmbasedpoliticalideology} on the topic of acknowledgment of JSDF in Japanese constitution}
      \label{fig:sub1}
    \end{subfigure}
    
    \begin{subfigure}{0.3\textwidth}
      \centering
      \includegraphics[width=1\linewidth]{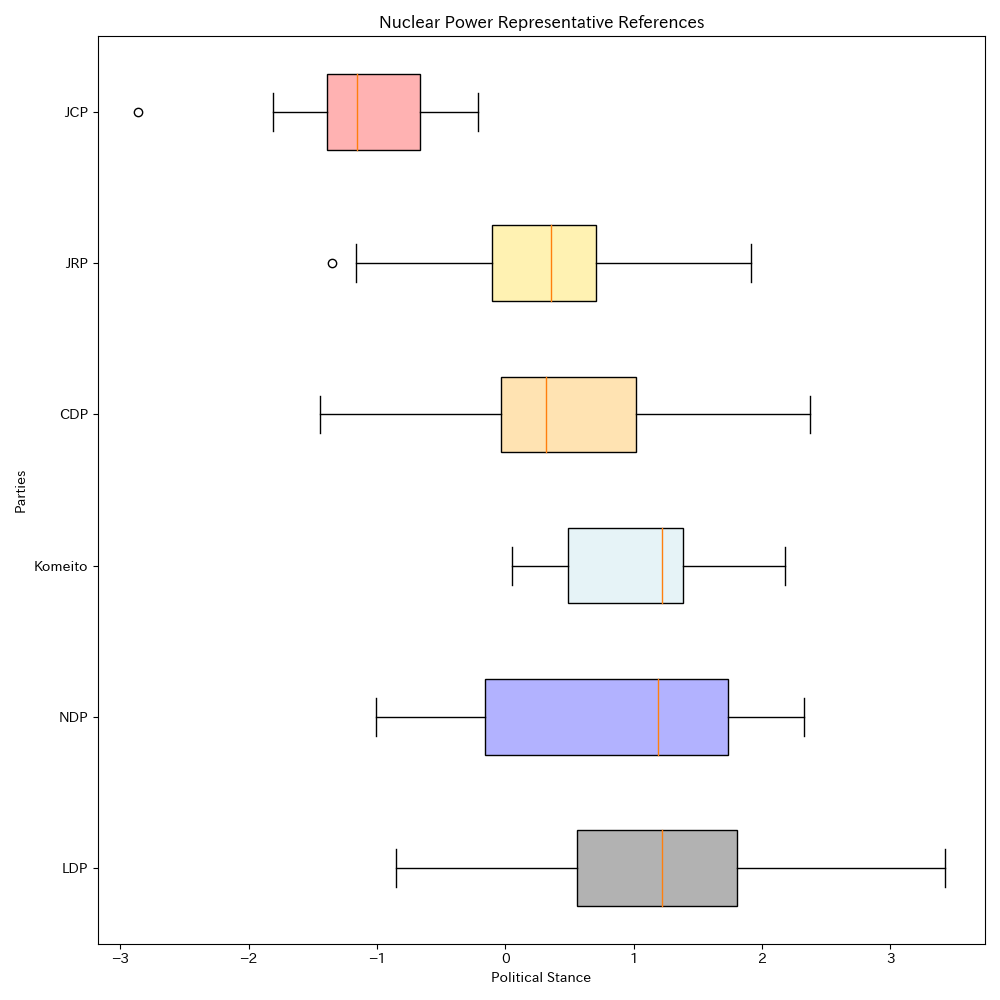}
      \caption{Box plot of result obtained by \citeauthor{kato2024lupinllmbasedpoliticalideology} on use of nuclear power in Japan}
      \label{fig:sub2}
    \end{subfigure}
\caption{Box plot results scaling politicians on different political controversies\citep{kato2024lupinllmbasedpoliticalideology}}
\label{fig: Previous box plots}
\end{figure}

\begin{figure}[h]
\centering
    \begin{subfigure}{0.3\textwidth}
      \centering
      \includegraphics[width=1\linewidth]{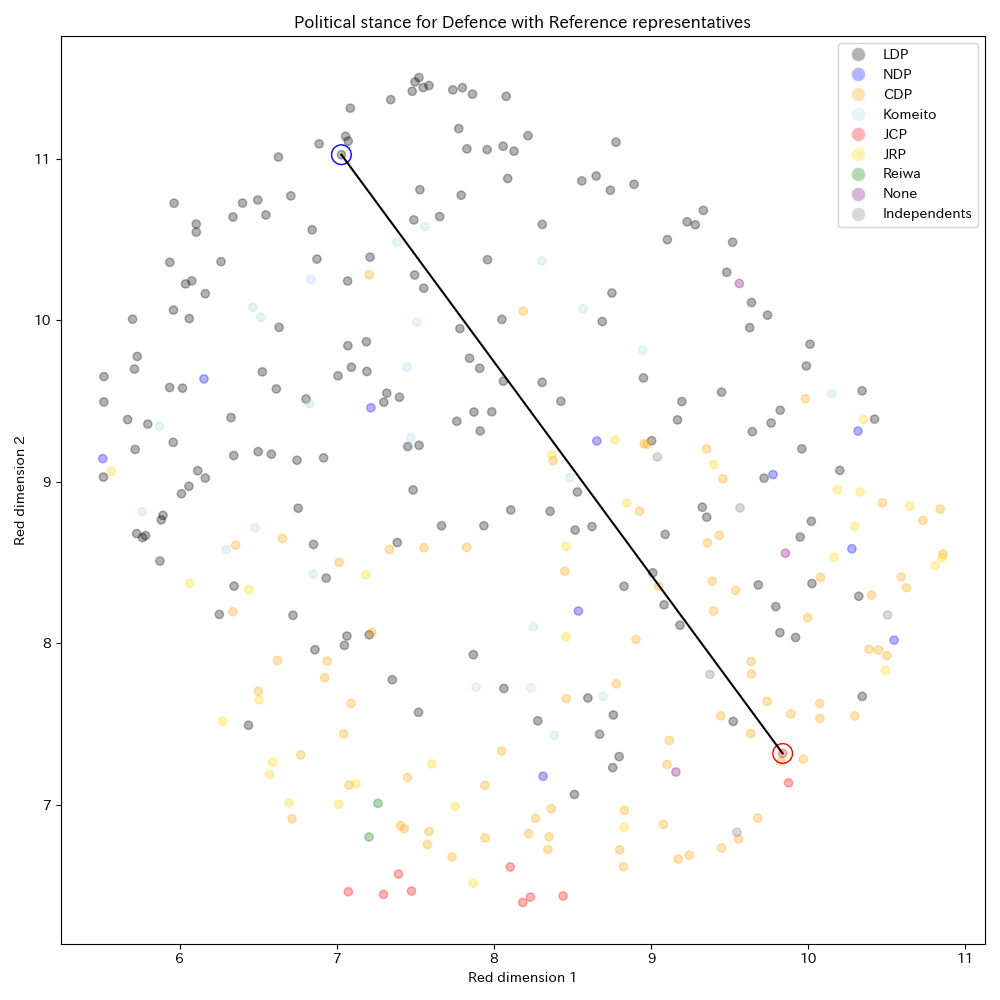}
      \caption{UMAP plot of topic of defence, black line connecting the politician references}
      \label{fig:sub1}
    \end{subfigure}
    
    \begin{subfigure}{0.3\textwidth}
      \centering
      \includegraphics[width=1\linewidth]{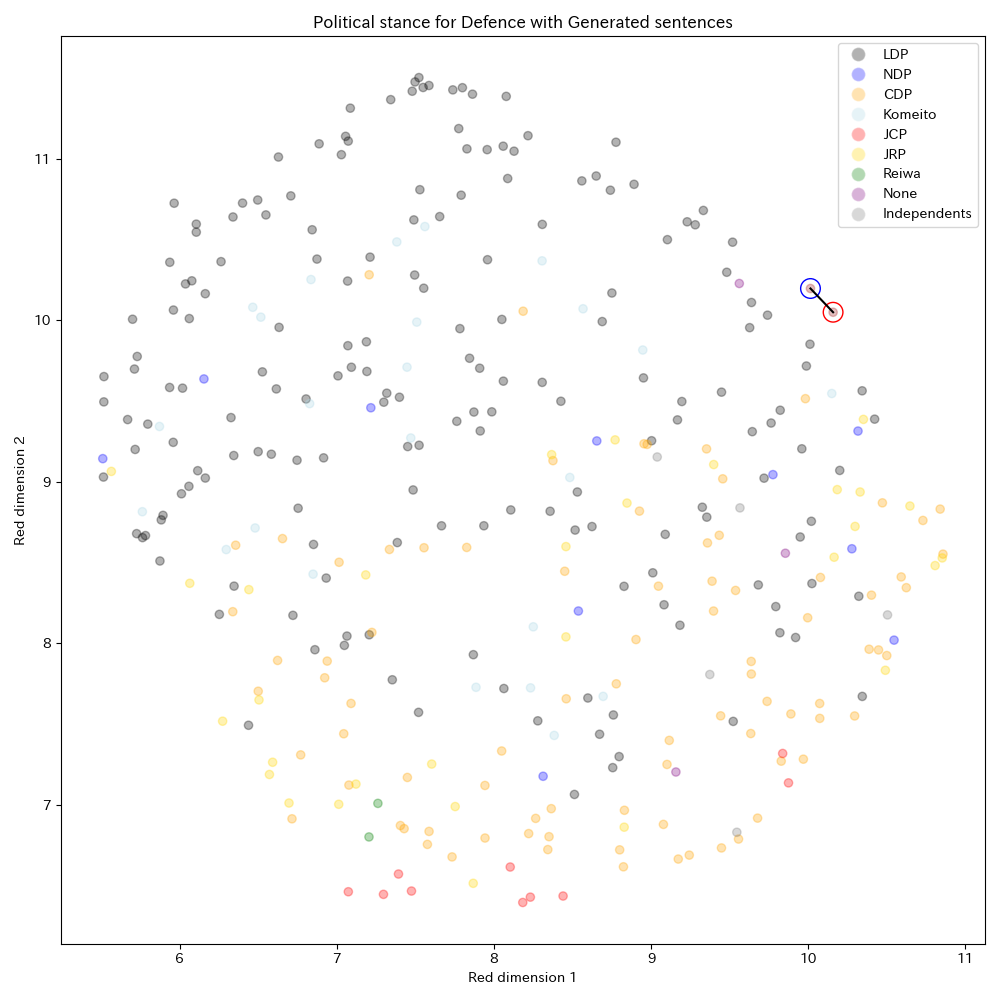}
      \caption{UMAP plot of topic of defence, black line connecting the generated references}
      \label{fig:sub2}
    \end{subfigure}
\caption{UMAP plot of the opinion embeddings of representatives on the topic of acknowledgement of JSDF in the constitution \citep{kato2024lupinllmbasedpoliticalideology}}
\label{fig: Previous umap}
\end{figure}

\FloatBarrier
\section{GPT prompts and replies}
\label{app:gpt-prompts}

\begin{figure}[htbp]
    \centering
    \begin{quote}
    \begin{CJK}{UTF8}{min}
      この政治家の発言を要約すると、彼の日本の防衛に対するスタンスは以下のように推察できます：

1. **専守防衛の重視**：この政治家は、日本の防衛政策は専守防衛に徹するべきであり、攻撃的な行動や武力行使は行わないという姿勢を堅持することを強調しています。

2. **集団的自衛権の批判**：政府が推進している集団的自衛権の行使を強く批判しており、これが日本を戦争に巻き込むリスクがあると懸念しています。憲法違反と見なされる可能性があり、日本の防衛政策の根幹を覆すものであるとしています。

3. **米軍との関係に対する警戒**：自衛隊が米軍と一体化して行動することに対して警戒感を表明し、特に米国の先制攻撃に日本が巻き込まれる可能性を問題視しています。

4. **外交の重視**：軍事行動よりも外交による紛争解決や平和的手段を模索するべきだとしており、特にイラン核合意への働きかけなどを通じた地域の緊張緩和を提唱しています。

5. **憲法改正への反対**：憲法九条に自衛隊を明記することが専守防衛の枠組みや集団的自衛権に関する解釈の拡張につながるとして反対の立場です。

この政治家は、日本の防衛政策が専守防衛を基盤とし、外交による解決を重視しつつ、集団的自衛権の行使を避けるべきとする平和志向のスタンスを堅持しています。
    \end{CJK}
    \end{quote}
    \caption{Example of formatted summary reply by GPT4o-mini}
    \label{fig:formatted reply GPT4o-mini summary}
\end{figure}
\begin{figure}[htbp]
    \centering
    \begin{quote}
    \begin{CJK}{UTF8}{min}
     麻生国務大臣の日本の防衛に対するスタンスは、防衛について非常に重視し、日本の安全保障を確保するために慎重かつ計画的に様々な施策を実行していることが伺えます。彼は、日本を取り巻く安全保障環境が厳しいと認識しており、それに対応するための実効的な防衛力の確保が重要だと考えています。また、日米安全保障条約の意義を強調し、米国との堅牢な関係維持が抑止力の確保に重要であるとみなしています。特に、北東アジアでの不安定要因や国際社会でのテロ、核拡散といった新たな脅威に対し、防衛力や外交努力による安定確保を最重要視しています。彼は、防衛政策の質と量の確保、また透明性の向上、地元負担の軽減と抑止力維持の両立など、多面的な要素を考慮した戦略を追求し続けていることが明確です。
    \end{CJK}
    \end{quote}
    \caption{Example of essay-style reply by GPT4o-mini}
    \label{fig:essay reply GPT4o-mini summary}
\end{figure}

\begin{figure}[htbp]
    \centering
    \begin{quote}
    \begin{CJK}{UTF8}{min}
        国会の議事録APIを使って「防衛」に関する政治家の発言を集めようと思います。その際に使うべき適切な検索単語をリストとして返してください。
    
        例：
        
        トピック：防衛
        
        検索単語：["自衛隊", "集団的自衛権", "安全保障", "軍事力", "防衛"]
        
        トピック：原発
        
        検索単語：["原子", "原発", "廃炉", "再稼働", "ベースロード"]
        
        トピック：経済対策
        
        検索単語：["経済", "景気", "デフレ", "財政政策", "金融政策", "インフレ", "中小企業", "雇用", "スタートアップ", "消費", "最低賃金"]
        
        結果はそのままコピーできるように[]で囲ってください。\newline
        \textbf{English Translation: }
        
        I would like to use the Parliamentary Proceedings API to collect statements from politicians regarding "defense". Please return a list of appropriate search terms that should be used for this purpose.
    
        Example:
        \newline
        Topic: Defense
        
        Search words: ["Self-Defense Forces", "collective self-defense", "security", "military power", "defense"].
        
        Topic: Nuclear Power
        
        Search words: ["atomic", "nuclear", "decommission", "restart", "baseload"].
        
        Topic: economic measures
        
        Search words: ["economy", "economic", "deflation", "fiscal policy", "monetary policy", "inflation", "small business", "employment", "start-up", "consumption", "minimum wage"]
        
        Please enclose the results in [] so that they can be copied verbatim.
        \end{CJK}
    \end{quote}
    \caption{Example of using the GPT4o-mini to generate topic-related search terms.}
    \label{fig:parliamentary_api_example}
\end{figure}

\begin{figure}[htbp]
    \centering
    \begin{quote}
        Below is a text summarizing the stance on TOPIC of several legislators extracted from the parliamentary proceedings.
        Please use this text as a basis for identifying issues related to this topic, and describe the polar opposing views on these issues.
        
        Response Format (Please be sure to respond in this format) 

        \textbf{Issue:} Outline of the issue
        
        \textbf{For:} Opinion in favor of the issue
       
        \textbf{Against:}Opinion in against the issue
        \newline
        
        \textbf{Issue: }Outline of the Issue
       
        \textbf{For:}Opinion in favor of the issue
        
        \textbf{Against:}Opinion in against the issue
        \newline \newline
        
        \textbf{Response Example:}
        
        Discussion Point: The Constitution should clearly state the Self-Defense Forces.
        
        Agree: The legal basis for the Self-Defense Forces is ambiguous. The Constitution should be amended to allow the SDF to operate internationally.
        
        Against: Affirming the existence of the SDF and specifying the SDF in the Constitution should be considered two different issues.
        \newline 
        \newline
        
        Discussion Point: Increasing the defense budget
        
        For: Reference is made to the responsibility to defend one's own country with one's own strength. Mentions the need to update equipment and improve the welfare of personnel.
        
        Against: Mentions the financial difficulties Japan is facing. Limited financial resources should be allocated to social security, education, and other areas.

        ---------------------------------------------------------------
        
       \textbf{ Summary text}

       $\langle$Summaries$\rangle$

       ---------------------------------------------------------------

    \end{quote}
    \caption{Prompting GPT4o-mini to detect axes of controversy within the topic of interest(Translated from Japanese)}
    \label{fig:axes detection}
\end{figure}

\FloatBarrier

\onecolumn
\section{Extracted political axes of debate}
\begin{figure}[h]
\centering
      \begin{lstlisting}[language=json,firstnumber=1]
    [
        {
            "topic": "Constitutional Specification of the Self-Defense Forces",
            "pro": "Since the existence of the Self-Defense Forces is already accepted by the public, their enshrining in the Constitution would strengthen their international credibility and legal stability.",
            "con": "The SDF does not need to be explicitly stated in the Constitution; the existing legal system is sufficient. Specifying it would create legal challenges and could undermine the spirit of exclusive defense."
        },
        {
            "topic": "Increase in defense budget",
            "pro": "In order to respond to changes in the security environment surrounding Japan, defense capabilities must be strengthened. Specifically, it is necessary to introduce advanced technologies such as missile defense and cyber security.",
            "con": "Increased defense spending could put pressure on social security and education budgets. In addition, it is difficult to gain public understanding to allocate reconstruction funds and tax increases to defense spending."
        },
        {
            "topic": "Exercise of the right of collective self-defense",
            "pro": "In order to respond to changes in the international security environment, the exercise of the right of collective self-defense should be clearly stated in the Constitution and defense policy should be modernized",
            "con": "The exercise of the right of collective self-defense deviates from the principle of exclusive defense, and constitutional amendments and changes in interpretation should be pursued with caution. It could increase the risk of war."
        },
        {
            "topic": "Strengthening Food Security",
            "pro": "Increasing food self-sufficiency and strengthening the agricultural production base is important for national security. We need to increase domestic production and stabilize people's livelihoods.",
            "con": "There is a fear that agricultural support may be put on the back burner due to the expansion of the defense budget. In addition, increasing self-sufficiency requires not only state initiative, but also private-sector strength."
        }
    ]
\end{lstlisting}

\caption{Extracted axes of political debate for the topic of defence}
\label{fig: extracted axes defense}
\end{figure}

\begin{figure}[h]
\centering
      \begin{lstlisting}[language=json,firstnumber=1]
[
    {
        "topic": "Strengthening Support for SMEs",
        "pro": "Since SMEs are the bloodstream of the economy and are responsible for the majority of employment, especially in Japan, some argue that policies to improve the business environment and enable wage increases are essential. Further support is also needed because the entire local economy is enriched when SMEs are revitalized.",
        "con": "Some are concerned that strengthening support for SMEs may not be fair to large companies. Others argue that the principle of competition in the market should be respected and excessive intervention should be avoided, regardless of the size of the company."
    },
    {
        "topic": "On Sales Tax Reductions",
        "pro": "A consumption tax cut is an immediate stimulus measure and is expected to stimulate the economy as a whole through stimulating consumption. It is also supported by the viewpoint that it is not socially fair because it is regressive and places a heavy burden on low-income individuals." ,
        "con": "The consumption tax is a stable source of funding for social security, and tax cuts are considered to be an obstacle to fiscal soundness. Concerns have also been expressed that the reduction in funding from a consumption tax cut could affect other social security and public services."
    },
    {
        "topic": "Acceptance of Foreign Workers",
        "pro": "Effective in industries with significant labor shortages, foreign workers, especially those with highly specialized skills, can contribute to economic growth. Some argue that by bringing in new skills and knowledge, they also contribute to improving Japan's international competitiveness." ,.
        "con": "There are concerns that a larger number of foreign workers will have a negative impact on the domestic labor market, as they may cause cultural friction and encourage low-wage work. Some argue that employment of domestic workers may be squeezed."
    }
    {
        "topic": "Review of monetary easing policy", {
        "pro": "Prolonged monetary easing policy should be carefully reviewed because of concerns about the risk of inflation and increased market credit costs. In particular, price-pushing factors such as higher import costs due to a weaker yen should also be taken into account." ,
        "con": "Some believe that monetary easing policy should be maintained until the economy fully recovers. There is concern that a change in policy at this point could hinder the end of deflation and economic growth."
    },
    {
        "topic": "On the government's aggressive intervention in the economy", {
        "pro": "Some argue that active government intervention is necessary for economic growth, the development of new industries, and regional development, especially efficient administrative and policy support." ,.
        "con": "Some are of the view that excessive government intervention will stifle free competition in the market and create unnecessary regulations. There also exists the opinion that government intervention should be prudent because it may affect the vitality and autonomy of the private sector."
    }
]
\end{lstlisting}

\caption{Extracted axes of political debate for the topic of economy}
\label{fig: extracted axes economy}
\end{figure}

\begin{figure}[h]
\centering
      \begin{lstlisting}[language=json,firstnumber=1]
[
    {
        "topic": "Resuming operations of nuclear power plants",
        "pro": "Nuclear power plants that meet the criteria should be restarted to ensure a stable energy supply and lower electricity prices.Important for achieving carbon neutrality.",
        "con": "Oppose restarting nuclear power plants because of the lessons learned from the Fukushima Daiichi nuclear accident, unresolved safety issues, and inadequate evacuation plans.We call for a transition to renewable energy."
    },
    {
        "topic": "New nuclear power plant construction and extension of operating periods", 
        "pro": "The introduction of new technologies and improvement of safety could contribute to a stable energy supply, and therefore the extension of operating periods should also be considered.",
        "con": "There are concerns about age-related deterioration and increased risk. We should aim for zero nuclear power plants and do not approve new construction or extension of operating periods."
    },
    {
        "topic": "Export of Nuclear Power Plants", 
        "pro": "Exports should be undertaken in order to utilize Japanese nuclear technology in the international market and maintain economic competitiveness." ,
        "con": "There are ethical concerns that risks should not be spread to other countries while safety cannot be ensured even domestically."
    },
    {
        "topic": "Abolition of nuclear power plants and introduction of renewable energy", 
        "pro": "Renewable energy is environmentally friendly and provides a sustainable energy supply. We should shift to this to reduce the risks posed by nuclear power.",
        "con": "Renewable energy alone may not be able to meet short-term demand, and the use of nuclear power should also be considered."
    },
    {
        "topic": "Victim support and compensation issues after the Fukushima Daiichi Nuclear Power Plant accident", 
        "pro": "Prompt and appropriate compensation and support for victims is necessary. The government should continue to take responsibility for the situation." ,
        "con": "It is inappropriate to place the responsibility solely on local governments and individual companies, as the burden should be considered by the entire population who benefited from electricity consumption."
    }
]
\end{lstlisting}

\caption{Extracted axes of political debate for the topic of nuclear power}
\label{fig: extracted axes nuclear power}
\end{figure}

\begin{figure}[h]
\centering
      \begin{lstlisting}[language=json,firstnumber=1]
[
    {
        "topic": "Employment and improving the employment environment to combat the declining birthrate",
        "pro": "Stable employment is important to combat the declining birthrate.
        Creating full-time jobs and raising wages will increase young people's  willingness to marry and raise children.",
        "con": "The employment environment alone will not solve the problem of declining birthrates.Comprehensive support measures are needed, and aspects other than just the economy should be strengthened."
    },
    {
        "topic": "Reducing the burden of education as a solution to the declining birthrate", 
        "pro": "Reducing the burden of education costs is a measure to counteract the declining birthrate.We need to promote free education from early childhood to higher education.",
        "con": "Free education requires a huge amount of financial resources and should be proceeded with caution because of the burden it will place on public finances."
    },
    {
        "topic": "Priorities for Defense Spending and the Budget for Low Fertility", 
        "pro": "The budget should be allocated to measures to combat the declining birthrate rather than defense spending. We need to support childcare for the future of the country." ,
        "con": "Defense spending is essential for security. The security of the nation is the only way to improve the child-rearing environment."
    },
    {
        "topic": "Expanding childcare leave and promoting participation", {
        "pro": "It is important to establish a system that makes it easier for both men and women to take childcare leave. Increased participation in childcare will lead to an increase in the birth rate." ,
        "con": "Expansion of the childcare leave system may become a burden on companies, resulting in a negative impact."
    }
]
\end{lstlisting}

\caption{Extracted axes of political debate for the topic of aging population}
\label{fig: extracted axes aging population}
\end{figure}

\begin{figure}[h]
\centering
      \begin{lstlisting}[language=json,firstnumber=1]
[
    {
        "topic": "Breaking away from dependence on fossil fuels and promoting renewable energy",
        "pro": "We should break away from dependence on fossil fuels and expand the use of renewable energy to the maximum extent possible.
        This will achieve a reduction in greenhouse gases and increase energy self-sufficiency." ,
        "con": "Rapidly reducing fossil fuel dependence is not realistic and may make it difficult to ensure a stable energy supply. The costs and technical constraints of renewable energy also need to be carefully evaluated."
    },
    {
        "topic": "The pros and cons of using nuclear power",
        "pro": "Nuclear power has low CO2 emissions and is important for a stable energy supply. Once its technical safety is ensured, it should be restarted and contribute to decarbonization." ,
        "con": "Dependence on nuclear power should be reduced and the spread of renewable energy should be promoted. Oppose new construction and increased dependence on nuclear power due to the risk of nuclear accidents and waste issues."
    },
    {
        "topic": "Specific measures to achieve the carbon neutrality goal", {
        "pro": "To achieve carbon neutrality, the introduction of renewable energy should be accelerated through technological innovation and government support. Utilization of hydrogen and ammonia and resource recycling are important." ,.
        "con": "Carbon neutrality goals should reflect economic and industrial realities and avoid abrupt shifts. A realistic approach that takes into account immediate economic effects and increased burdens is needed."
    },
    {
        "topic": "The role of local and national governments in combating climate change", {
        "pro": "Local governments should promote local production and local consumption of renewable energy by taking advantage of the characteristics of local resources, and the national government should prepare financial support and policies to encourage this." ,
        "con": "There are limits to what local governments can do to combat climate change on their own, and a unified and comprehensive policy should be promoted under the leadership of the national government. Local initiatives should only play a complementary role."
    }
]
\end{lstlisting}

\caption{Extracted axes of political debate for the topic of climate change}
\label{fig: extracted axes climate change}
\end{figure}

\FloatBarrier
\clearpage
\onecolumn
\section{Original Japanese noun phrases with high PMI for the topics discussed section \ref{section: qual-eval}}

\begin{table*}[htbp]
\begin{CJK}{UTF8}{min}
\centering
\renewcommand{\arraystretch}{1.5}%
\begin{tabularx}{\textwidth}{|>{\centering\arraybackslash}X|>{\centering\arraybackslash}X|>{\centering\arraybackslash}X|}
\hline
\textbf{Segment 1(against)} & \textbf{Segment 2(middle)} & \textbf{Segment 3(for)} \\ \hline
\begin{tabular}[c]{@{}l@{}}
	自然エネルギー重視 \\ 高コスト \\ 原子力艦船 \\ 林業再生 \\ その準備 \\ 教科書 \\ 控訴 \\ 原発関連企業 \\ 秘密主義 \\ 産業事故 \\ 三倍 \\ 賠償スキーム \\ 適切に機能していなかったこと \\ 反原発 \\ 日米同盟 \\ 企業文化 \\ 地震安全対策 \\ 賛成・最近 \\ 核問題 \\ 官邸 \\ 志賀原子力発電所 \\ 疑問を持つ・福島原発 \\ 教訓を無視していると指摘 \\ 情報隠蔽 \\ 復興大臣 \\ 懸念・現行
\end{tabular} 
& 
\begin{tabular}[c]{@{}l@{}}

	利権 \\ 不透明感 \\ 必要な人材 \\ 輸入増加 \\ エネルギー自給率 \\ 裁判 \\ 裁判例 \\ 仮設住宅 \\ 工程表 \\ 原子力施設 \\ 具体的な取り組み \\ 小型原子炉 \\ その他 \\ 産業推進 \\ カーボンニュートラル \\ 汚染水処理 \\ 避難所 \\ 消防団 \\ 地域原子力防災協議会 \\ 浜岡原発 \\ 警察 \\ 一自治体 \\ 拠点 \\ 新しい産業 \\ 地域振興策
\end{tabular} 
& 
\begin{tabular}[c]{@{}l@{}}
	産業政策 \\ やりがい \\ 電力部門 \\ 独自性 \\ 地元企業 \\ 革新原子力技術 \\ 原子力損害賠償制度 \\ CO2排出削減 \\ 廃炉事業 \\ 安全性を最優先 \\ 安全性を最重要 \\ 脱炭素化 \\ 北海道 \\ 原発産業 \\ 依存度を下げる必要 \\ 世界中 \\ 原発運転期間 \\ 廃炉完了 \\ 核廃絶 \\ 小型モジュール炉 \\ 二酸化炭素排出量 \\ 原子力技術者 \\ 選択肢を広げること \\ 地域防災計画 \\ 住民支援
\end{tabular} \\ \hline
\end{tabularx}
\caption{Example of noun phrases that have high PMI with different clusters of representatives for the topic: Restarting Nuclear Power Plants in Japan}
\label{table:PMI-nuclear-jp}
\end{CJK}
\end{table*}

\begin{table*}[htbp]
\begin{CJK}{UTF8}{min}
\centering
\renewcommand{\arraystretch}{1.5}%
\begin{tabularx}{\textwidth}{|>{\centering\arraybackslash}X|>{\centering\arraybackslash}X|>{\centering\arraybackslash}X|}
\hline
\textbf{Segment 1(against)} & \textbf{Segment 2(middle)} & \textbf{Segment 3(for)} \\ \hline
\begin{tabular}[c]{@{}l@{}}
	自然エネルギー重視 \\ 高コスト \\ 原子力艦船 \\ 林業再生 \\ その準備 \\ 教科書 \\ 控訴 \\ 原発関連企業 \\ 秘密主義 \\ 産業事故 \\ 三倍 \\ 賠償スキーム \\ 適切に機能していなかったこと \\ 反原発 \\ 日米同盟 \\ 企業文化 \\ 地震安全対策 \\ 賛成・最近 \\ 核問題 \\ 官邸 \\ 志賀原子力発電所 \\ 疑問を持つ・福島原発 \\ 教訓を無視していると指摘 \\ 情報隠蔽 \\ 復興大臣 \\ 懸念・現行
\end{tabular} 
& 
\begin{tabular}[c]{@{}l@{}}
	有害廃棄物 \\ 土地利用規制法案 \\ この法案 \\ 反対・政府 \\ 財政再建 \\ イラク \\ 慎重・政府 \\ 国際的な平和協力活動 \\ 関係省庁間 \\ 保育 \\ 追加財源 \\ 特別委員会 \\ 青天井 \\ フルスペック \\ ドイツ \\ 憲法九条 \\ 代わり \\ 手法 \\ 公文書 \\ 現政権 \\ 解釈改憲 \\ 反撃能力を持つこと \\ アジア諸国 \\ 憲法解釈変更
\end{tabular} 
& 
\begin{tabular}[c]{@{}l@{}}
	イギリス \\ 無人アセット防衛能力 \\ アナログ通信 \\ 公共交通 \\ 航空機製造産業 \\ 民生部門 \\ 無人装備 \\ 海洋調査 \\ 民生技術 \\ 倫理規程 \\ メタンハイドレート \\ 技術安全保障 \\ 広義 \\ 競争力強化 \\ 同志国 \\ 人間 \\ 多国間連携 \\ 安定確保 \\ 重要物資 \\ ない世界 \\ AI \\ 核抑止 \\ 脆弱性 \\ 武器輸出三原則 \\ 経済活性化 \\ 海洋資源 \\ 詳しくない方 \\ 生産能力 \\ 核開発
\end{tabular} \\ \hline
\end{tabularx}
\caption{Example of noun phrases that have high PMI with different clusters of representatives for the topic: Changing the Japanese constitution to acknowledge the JSDF}
\label{table:PMI-constitution-jp}
\end{CJK}
\end{table*}

\begin{table*}[htbp]
\begin{CJK}{UTF8}{min}
\centering
\renewcommand{\arraystretch}{1.5}%
\begin{tabularx}{\textwidth}{|>{\centering\arraybackslash}X|>{\centering\arraybackslash}X|>{\centering\arraybackslash}X|}
\hline
\textbf{Segment 1(against)} & \textbf{Segment 2(middle)} & \textbf{Segment 3(for)} \\ \hline
\begin{tabular}[c]{@{}l@{}}
	公共放送 \\ 公立学校 \\ ケア労働 \\ 生活資金 \\ 雇用制度改革 \\ 若い船員 \\ 駐留軍 \\ 調達制度 \\ 就労確保 \\ 軍事優先 \\ 同等以上 \\ 造船業 \\ 土地取得 \\ 防衛産業 \\ ケア \\ 防衛省 \\ 専門職 \\ アフリカ \\ 医療従事者 \\ 財政安定 \\ 丁寧な説明 \\ 国土交通省 \\ 信用保証制度 \\ 投資協定 \\ 医療機関 \\ カジノ解禁 \\ 共生社会 \\ 男性 \\ 災害復旧
\end{tabular} 
& 
\begin{tabular}[c]{@{}l@{}}
	給付型奨学金 \\ 賛成・国際情勢 \\ 労働契約法 \\ 技術移転 \\ 地方都市 \\ 経済安保 \\ 問題解決 \\ 賛成・地方経済 \\ 中小零細企業 \\ 貧困問題 \\ ふるさと納税制度 \\ ユニバーサルサービス \\ 受動喫煙 \\ 定額給付金 \\ 国産ワクチン \\ 民間主導 \\ その際 \\ 低所得者支援 \\ 私立高校 \\ 運転手 \\ 国際的な協力 \\ 感染症法 \\ 与党 \\ 環境再生 \\ 地域特性 \\ 補正予算案 \\ ウクライナ情勢
\end{tabular} 
& 
\begin{tabular}[c]{@{}l@{}}
	アニマルウエルフェア \\ そこ \\ 多子加算 \\ 若者・子育て世代 \\ 庶民 \\ 国内消費 \\ ジビエ \\ 需要拡大 \\ 輸入価格 \\ 食品表示 \\ 資産デフレ \\ カーボンプライシング \\ エネルギーコスト上昇 \\ 発言をもと \\ 新たなビジネスモデル \\ 燃料価格 \\ 両面 \\ 消費行動 \\ 消費者物価 \\ アフターコロナ \\ 事業承継税制 \\ 両者 \\ 官民連携 \\ 企業間 \\ 発言を基 \\ 融資制度 \\ メリット \\ エネルギーコスト \\ 中堅企業
\end{tabular} \\ \hline
\end{tabularx}
\caption{Example of noun phrases that have high PMI with different clusters of representatives for the topic: Reducing the consumption tax rate}
\label{table:PMI-taxrate-jp}
\end{CJK}
\end{table*}

\clearpage
\twocolumn

\section{Statistics related to the diachronic analysis}
\begin{table}[h]
	\begin{CJK}{UTF8}{min}

    \centering
    \begin{tabular}{c|ccc}
        \hline
        Year & JCP & LDP & Komeito \\
        \hline
        2000 & 59 / 6 & 159 / 18 & 34 / 7 \\
        2001 & 3 / 2 & 55 / 16 & 11 / 4 \\
        2002 & 77 / 5 & 122 / 32 & 33 / 7 \\
        2003 & 24 / 3 & 63 / 22 & 14 / 6 \\
        2004 & 18 / 4 & 131 / 30 & 23 / 6 \\
        2005 & 22 / 4 & 170 / 25 & 17 / 3 \\
        2006 & 19 / 6 & 125 / 21 & 10 / 4 \\
        2007 & 50 / 8 & 221 / 34 & 19 / 8 \\
        2008 & 32 / 7 & 121 / 29 & 9 / 5 \\
        2009 & 13 / 3 & 114 / 29 & 12 / 6 \\
        2010 & 49 / 7 & 79 / 26 & 29 / 4 \\
        2011 & 565 / 15 & 1864 / 144 & 415 / 30 \\
        2012 & 314 / 14 & 1199 / 84 & 144 / 22 \\
        2013 & 273 / 16 & 1204 / 106 & 146 / 24 \\
        2014 & 374 / 18 & 1416 / 104 & 223 / 30 \\
        2015 & 318 / 23 & 764 / 78 & 121 / 19 \\
        2016 & 274 / 20 & 634 / 74 & 75 / 17 \\
        2017 & 408 / 22 & 733 / 60 & 106 / 20 \\
        2018 & 247 / 17 & 400 / 49 & 63 / 16 \\
        2019 & 166 / 12 & 415 / 53 & 24 / 13 \\
        2020 & 113 / 14 & 339 / 58 & 47 / 16 \\
        2021 & 191 / 14 & 548 / 57 & 58 / 11 \\
        2022 & 166 / 14 & 636 / 74 & 74 / 17 \\
        2023 & 256 / 13 & 883 / 75 & 119 / 18 \\
        \hline
    \end{tabular}
    \caption{Number of Speech Segments / Number of Representatives on Nuclear Power}
    \label{tab:nuclear_speech}
	\end{CJK}
\end{table}

\begin{table}[h]
\begin{CJK}{UTF8}{min}

    \centering
    \begin{tabular}{c|ccc}
        \hline
        Year & JCP & LDP & Komeito \\
        \hline
        2000 & 40 / 8 & 503 / 42 & 10 / 5 \\
        2001 & 207 / 18 & 1231 / 54 & 77 / 9 \\
        2002 & 141 / 14 & 1111 / 64 & 101 / 12 \\
        2003 & 210 / 14 & 1930 / 64 & 222 / 13 \\
        2004 & 227 / 14 & 2115 / 99 & 201 / 19 \\
        2005 & 116 / 13 & 1727 / 98 & 113 / 17 \\
        2006 & 250 / 14 & 2072 / 99 & 208 / 21 \\
        2007 & 234 / 16 & 1764 / 99 & 136 / 18 \\
        2008 & 150 / 11 & 1232 / 93 & 78 / 20 \\
        2009 & 139 / 11 & 976 / 91 & 74 / 16 \\
        2010 & 98 / 11 & 826 / 87 & 139 / 15 \\
        2011 & 47 / 9 & 816 / 80 & 152 / 17 \\
        2012 & 73 / 12 & 750 / 86 & 97 / 17 \\
        2013 & 145 / 13 & 1874 / 149 & 150 / 28 \\
        2014 & 291 / 17 & 2549 / 127 & 222 / 29 \\
        2015 & 477 / 22 & 3541 / 136 & 421 / 29 \\
        2016 & 245 / 23 & 1113 / 102 & 71 / 25 \\
        2017 & 311 / 19 & 1455 / 110 & 104 / 26 \\
        2018 & 152 / 18 & 1429 / 121 & 105 / 24 \\
        2019 & 138 / 20 & 1245 / 97 & 59 / 15 \\
        2020 & 91 / 16 & 865 / 94 & 72 / 24 \\
        2021 & 107 / 14 & 1157 / 126 & 125 / 29 \\
        2022 & 214 / 20 & 2721 / 177 & 231 / 38 \\
        2023 & 255 / 20 & 2913 / 184 & 324 / 40 \\
        \hline
    \end{tabular}
    \caption{Number of Speech Segments / Number of Representatives on Defense}
    \label{tab:defense_speech}
	\end{CJK}
\end{table}

\FloatBarrier
\section{Party Names and Abbreviations}
\begin{CJK}{UTF8}{min}
\begin{table}[htbp] 
\centering
\caption{Party names and abbreviations}\label{tab:Parties and abbrev}
\begin{tabularx}{.45\textwidth}{ c|X|X } 
\hline
 Party Name(JP) & Party Name(EN) & Abbreviation\\  
 \hline
  自由民主党  & Liberal Democratic Party & LDP \\
  \hline
  国民民主党  & National Democratic Party & NDP \\
  \hline
  立憲民主党  & Constitutional Democratic Party & CDP \\
  \hline
  日本共産党  & Japanese Communist Party & JCP \\
  \hline
  公明党  & Komeito & Komeito \\
   \hline
  日本維新の会  & Japan Restoration Party & JRP \\
  \hline
  れいわ新選組  & Reiwa Shinsengumi & Reiwa \\
  \hline
  日本保守党  & Conservative Party of Japan & CPJ \\
  \hline
  参政党  & Sanseito & Sanseito \\
  \hline
  
\end{tabularx}
\end{table}
\end{CJK}

\clearpage
\bibliographystyle{elsarticle-harv} 
\bibliography{references}

\begin{thebibliography}{25}
\expandafter\ifx\csname natexlab\endcsname\relax\def\natexlab#1{#1}\fi
\providecommand{\url}[1]{\texttt{#1}}
\providecommand{\href}[2]{#2}
\providecommand{\path}[1]{#1}
\providecommand{\DOIprefix}{doi:}
\providecommand{\ArXivprefix}{arXiv:}
\providecommand{\URLprefix}{URL: }
\providecommand{\Pubmedprefix}{pmid:}
\providecommand{\doi}[1]{\href{http://dx.doi.org/#1}{\path{#1}}}
\providecommand{\Pubmed}[1]{\href{pmid:#1}{\path{#1}}}
\providecommand{\bibinfo}[2]{#2}
\ifx\xfnm\relax \def\xfnm[#1]{\unskip,\space#1}\fi
\bibitem[{Catalinac(2018)}]{CATALINAC_2018}
\bibinfo{author}{Catalinac, A.}, \bibinfo{year}{2018}.
\newblock \bibinfo{title}{Positioning under alternative electoral systems:
  Evidence from japanese candidate election manifestos}.
\newblock \bibinfo{journal}{American Political Science Review}
  \bibinfo{volume}{112}, \bibinfo{pages}{31–48}.
\newblock \DOIprefix\doi{10.1017/S0003055417000399}.
\bibitem[{Church and Hanks(1990)}]{church-hanks-1990-word}
\bibinfo{author}{Church, K.W.}, \bibinfo{author}{Hanks, P.},
  \bibinfo{year}{1990}.
\newblock \bibinfo{title}{Word association norms, mutual information, and
  lexicography}.
\newblock \bibinfo{journal}{Computational Linguistics} \bibinfo{volume}{16},
  \bibinfo{pages}{22--29}.
\newblock \URLprefix \url{https://aclanthology.org/J90-1003/}.
\bibitem[{Curini et~al.(2020)Curini, Hino and Osaka}]{Curini_Hino_Osaka_2020}
\bibinfo{author}{Curini, L.}, \bibinfo{author}{Hino, A.},
  \bibinfo{author}{Osaka, A.}, \bibinfo{year}{2020}.
\newblock \bibinfo{title}{The intensity of government–opposition divide as
  measured through legislative speeches and what we can learn from it: Analyses
  of japanese parliamentary debates, 1953–2013}.
\newblock \bibinfo{journal}{Government and Opposition} \bibinfo{volume}{55},
  \bibinfo{pages}{184–201}.
\newblock \DOIprefix\doi{10.1017/gov.2018.15}.
\bibitem[{Diaf et~al.(2022)Diaf, Döpke, Fritsche and
  Rockenbach}]{sharks-minnows}
\bibinfo{author}{Diaf, S.}, \bibinfo{author}{Döpke, J.},
  \bibinfo{author}{Fritsche, U.}, \bibinfo{author}{Rockenbach, I.},
  \bibinfo{year}{2022}.
\newblock \bibinfo{title}{Sharks and minnows in a shoal of words: Measuring
  latent ideological positions based on text mining techniques}.
\newblock \bibinfo{journal}{European Journal of Political Economy}
  \bibinfo{volume}{75}, \bibinfo{pages}{102179}.
\newblock \DOIprefix\doi{https://doi.org/10.1016/j.ejpoleco.2022.102179}.
\bibitem[{Kato et~al.(2024)Kato, Purnomo, Cochrane and
  Saqur}]{kato2024lupinllmbasedpoliticalideology}
\bibinfo{author}{Kato, K.}, \bibinfo{author}{Purnomo, A.},
  \bibinfo{author}{Cochrane, C.}, \bibinfo{author}{Saqur, R.},
  \bibinfo{year}{2024}.
\newblock \bibinfo{title}{L(u)pin: Llm-based political ideology nowcasting}.
\newblock \URLprefix \url{https://arxiv.org/abs/2405.07320},
  \href{http://arxiv.org/abs/2405.07320}{{\tt arXiv:2405.07320}}.
\bibitem[{KENDALL(1938)}]{kendall1938}
\bibinfo{author}{KENDALL, M.G.}, \bibinfo{year}{1938}.
\newblock \bibinfo{title}{A new measure of rank correlation}.
\newblock \bibinfo{journal}{Biometrika} \bibinfo{volume}{30},
  \bibinfo{pages}{81--93}.
\newblock \URLprefix \url{https://doi.org/10.1093/biomet/30.1-2.81},
  \DOIprefix\doi{10.1093/biomet/30.1-2.81},
  \href{http://arxiv.org/abs/https://academic.oup.com/biomet/article-pdf/30/1-2/81/423380/30-1-2-81.pdf}{{\tt
  arXiv:https://academic.oup.com/biomet/article-pdf/30/1-2/81/423380/30-1-2-81.pdf}}.
\bibitem[{van Lange and
  Futselaar(2019)}]{Debating-Evil:-Using-Word-Embeddings-to-Analyze-Parliamentary-Debates}
\bibinfo{author}{van Lange, M.}, \bibinfo{author}{Futselaar, R.},
  \bibinfo{year}{2019}.
\newblock \bibinfo{title}{Debating evil: Using word embeddings to analyze
  parliamentary debates on war criminals in the netherlands}.
\newblock \bibinfo{journal}{Contributions to Contemporary History}
  \bibinfo{volume}{59}, \bibinfo{pages}{140--156}.
\newblock \DOIprefix\doi{10.51663/pnz.59.1.07}.
\bibitem[{Laver et~al.(2003)Laver, Benoit and Garry}]{LAVER_BENOIT_GARRY_2003}
\bibinfo{author}{Laver, M.}, \bibinfo{author}{Benoit, K.},
  \bibinfo{author}{Garry, J.}, \bibinfo{year}{2003}.
\newblock \bibinfo{title}{Extracting policy positions from political texts
  using words as data}.
\newblock \bibinfo{journal}{American Political Science Review}
  \bibinfo{volume}{97}, \bibinfo{pages}{311–331}.
\newblock \DOIprefix\doi{10.1017/S0003055403000698}.
\bibitem[{Li et~al.(2024)Li, Li, Chen, Gui, Yang, Yu, Wang, Cai, Zhou, Shen,
  Qian, Chen, Xue, Sun, He, Chen, Ding, Du, Mu, Pei, Zhao, Swayamdipta,
  Neiswanger, Wei, Hu, Zhu, Chen, Lu, Shi, Qin, Fu, Tu, Yang, Yoo, Zhang,
  Rossi, Zhan, Zhao, Ferrara, Liu, Huang, Zhang, Rothenberg, Ji, Yu, Zhao and
  Dong}]{li2024politicalllmlargelanguagemodels}
\bibinfo{author}{Li, L.}, \bibinfo{author}{Li, J.}, \bibinfo{author}{Chen, C.},
  \bibinfo{author}{Gui, F.}, \bibinfo{author}{Yang, H.}, \bibinfo{author}{Yu,
  C.}, \bibinfo{author}{Wang, Z.}, \bibinfo{author}{Cai, J.},
  \bibinfo{author}{Zhou, J.A.}, \bibinfo{author}{Shen, B.},
  \bibinfo{author}{Qian, A.}, \bibinfo{author}{Chen, W.}, \bibinfo{author}{Xue,
  Z.}, \bibinfo{author}{Sun, L.}, \bibinfo{author}{He, L.},
  \bibinfo{author}{Chen, H.}, \bibinfo{author}{Ding, K.}, \bibinfo{author}{Du,
  Z.}, \bibinfo{author}{Mu, F.}, \bibinfo{author}{Pei, J.},
  \bibinfo{author}{Zhao, J.}, \bibinfo{author}{Swayamdipta, S.},
  \bibinfo{author}{Neiswanger, W.}, \bibinfo{author}{Wei, H.},
  \bibinfo{author}{Hu, X.}, \bibinfo{author}{Zhu, S.}, \bibinfo{author}{Chen,
  T.}, \bibinfo{author}{Lu, Y.}, \bibinfo{author}{Shi, Y.},
  \bibinfo{author}{Qin, L.}, \bibinfo{author}{Fu, T.}, \bibinfo{author}{Tu,
  Z.}, \bibinfo{author}{Yang, Y.}, \bibinfo{author}{Yoo, J.},
  \bibinfo{author}{Zhang, J.}, \bibinfo{author}{Rossi, R.},
  \bibinfo{author}{Zhan, L.}, \bibinfo{author}{Zhao, L.},
  \bibinfo{author}{Ferrara, E.}, \bibinfo{author}{Liu, Y.},
  \bibinfo{author}{Huang, F.}, \bibinfo{author}{Zhang, X.},
  \bibinfo{author}{Rothenberg, L.}, \bibinfo{author}{Ji, S.},
  \bibinfo{author}{Yu, P.S.}, \bibinfo{author}{Zhao, Y.},
  \bibinfo{author}{Dong, Y.}, \bibinfo{year}{2024}.
\newblock \bibinfo{title}{Political-llm: Large language models in political
  science}.
\newblock \URLprefix \url{https://arxiv.org/abs/2412.06864},
  \href{http://arxiv.org/abs/2412.06864}{{\tt arXiv:2412.06864}}.
\bibitem[{Linegar et~al.(2023)Linegar, Kocielnik and
  Alvarez}]{LLMs-and-political-science}
\bibinfo{author}{Linegar, M.}, \bibinfo{author}{Kocielnik, R.},
  \bibinfo{author}{Alvarez, R.M.}, \bibinfo{year}{2023}.
\newblock \bibinfo{title}{Large language models and political science}.
\newblock \bibinfo{journal}{Frontiers in Political Science}
  \bibinfo{volume}{5}.
\newblock \URLprefix
  \url{https://www.frontiersin.org/articles/10.3389/fpos.2023.1257092},
  \DOIprefix\doi{10.3389/fpos.2023.1257092}.
\bibitem[{McInnes et~al.(2018)McInnes, Healy, Saul and
  Grossberger}]{mcinnes2018umap-software}
\bibinfo{author}{McInnes, L.}, \bibinfo{author}{Healy, J.},
  \bibinfo{author}{Saul, N.}, \bibinfo{author}{Grossberger, L.},
  \bibinfo{year}{2018}.
\newblock \bibinfo{title}{Umap: Uniform manifold approximation and projection}.
\newblock \bibinfo{journal}{The Journal of Open Source Software}
  \bibinfo{volume}{3}, \bibinfo{pages}{861}.
\bibitem[{Mens and Gallego(2023)}]{Scaling-Political-Texts-with-ChatGPT}
\bibinfo{author}{Mens, G.L.}, \bibinfo{author}{Gallego, A.},
  \bibinfo{year}{2023}.
\newblock \bibinfo{title}{Scaling political texts with chatgpt}.
\newblock \href{http://arxiv.org/abs/2311.16639}{{\tt arXiv:2311.16639}}.
\bibitem[{{Mielka}(2024)}]{Mielka}
\bibinfo{author}{{Mielka}}, \bibinfo{year}{2024}.
\newblock \bibinfo{title}{{Policy Comparison}}.
\newblock \URLprefix \url{https://japanchoice.jp/policy-comparison}.
\bibitem[{Nanni et~al.(2021)Nanni, Glavas, Rehbein, Ponzetto and
  Stuckenschmidt}]{Political-Text-Scaling-Meets-Computational-Semantics}
\bibinfo{author}{Nanni, F.}, \bibinfo{author}{Glavas, G.},
  \bibinfo{author}{Rehbein, I.}, \bibinfo{author}{Ponzetto, S.P.},
  \bibinfo{author}{Stuckenschmidt, H.}, \bibinfo{year}{2021}.
\newblock \bibinfo{title}{Political text scaling meets computational
  semantics}.
\newblock \href{http://arxiv.org/abs/1904.06217}{{\tt arXiv:1904.06217}}.
\bibitem[{{National Diet Library}(2019)}]{kokkai_diet_api}
\bibinfo{author}{{National Diet Library}}, \bibinfo{year}{2019}.
\newblock \bibinfo{title}{Diet proceedings search system api specifications for
  search}.
\newblock \bibinfo{howpublished}{\url{https://kokkai.ndl.go.jp/api.html}}.
\newblock \URLprefix \url{https://kokkai.ndl.go.jp/api.html}.
  \bibinfo{note}{accessed: 2025-01-15}.
\bibitem[{Poole and Rosenthal(1985)}]{spatial-model-for-legislative-roll-call}
\bibinfo{author}{Poole, K.T.}, \bibinfo{author}{Rosenthal, H.},
  \bibinfo{year}{1985}.
\newblock \bibinfo{title}{A spatial model for legislative roll call analysis}.
\newblock \bibinfo{journal}{American Journal of Political Science}
  \bibinfo{volume}{29}, \bibinfo{pages}{357--384}.
\newblock \URLprefix \url{http://www.jstor.org/stable/2111172}.
\bibitem[{Poole and Rosenthal(2001)}]{dw-nominate}
\bibinfo{author}{Poole, K.T.}, \bibinfo{author}{Rosenthal, H.},
  \bibinfo{year}{2001}.
\newblock \bibinfo{title}{D-nominate after 10 years: A comparative update to
  congress: A political-economic history of roll-call voting}.
\newblock \bibinfo{journal}{Legislative Studies Quarterly}
  \bibinfo{volume}{26}, \bibinfo{pages}{5--29}.
\newblock \URLprefix \url{http://www.jstor.org/stable/440401}.
\bibitem[{Pujari and
  Goldwasser(2021)}]{Understanding-Politics-via-Contextualized}
\bibinfo{author}{Pujari, R.}, \bibinfo{author}{Goldwasser, D.},
  \bibinfo{year}{2021}.
\newblock \bibinfo{title}{Understanding politics via contextualized discourse
  processing}.
\newblock \href{http://arxiv.org/abs/2012.15784}{{\tt arXiv:2012.15784}}.
\bibitem[{Rashed et~al.(2022)Rashed, Kutlu, Darwish, Elsayed and
  Bayrak}]{Embeddings-Based-Clustering-for-Target-Specific-Stances}
\bibinfo{author}{Rashed, A.}, \bibinfo{author}{Kutlu, M.},
  \bibinfo{author}{Darwish, K.}, \bibinfo{author}{Elsayed, T.},
  \bibinfo{author}{Bayrak, C.}, \bibinfo{year}{2022}.
\newblock \bibinfo{title}{Embeddings-based clustering for target specific
  stances: The case of a polarized turkey}.
\newblock \href{http://arxiv.org/abs/2005.09649}{{\tt arXiv:2005.09649}}.
\bibitem[{Reimers and Gurevych(2019)}]{reimers2019sentencebert}
\bibinfo{author}{Reimers, N.}, \bibinfo{author}{Gurevych, I.},
  \bibinfo{year}{2019}.
\newblock \bibinfo{title}{Sentence-bert: Sentence embeddings using siamese
  bert-networks}.
\newblock \href{http://arxiv.org/abs/1908.10084}{{\tt arXiv:1908.10084}}.
\bibitem[{Rheault and
  Cochrane(2020)}]{Word-embeddings-for-analysis-of-ideological-placement}
\bibinfo{author}{Rheault, L.}, \bibinfo{author}{Cochrane, C.},
  \bibinfo{year}{2020}.
\newblock \bibinfo{title}{Word embeddings for the analysis of ideological
  placement in parliamentary corpora}.
\newblock \bibinfo{journal}{Political Analysis} \bibinfo{volume}{28},
  \bibinfo{pages}{112–133}.
\newblock \DOIprefix\doi{10.1017/pan.2019.26}.
\bibitem[{Slapin and Proksch(2008)}]{wordfishing}
\bibinfo{author}{Slapin, J.B.}, \bibinfo{author}{Proksch, S.O.},
  \bibinfo{year}{2008}.
\newblock \bibinfo{title}{A scaling model for estimating time-series party
  positions from texts}.
\newblock \bibinfo{journal}{American Journal of Political Science}
  \bibinfo{volume}{52}, \bibinfo{pages}{705--722}.
\newblock \URLprefix \url{http://www.jstor.org/stable/25193842}.
\bibitem[{Spearman(1904)}]{spearman1904}
\bibinfo{author}{Spearman, C.}, \bibinfo{year}{1904}.
\newblock \bibinfo{title}{The proof and measurement of association between two
  things}.
\newblock \bibinfo{journal}{American Journal of Psychology}
  \bibinfo{volume}{15}, \bibinfo{pages}{72--101}.
\newblock \DOIprefix\doi{10.2307/1412159}.
\bibitem[{{Tohoku NLP Group}(2023)}]{Tohoku}
\bibinfo{author}{{Tohoku NLP Group}}, \bibinfo{year}{2023}.
\newblock \bibinfo{title}{tohoku-nlp/bert-base-japanese-v3}.
\newblock \URLprefix
  \url{https://huggingface.co/tohoku-nlp/bert-base-japanese-v3}.
\bibitem[{Wu et~al.(2023)Wu, Nagler, Tucker and
  Messing}]{llm-latent-position-of-politicians}
\bibinfo{author}{Wu, P.Y.}, \bibinfo{author}{Nagler, J.},
  \bibinfo{author}{Tucker, J.A.}, \bibinfo{author}{Messing, S.},
  \bibinfo{year}{2023}.
\newblock \bibinfo{title}{Large language models can be used to estimate the
  latent positions of politicians}.
\newblock \href{http://arxiv.org/abs/2303.12057}{{\tt arXiv:2303.12057}}.

\end{thebibliography}
\end{document}